\documentclass[12pt]{article}
\usepackage{fullpage,amsmath,amssymb,mathtools,natbib,hyperref}
\usepackage{titling,enumitem}
\usepackage[skip=0pt]{caption} 
\usepackage[T1]{fontenc}
\usepackage{setspace}
\usepackage{datetime,bigints}
\usepackage[dvipsnames]{xcolor}
\usepackage[most]{tcolorbox}
\usdate
\usepackage{multirow} 
\usepackage{float}
\usepackage[boxruled]{algorithm2e}
\usepackage{arydshln}
\usepackage[section]{placeins}
\usepackage[titletoc,toc,title]{appendix}
\usepackage{fancyvrb} 
\usepackage{listings,enumitem}
\usepackage{bm}
\usepackage{xcolor,colortbl} 
\newcommand{\notes}[1]{%
    \linespread{0.2}\vspace{0.1em}%
    \captionsetup{justification=justified}%
    \caption*{\footnotesize #1}%
}

\definecolor{lightgray}{gray}{0.9}
\xdefinecolor{gray}{rgb}{0.8,0.8,0.8}
\xdefinecolor{blue}{RGB}{58,95,205}% R's royalblue3; #3A5FCD
\DeclareCaptionFormat{listing}{\rule{\dimexpr\textwidth+17pt\relax}{0.4pt}\vskip1pt#1#2#3}
\definecolor{gray}{gray}{0.85}
\definecolor{LightCyan}{rgb}{0.88,1,1}

\newcolumntype{a}{>{\columncolor{gray}}c}
\newcolumntype{b}{>{\columncolor{white}}c}

\usepackage{xr,caption}
\renewenvironment{thebibliography}[1]{
  \begin{oldthebibliography}{#1}
    \setlength{\itemsep}{0em}
    \setlength{\parskip}{0em}
}
{
  \end{oldthebibliography}
}

\newcommand{\argmin}{\text{argmin}}

\usepackage{thmtools} 
\usepackage{amsthm}
{
      \theoremstyle{plain}
      
      \newtheorem{theorem}{Theorem}
      \newtheorem{example}{Example}
      \newtheorem{proposition}{Proposition}
      \newtheorem{lemma}{Lemma}
      
      \newtheorem{assumption}{Assumption}
      
  }

\renewcommand{\arraystretch}{1.5}

\def\lQ{\scalebox{-1}[1]{''}}

\makeatletter
\renewenvironment{abstract}{%
    \if@twocolumn
      \section*{\abstractname}%
    \else %% <- here I've removed \small
      \begin{center}%
        {\bfseries \normalsize\abstractname\vspace{\z@}}%  %% <- here I've added \Large
      \end{center} \vspace{-0.5cm}%
      \quotation
    \fi}
    {\if@twocolumn\else\endquotation\fi}
\makeatother

\begin{document}

  %\begin{frontmatter}
  \title{Convexity Not Required:\\ Estimation of Smooth Moment Condition Models}
  \author{ Jean-Jacques Forneron\thanks{Department of Economics, Boston University, 270 Bay State Road, Boston, MA 02215 USA.\newline Email: \href{mailto:jjmf@bu.edu}{jjmf@bu.edu}, Website: \href{http://jjforneron.com}{http://jjforneron.com}. } \and Liang Zhong\thanks{Faculty of Business and Economics, The University of Hong Kong, Pokfulam Road, Hong Kong.\newline Email: \href{samzl@hku.hk}{samzl@hku.hk}, Website: \href{https://samzl1.github.io/}{https://samzl1.github.io/}. \newline This paper was written while the second author was a doctoral student at Boston University.
  The authors would like to thank Jessie Li for suggesting to look at misspecified models and Hiro Kaido, David Lakagos, Bernard Salani\'e and participants at the NY Camp Econometrics Conference for useful comments.} 
  } \date{\today}
  \maketitle 

  \begin{abstract}  %\normalsize
    Generalized and Simulated Method of Moments are often used to estimate structural Economic models. Yet, it is commonly reported that optimization is challenging because the corresponding objective function is non-convex. For smooth problems, this paper shows that convexity is not required: under conditions involving the Jacobian of the moments, certain algorithms are globally convergent. These include a gradient-descent and a Gauss-Newton algorithm with appropriate choice of tuning parameters. The results are robust to 1) non-convexity, 2) one-to-one moderately non-linear reparameterizations, and 3) moderate misspecification. The conditions preclude non-global optima. Numerical and empirical examples illustrate the condition, non-convexity, and convergence properties of different optimizers.
  \end{abstract}
  

  \baselineskip=18.0pt
  \thispagestyle{empty}
  \setcounter{page}{0}
  
  %\newpage \tableofcontents
\newpage

\section{Introduction}
The Generalized and Simulated Method of Moments (GMM, SMM) are commonly used to estimate structural Economic models. To find estimates, modern computer software provides researchers with a large set of free and non-free numerical optimizers, which, after inputting some tuning parameters, return a guess for the parameters of interest. While sampling properties of estimators are often derived, their practical implementation often receives a less detailed treatment. There is now a vast literature on statistical learning with a convex loss function, using stochastic gradient-descent. However, these results need not directly apply to GMM, as it often involves non-convex minimizations. 
A number of authors have pointed out the lack of robustness of off-the-shelf methods, and \citet{knittel2014} illustrate this in the context of demand estimation. This is perhaps not surprising since non-convex optimization is subject to a curse of dimensionality \citep[][Section 2]{andrews1997} and becomes increasingly challenging when the number of parameters is moderate or large.

The main contribution of the paper is to show that \textit{convexity is not required} for some methods to perform well in GMM estimation specifically: some algorithms are globally convergent under a global rank condition involving the Jacobian of the moments and the weighting matrix. This defines a class of non-convex problems that is as hard as convex problems for optimization. Since this is perhaps surprising, the following gives some intuition behind the result. Given sample moments $\overline{g}_n(\theta)$ with Jacobian $G_n(\theta)$, one can minimize the GMM objective function $Q_n(\theta) = 1/2 \overline{g}_n(\theta)^\prime W_n \overline{g}_n(\theta)$ iteratively, by minimizing successive quadratic approximations. To this end, convex optimizers rely on a quadratic expansion of $Q_n(\theta)$ using its gradient and Hessian. This quadratic approximation yields a proper minimization problem only if the Hessian is strictly positive definite, i.e. $Q_n$ is convex.%The Hessian can be singular, or non-definite, depending on the last term.\footnote{$\otimes$ is the kronecker product, $\text{vec}$ vectorizes the matrix to a column vector, $\partial_\theta \text{vec}( G_n^\prime (\theta) )$ is the Jacobian of the vectorized Jacobian. } When this is the case, these methods can be unstable.

Another approach, is to linearly expand the sample moments using the Jacobian and plug the linearized moments into the GMM objective. Since the approximate moments are linear, this yields a proper minimization problem as long as $G_n$ has full rank. Gauss-Newton (\textsc{gn}) relies on this approach. Gradient-descent can be motivated by either the quadratic of linear approximation. In the just-identified case, it is well known that \textsc{gn} is \textit{locally convergent} when $G_n$ has full rank around the solution. This paper goes further by showing that \textsc{gn} and gradient-descent are \textit{globally convergent} when the product of $G_n$, $W_n$, and an average of $G_n$ has full rank everywhere. Unlike existing results, this applies to just and over-identified moments. The condition can be relaxed for the product to only be non-singular in a specific direction, towards the global minimizer. Under this weaker condition, \textsc{gn} with a Levenberg-Marquardt regularization and gradient descent are globally convergent. 
Importantly, for correctly specified models, the conditions imply that there are no local optima, besides the global minimizer; a necessary condition for global convergence of gradient-based optimizers. It is shown that these convergence results are robust to 1) moderate misspecification, and 2) moderately non-linear reparameterizations. However, the results may or may not hold depending on the choice of weighting matrix. In particular, when $W_n$ is ill-conditioned, convergence can be significantly slower.

Several conditions found in the convex and non-convex optimization literatures imply the weaker condition introduced of this paper. These include strong, star, and quasar convexity of the objective function. It also relates to the Polyak-Łojasiewicz condition, an important inequality which has gathered much interest in machine learning to prove convergence of gradient-descent. Strong monotonicity of the moments, a condition for solving just-determined system of non-linear equations, and strong injectivity, introduced here for just and over-identified models, also imply the weaker condition. Hence, the condition introduced in this paper is a common denominator of several existing conditions. In terms of econometrics properties, the conditions are sufficient for the parameters to be both locally and globally identified, when the model is correct or moderately misspecified.

A simple MA(1) estimation from \citet{gourieroux1996} illustrates the results analytically and numerically. The problem is non-convex: the scalar Hessian can be positive, negative, or zero; yet the conditions hold. As predicted, the recommended \textsc{gn} algorithm converges. Newton-Raphson provably diverges, and off-the-shelf optimizers can be unstable. When the model is moderately misspecified, \textsc{gn} remains globally convergent. In line with theory, significant misspecification can produce non-global optima which hinder the global convergence of gradient-descent and Gauss-Newton. 

Two empirical applications further illustrate the results. The first application revisits the numerical results of \citet{knittel2014} for estimating random coefficient demand models on Nevo's generated cereal data. The same \textsc{gn} algorithm systematically converges from a wide range of starting values. 
In contrast, R's more sophisticated built-in optimizers can be inaccurate and often crash without additional error-handling. The second application estimates a small New Keynesian model with endogenous total factor productivity by impulse response matching. Matlab's built-in optimizers have better error-handling so that crashes are less problematic. Nonetheless, these optimizers' performance can be mixed whereas \textsc{gn} performs well for nearly all starting values. 

Numerically, in all three applications, the GMM objective is non-convex at most values. The strong injectivity condition holds at most values, an indication that \textsc{gn} and gradient-descent are appropriate. The later converges very slowly, however.  These findings explain the good performance of \textsc{gn} relative to more commonly used methods. 
The main takeaway is that non-convexity need not be a deterrent to structural estimation: simple algorithms can converge quickly and globally under alternative conditions. %Should the rank condition fail, \citet{Forneron2022b} builds an algorithm that is globally convergent under standard econometric assumptions and allows for non-smooth sample moments. %A finite-sample analysis provides global convergence rates, derived using only standard econometric assumptions.

\paragraph{Structure of the paper.}  Section \ref{sec:Main} contains the main assumptions and results. Section \ref{sec:lit_char} reviews existing conditions found in the literature and relates the main assumptions with these conditions. Section \ref{sec:recs} suggests a numerical procedure to check whether the main assumption holds or not and a way to set the tuning parameter. Section \ref{sec:empirics} illustrates the results with one numerical and two empirical applications. Appendices \ref{apx:proofs} and \ref{apx:lit_char} give the proofs to the main results and additional results. The Supplemental Material consists of: Appendices \ref{apx:local}-\ref{apx:optizs}. Appendix \ref{apx:local} provides additional local convergence results, which complement the main global convergence results in the paper. Appendix \ref{sec:survey_properties} provides of survey of empirical practice in the American Economic Review between 2016 and 2018.  Appendix \ref{apx:Rcode} gives R code to replicate the numerical MA(1) example. Appendix \ref{apx:addise} provides additional simulation and empirical results. Appendix \ref{apx:optizs} gives additional details about the methods found in the survey of Appendix \ref{sec:survey_properties}.

\paragraph{Notation:} In the following $\lambda_{\min}$, $\lambda_{\max}$ return the smallest and largest eigenvalues of a square positive semidefinite matrix. For an arbitrary rectangular matrix $A$ of size $d_g \times d_\theta$ with $d_g \geq d_\theta$,  $\sigma_{\min}$, $\sigma_{\max}$ are the smallest and largest singular values of $A$ defined as $\sigma_{\min}(A) = \sqrt{ \lambda_{\min}(A^\prime A) }$ and $\sigma_{\max}(A) = \sqrt{ \lambda_{\max}(A^\prime A) }$; $A$ has full rank if, and only if, $\sigma_{\min}(A) > 0$.

\section{GMM Estimation without Convexity} \label{sec:Main}

Let $\overline{g}_n(\theta) = 1/n \sum_{i=1}^n g(\theta;x_i)$ be the sample moments and $G_n(\theta) = \partial_\theta \overline{g}_n(\theta)$ their Jacobian. Their population counterparts are $g(\theta) = \mathbb{E}[g(\theta;x_i)]$ and $G(\theta) = \partial_\theta g(\theta)$. $W_n$ is a weighting matrix which, for simplicity, does not depend on $\theta$ -- this excludes continuously-updated estimations. The sample GMM objective function is:
\[ Q_n(\theta) = \frac{1}{2}\overline{g}_n(\theta)^\prime W_n \overline{g}_n(\theta), \]
and the goal is to find the global minimizer $\hat{\theta}_n$ of $Q_n$ in $\mathbb{R}^{d_\theta}$. The population objective $Q(\theta) = \frac{1}{2} g(\theta)^\prime W g(\theta)$, defined similarly using the limit $W$ of $W_n$, has a global minimizer $\theta^\dagger$. Throughout, it will be assumed that the sample $Q_n$ is continuously differentiable. More specifically, this paper considers derivative-based optimizers of the form: 
\begin{align}
  \theta_{k+1} = \theta_k - \gamma P_{k,n} G_n(\theta_k)^\prime W_n \overline{g}_n(\theta_k), \label{eq:update}
\end{align}
for $k = 0,1,\dots$, some staring value $\theta_0 \in \mathbb{R}^{d_\theta}$ and a matrix $P_{k,n}$, called conditioning matrix, assumed to be symmetric. The tuning parameter $\gamma \in (0,1]$ is called the learning rate. There are several ways to motivate (\ref{eq:update}) as a minimization algorithm in the context of GMM estimation. They are conceptually similar but implicitly rely on a different set of assumptions. The first is to consider a quadratic approximation of the GMM objective function $Q_n$:
\[ Q_n(\theta) \simeq Q_n(\theta_k) + \partial_\theta Q_n(\theta_k)(\theta - \theta_k) + \frac{1}{2 \gamma} (\theta - \theta_k)^\prime \partial^2_{\theta,\theta^\prime} Q_n(\theta_k) (\theta - \theta_k), \] 
here $\gamma$ penalizes the quality of the quadratic approximation. For linear models, such as OLS and IV regressions, $Q_n$ is quadratic so that $\gamma = 1$ is feasible. For non-linear models, the approximation is inexact, and $\gamma < 1$ is generally required. Minimizing the right-hand-side with respect to $\theta$ yields a Newton-Raphson (\textsc{nr}) iteration: $\theta_{k+1} = \theta_k - \gamma [\partial^2_{\theta,\theta^\prime} Q_n(\theta_k)]^{-1}\partial_\theta Q_n(\theta_k)$ with $ \partial_\theta Q_n(\theta_k) = G_n(\theta_k)^\prime W_n \overline{g}_n(\theta_k)$ and $P_{k,n} = [\partial^2_{\theta,\theta^\prime} Q_n(\theta_k)]^{-1}$. A quasi-Newton (\textsc{qn}) iterations replaces the Hessian matrix $\partial^2_{\theta,\theta^\prime} Q_n(\theta_k)$ with an approximation computed sequentially over $k$. The most popular \textsc{qn} software implementation is called \textsc{bfgs}. 
Importantly, the quadratic approximation implicitly requires that is $H_n(\theta) = \partial^2_{\theta,\theta^\prime} Q_n(\theta_k)$ strictly positive definite around $\theta_k$ so that (\ref{eq:update}) yields a minimizer of the quadratic approximation. %When $H_n$ is non-definite, (\ref{eq:update}) is not the minimizer.

Another way to motivate (\ref{eq:update}) is to consider a linear approximation of the moments and plug it into the GMM objective function:
\begin{align*} \overline{g}_n(\theta) &\simeq \phantom{\Big[}\overline{g}_n(\theta_k) + \frac{1}{\gamma} G_n(\theta_k)(\theta-\theta_k),\\ Q_n(\theta) &\simeq \frac{1}{2}\Big[ \overline{g}_n(\theta_k) + \frac{1}{\gamma} G_n(\theta_k)(\theta-\theta_k) \Big]^\prime W_n \Big[ \overline{g}_n(\theta_k) + \frac{1}{\gamma} G_n(\theta_k)(\theta-\theta_k) \Big], \end{align*}
where now $\gamma$ penalizes the quality of the linear approximation.
Take the first order condition in the last display to find (\ref{eq:update}) with $P_{k,n} = (G_n(\theta_k)^\prime W_n G_n(\theta_k))^{-1}$, a Gauss-Newton (\textsc{gn}) iteration. The quadratic approximation requires the Hessian $H_n$ of $Q_n$ to be strictly positive definite at $\theta_k$. A \textsc{gn} iteration minimizes the linear approximation as long as the Jacobian $G_n$ of $\overline{g}_n$ has full rank at $\theta_k$ so that $G_n(\theta_k)^\prime W_n G_n(\theta_k)$ is strictly positive definite. Standard regularity condition imply local convexity around $\hat{\theta}_n$. Still, convexity is more challenging to satisfy away from the solution since $\|\overline{g}_n(\theta_k)\| \gg 0$ can result in a non-definite Hessian $H_n(\theta_k) = G_n(\theta_k)^\prime W_n G_n(\theta_k) + (\overline{g}_n(\theta_k)^\prime W_n \otimes I_d) \partial_\theta \text{vec}[ G_n(\theta_k)^\prime ]$, depending on the last term. This suggests that quadratic-based methods (\textsc{nr}, \textsc{bfgs}) and linear-based methods (\textsc{gn}) can behave differently when $Q_n$ is globally non-convex. Gradient-Descent (\textsc{gd}) can be motivated by either a linear or a quadratic approximation. The following summarizes the choice of $P_{k,n}$ for each algorithm:
\begin{table}[H] \caption{Optimizers considered in (\ref{eq:update})} \label{tab:methods} \centering \renewcommand{\arraystretch}{0.9} {
\begin{tabular}{lrl}
  \hline \hline
  1. & Gradient-Descent (\textsc{gd}) & $P_{k,n} = I_d$,\\
  2. & Newton-Raphson (\textsc{nr}) & $P_{k,n} = [ \partial^2_{\theta,\theta^\prime} Q_n(\theta_k) ]^{-1}$,\\
  3. & quasi-Newton (\textsc{qn}) & $P_{k,n}$ approximates $[ \partial^2_{\theta,\theta^\prime} Q_n(\theta_k) ]^{-1}$,\\
  4. & Gauss-Newton (\textsc{gn}) & $P_{k,n} = [G_n(\theta_k)^\prime W_n G_n(\theta_k)]^{-1}$.\\  \hline \hline
\end{tabular} }
\end{table}
\subsection{Main Assumptions} \label{sec:ass}

The following gives the main assumptions on the population moments used to describe the large sample properties of the estimator $\hat{\theta}_n$ and optimization algorithms. %When these assumptions hold, the sample moments have similar properties, this is shown in Lemmas \ref{lemma:pop}, \ref{lemma:conds_n}.

\begin{assumption} \label{ass:1prim} The observations $x_i$ are iid and:
  \begin{enumerate}[topsep=0pt,itemsep=-1ex,partopsep=1ex,parsep=1ex]
    \item[(i)] $Q(\theta) = 1/2\|g(\theta)\|^2_{W}$ has a unique minimizer $\theta^{\dagger} \in \mathbb{R}^{d_\theta}$,
    \item[(ii)] $g(\theta;x_i)$ and $g(\theta) = \mathbb{E}[g(\theta;x_i)]$ are continuously differentiable on $\mathbb{R}^{d_\theta}$,
    \item[(iii)] for all $\theta \in \mathbb{R}^{d_\theta}$: $\mathbb{E}[\|G(\theta;x_i)\|^2]<\infty$, $\mathbb{E}[\|g(\theta;x_i)\|^2]<\infty$, $\sigma_{\max}[G(\theta)] < \overline{\sigma} < \infty$;\\ there exists $\bar{L}(\cdot) \geq 0$ such that $\mathbb{E}[\bar{L}(x_i)] < L <\infty$, $\mathbb{E}[|\bar{L}(x_i)|^2] < \infty$, and\\ for all $\theta_1, \theta_2 \in \mathbb{R}^{d_\theta}$: $\|G(\theta_1; x_i)-G(\theta_2; x_i)\| \leq \bar{L}(x_i) \|\theta_1-\theta_2\|$,
    \item[(iv)] there exists $R_G > 0$ such that $\sigma_{\min}[G(\theta)] > \underline{\sigma} >0$ for all $\|\theta - \theta^\dagger\| <  R_G$,
    \item[(v)]  there exists $\bar{M}(\cdot)$ such that $\mathbb{E}[|\bar{M}(x_i)|^2] < \infty$, $\mathbb{E}[\bar{M}(x_i)] < M <\infty$, and for any $R>0$, $\|G(\theta;x_i)-G(\theta_R;x_i)\| \leq \bar{M}(x_i)/(1+R)$, where $\theta_R = \frac{R}{\|\theta\|} \theta$ if $\|\theta\|>R$, $\theta_R = \theta$ otherwise, 
    \item[(vi)] $W_n \overset{p}{\to} W$, $0 < \underline{\lambda}_W < \lambda_{\min}(W) \leq \lambda_{\max}(W) < \overline{\lambda}_W < \infty$.
  \end{enumerate}
\end{assumption}

Assumption \ref{ass:1prim} consists mainly of standard conditions to derive asymptotic properties for $\hat{\theta}_n$. The \textit{iid} assumption can be relaxed to allow for time-series dependence. The parameter space is unbounded to accommodate the unconstrained optimization. The technical condition (v) and the next Assumption imply that $Q_n$ has a strictly quadratic lower bound. This ensures consistency without assuming compactness or uniform consistency of the sample moments.  
The quantity $\sigma_{\min}[G(\theta)]$ in the local identification condition refers to the smallest singular value of $G(\theta)$. The main Assumption \ref{ass:conds} below will rely on the following quantities: %\footnote{For a rectangular matrix $G$ of size $n \times m$, $m < n$, the singular values are given by $\sigma_{j}[G] = \sqrt{\lambda_{j}(G^\prime G)} \geq 0$, where $\lambda_{j}$ are eigenvalues; $G^\prime G$ is a square matrix of size $m \times m$.}
\[ \overline{G}(\theta) = \int_0^1  G(\omega \theta + (1-\omega)\theta^\dagger) d\omega, \quad \overline{G}(\theta_1,\theta_2) = \int_0^1  G(\omega \theta_1 + (1-\omega)\theta_2) d\omega. \] 
The matrix $\overline{G}(\theta)$ is an average derivative over the path from $\theta$ to the solution $\theta^\dagger$. The matrix plays a role in the mean-value identity: $g(\theta_1)-g(\theta_2) = \overline{G}(\theta_1,\theta_2)(\theta_1-\theta_2)$  (see Lemma \ref{lem:OMV}).
\begin{assumption} \label{ass:conds} There exists $0< \rho < \underline{\sigma}\underline{\lambda}_W/2$ such that, for all $\theta \in \mathbb{R}^{d_\theta}$, either:
  \begin{enumerate}[topsep=0pt,itemsep=-1ex,partopsep=1ex,parsep=1ex]
    \item[(a)] $\sigma_{\min}[G(\theta)^\prime W \overline{G}(\theta)] > \rho \underline{\sigma}$, or 
    \item[(b)] $\|G(\theta)^\prime W \overline{G}(\theta)(\theta-\theta^\dagger) \| > \rho \underline{\sigma} \|\theta-\theta^\dagger\|$.
  \end{enumerate}
\end{assumption}
%\textcolor{red}{[SIMPLIFY: Condition (ii) the other two (i) when $W$ is invertible and $G$ is bounded.]}

%\textcolor{red}{[POSSIBILITY: Add a sample version of Assumptions 2, 2'. Call these Assumption 2n (a) and (b) Re-write the Theorems assuming Assumption 2n (b).] Assumptions 1+2 imply Assumption 2n (a,b). Question: does Assumption 2' + Local Identification Imply Assumption 2n (b)? ADD: Local Identification + continuity + necessary conditions for GD to be globally CV $\Rightarrow$ Assumption 2'.}

Assumption \ref{ass:conds} gives the main conditions used in this paper for global GMM estimation of just and over-identified models.\footnote{The factor $\rho$ is assumed to be set, without loss of generality, such that $\sigma_{\min}[\overline{G}(\theta)] > \underline{\sigma}$ under (a) and $\|\overline{G}(\theta)(\theta-\theta^\dagger)\| > \underline{\sigma} \|\theta-\theta^\dagger\|$ under (b) for $\underline{\sigma}$ found in Assumption \ref{ass:1prim}.} Assumption \ref{ass:conds} (a) replaces the convexity condition $0 < \underline{\lambda}_H \leq \lambda_{\min}[H_n(\theta)] \leq \lambda_{\max}[H_n(\theta)] < \overline{\lambda}_H < \infty$ used to derive convergence results for \textsc{gd}, \textsc{nr} and \textsc{qn}.\footnote{See \citet[pp33-35]{Nesterov2018}, especially equations (1.2.25), (1.2.27) and Theorem 1.2.4 for \textsc{gd}.}, which may not hold for GMM. A sufficient, but restrictive, condition for Assumption \ref{ass:conds} (a) is that $g$ is the derivative of a convex function, for instance a Probit log-likelihood function. Further sufficient conditions are listed in Section \ref{sec:lit_char}. 
Assumption \ref{ass:conds} (a) implies Assumption \ref{ass:conds} (b); the latter is the weaker condition. Assumption \ref{ass:conds} (a) implies that $G(\theta)$ has full rank for all $\theta$, Assumption \ref{ass:conds} (b) only requires $G(\theta)^\prime W \overline{G}(\theta)$ to be non-singular in the relevant direction $(\theta-\theta^\dagger)$. For over-identified models, both conditions (a) and (b) depend on the choice of weighting matrix $W$. Indeed, unlike square matrices, the product of full rank rectangular matrices does not automatically have full rank,\footnote{Take $G(\theta_1)^\prime = (1, 0)$ and $G(\theta_2)^\prime = (0, 1)$, both have full rank and yet $G(\theta_1)^\prime G(\theta_2)=0$ is singular.} and the weighting matrix changes the way $G$ and $\overline{G}$ are multiplied. It is possible for the product to be singular even when $G$ and $\overline{G}$ have full rank. Importantly, Assumption \ref{ass:conds} may or may not hold depending on the choice of weighting matrix $W$. If Assumption \ref{ass:conds} is not satisfied using the preferred weighting matrix, the algorithm remains locally convergent. A two-step estimation, with a weighting matrix for which Assumption \ref{ass:conds} holds in the first step, would provide a valid estimation strategy in that case. Assumption \ref{ass:conds} is invariant to some one-to-one reparameterizations, this is shown in the next section.

Under Assumption \ref{ass:conds}, the parameters are both locally and globally identified (i.e. Assumption \ref{ass:1prim} (iv) and (i)).  Conversely, Assumption \ref{ass:1prim} (iv) implies that Assumption \ref{ass:conds} (a) holds locally around $\theta^\dagger$. The condition requires that it holds globally rather than locally.\footnote{See Lemmas \ref{lem:ass1_2}, \ref{lem:ass2_1} and Propositions \ref{prop:PL}, \ref{prop:PLmis}.}
Under Assumptions \ref{ass:1prim} and \ref{ass:conds}, a sample analog of Assumption \ref{ass:conds} holds for the following quantities:
\[ \overline{G}_n(\theta) = \int_0^1 G_n(\omega \theta + (1-\omega)\hat{\theta}_n) d\omega, \quad \overline{G}_n(\theta_1,\theta_2) = \int_0^1 G_n(\omega \theta_1 + (1-\omega)\theta_2) d\omega, \] 
with probability approaching 1, this is shown in Lemma \ref{lemma:conds_n}. When Assumption \ref{ass:conds} cannot be verified analytically, a related condition which does not involve the minimizer can be checked numerically on the sample moments and their Jacobian. This is considered in Section \ref{sec:check}.

\begin{lemma} \label{lem:consistency} Suppose Assumptions \ref{ass:1prim} and \ref{ass:conds} hold, then $\hat{\theta}_n \overset{p}{\to} \theta^\dagger$ and $Q_n(\hat{\theta}_n)\overset{p}{\to} Q(\theta^\dagger)$.
\end{lemma}
Lemma \ref{lem:consistency} shows that, although the parameter space is unbounded and $Q_n$ is non-convex, $\hat{\theta}_n$ is a consistent estimator under Assumptions \ref{ass:1prim} and \ref{ass:conds}.

%Primitive conditions for Assumption \ref{ass:1} are given in Appendix \ref{apx:pop}. The uniqueness of the arg-minimizer $\hat\theta_n$ ensures that the optimization problem has a unique, well-defined solution. Without loss of generality, $R_G$ is such that the closed ball around $\hat\theta_n$ of radius $R_G$ is a subset of $\Theta$, with probability approaching 1. 
%\textsc{gd} iterations are always well defined, whereas \textsc{nr} requires the Hessian  $\partial^2_{\theta,\theta^\prime}Q_n$, and \textsc{gn} the Jacobian $G_n$ to be non-singular. 
\begin{assumption} \label{ass:2} With probability approaching 1:
  $P_{k,n}$ is symmetric and such that:\\ $0 < \underline{\lambda}_P \leq \lambda_{\min}(P_{k,n}) \leq \lambda_{\max}(P_{k,n}) \leq \overline{\lambda}_P < \infty$.
\end{assumption}

Assumption \ref{ass:2} requires $P_{k,n}$ to be finite and strictly positive definite. This is always the case for \textsc{gd} since $P_{k,n} = I_d$, and holds for \textsc{gn} under Assumption \ref{ass:conds} (a). If the moments only satisfy Assumption \ref{ass:conds} (b),  Assumption \ref{ass:2} does not necessarily hold for \textsc{gn} since the Jacobian $G_n(\theta_k)$ can be singular, but it remains valid for \textsc{gd}. When Assumption \ref{ass:2} fails, one approach is to regularize the inverse using the so-called Levenberg-Marquardt (LM) algorithm to \textsc{gn} by setting $P_{k,n} = (G_n(\theta_k)^\prime W_n G_n(\theta_k) + \lambda I_d )^{-1}$ so that $\overline{\lambda}_P < \lambda^{-1} < \infty$ and $P_{k,n}$ is finite. Note that Assumption \ref{ass:2} does hold for \textsc{gn} under strong injectivity conditions introduced in the next Section. \citet[Ch3.4]{nocedal-wright:06} list several additional approaches to enforce Assumption \ref{ass:2}, mainly for convex optimizers.%Global convergence is not guaranteed, however, since non-global local optima may exist since the PL inequality can fail.%

 \subsection{Global Convergence Results} \label{sec:global_cv}
 The following provides the main results: the global convergence properties of gradient-based algorithms. In the following, the initial value $\theta_0$ is taken from $\Theta$, a compact subset of $\mathbb{R}^{d_\theta}$. This is a technical assumption; although the optimization is unconstrained, the sample moments are not uniformly consistent on $\mathbb{R}^{d_\theta}$ which complicates the analysis. The following shows global convergence, uniformly over $\theta_0 \in \Theta$. The main idea is to show that, with probability approaching $1$, the optimization path $(\theta_k)_{k\geq 0}$ is restricted to a compact set, determined by $\theta_0$, where the sample moments are uniformly consistent. Without loss of generality, $\Theta$ is assumed convex and large enough that $\theta^\dagger \in \text{interior}(\Theta)$. In addition, local convergence results can be found in Appendix \ref{apx:local}, those results are new in the case of overidentified and misspecified models as they allow for $\|\overline{g}_n(\hat{\theta}_n)\|_{W_n}\neq 0$.

%For $\textsc{gn}$, Proposition \ref{prop:local_cv} implies that for any $\gamma\in(0,1)$ there is a local neighborhood of fast local convergence under Assumption \ref{ass:1}. 

%Because $L = 0$ implies $R_n = \infty$, which is only true for linear models, e.g. OLS, Proposition \ref{prop:local_cv} appears to imply that global convergence is only guaranteed for linear models. However, the following Theorem shows that whenever $G_n$ has full rank everywhere, (\ref{eq:update}) will be globally convergent for a well chosen value of $\gamma$.

%\subsubsection{Just-Identified Models} The Theorem below proves global convergence for $\gamma \in (0,1)$ sufficiently small.% -- under additional restrictions on $G_n$.

\begin{theorem}[Correctly Specified] \label{th:global_cv_cs} Suppose Assumptions \ref{ass:1prim}, \ref{ass:conds}, \ref{ass:2} hold and $Q(\theta^\dagger) = 0$. Then, for $\gamma$ small enough, there exists $\overline{\gamma} \in (0,1)$, $0 < \underline{\lambda} \leq \overline{\lambda} < +\infty$, and $C \geq 0$ such that: 
\[ \|\theta_{k+1}-\hat{\theta}_n\| \leq (1-\overline{\gamma})^{k+1}\frac{\sqrt{\overline{\lambda} + C \|\overline{g}_n(\hat{\theta}_n)\|_{W_n}}}{\sqrt{\underline{\lambda} - C \|\overline{g}_n(\hat{\theta}_n)\|_{W_n}}}\|\theta_0-\hat{\theta}_n\|,\]
  for any starting value $\theta_0 \in \Theta$, with probability approaching $1$. 
\end{theorem}

Theorem \ref{th:global_cv_cs} provides global convergence results that are comparable to the convex case. Because the factor $(1-\overline{\gamma})$ is less than $1$, the distance to the solution $\|\theta_{k+1}-\hat{\theta}_n\|$ decreases exponentially fast with $k$, as in the convex case. Several factors affect convergence. The constants $\underline{\lambda}$, $\overline{\lambda}$ coincide with $C_2 = 1/2 \rho^2 \underline{\sigma}^2/[\overline{\sigma}^2 \overline{\lambda}_W]$, $C_3 = 1/2 \overline{\sigma}^2 \overline{\lambda}_W$ in Proposition \ref{prop:PL} below. The convergence rate $1-\overline{\gamma}$ depends on $C_1 = 1/2 \rho^2 \underline{\sigma}^2 / [\overline{\sigma}^2 \overline{\lambda}_W]$, from the same Proposition. 

Through these constants, it appears that identification strength - here measured by $\rho \underline{\sigma}$ - and the choice of weighting matrix $W_n$ affect the convergence properties. In particular, a weighting matrix that is ill-conditioned can lead to slower convergence. This can make optimization challenging. When the sample moments are highly correlated, the optimal weighting matrix can be ill-conditioned. Using equal weighting, a diagonal weighting matrix, or regularizing the optimal weighting matrix with $W_n = ( \hat{V}_n + \lambda I_d )^{-1}$, where  $\hat{V}_n$ estimates the variance of $\sqrt{n}\overline{g}_n(\theta^\dagger)$, could improve numerical stability.

The size of $\|\overline{g}_n(\hat{\theta}_n)\|_{W_n}$ further affects convergence. The constant $C$ coincides with $C_4 = \overline{\lambda}_W^{1/2} L$ in Proposition \ref{prop:PLmis} below. The constant $L$ measures the non-linearity of the sample moments, $L=0$ corresponds to linear models. For linear models, $C=0$ implies that $\|\overline{g}_n(\hat{\theta}_n)\|_{W_n}$ does not affect convergence. Non-linear models have $L>0$ which makes optimization more sensitive to $\|\overline{g}_n(\hat{\theta}_n)\|_{W_n}$ for overidentified models. 

In applications, $\|\overline{g}_n(\hat\theta_n)\|_{W_n}$ can be relatively large so that misspecification becomes a concern. Understanding the robustness of Theorem \ref{th:global_cv_cs} to non-negligible deviations from $Q(\theta^\dagger)=0$ is then empirically relevant. The following considers models where the quantity:
\[ Q_n(\hat{\theta}_n) \overset{p}{\to} Q(\theta^\dagger) := \varphi/2 >0 \] 
does not vanish asymptotically which implies that $\|\overline{g}_n(\hat{\theta}_n)\|_{W_n}$ matters for convergence, even in large samples. 
Since $G_n$ cannot be full rank at $\theta=\hat\theta_n$ when the model is both just-identified and misspecified, the results presented here solely consider over-identified models.\footnote{The solution $\hat\theta_n$ is s.t. $G_n(\hat\theta_n)^\prime W_n \overline{g}_n(\hat\theta_n)=0$, misspecification implies $\overline{g}_n(\hat\theta_n) \neq 0$, and since $W_n$ has full rank, it must be that $G_n(\hat\theta_n)$ is singular for just-identified models. For over-identified models, $\overline{g}_n(\hat\theta_n)$ is in the null space of $G_n(\hat\theta_n)^\prime W_n$, which allows $G_n(\hat\theta_n)$ to be full rank.}%Having $\|\overline{g}_n(\hat{\theta}_n)\|_{W_n} = o_p(1)$ under correct specification implies that optimization behaves similarly to the just-identified case when the sample size is large.

\begin{theorem}[Misspecified] \label{th:global_cv_ms} Suppose Assumptions \ref{ass:1prim}, \ref{ass:conds}, \ref{ass:2} hold and $Q(\theta^\dagger) = \varphi/2 >0$, such that:
\begin{align}
  \sqrt{\varphi} < \min \left( \frac{\rho \underline{\sigma}}{\sqrt{\overline{\lambda}_W} L}, \frac{1}{2}\frac{\rho^2 \underline{\sigma}^2}{\overline{\lambda}_W^{3/2} \overline{\sigma}^2 L} \right), \label{eq:phi_conds}
\end{align}
then, for $\gamma$ small enough, there exists $\overline{\gamma} \in (0,1)$, $0 < \underline{\lambda} \leq \overline{\lambda} < +\infty$ and $C>0$ such that $\underline{\lambda} - C \|\overline{g}_n(\hat{\theta}_n)\|_{W_n} \overset{p}{\to} \underline{\lambda} - C \sqrt{\varphi} > 0$, and: \[\|\theta_{k+1}-\hat{\theta}_n\| \leq (1-\overline{\gamma})^{k+1}\frac{\sqrt{\overline{\lambda} + C \|\overline{g}_n(\hat{\theta}_n)\|_{W_n}}}{\sqrt{\underline{\lambda} - C \|\overline{g}_n(\hat{\theta}_n)\|_{W_n}}}\|\theta_0-\hat{\theta}_n\|,\]
for any starting value $\theta_0 \in \Theta$, with probability approaching $1$. 
\end{theorem}

Theorem \ref{th:global_cv_ms} shows that convergence is robust to `moderate' amounts of misspecification. For linear models, $L=0$ implies that (\ref{eq:phi_conds}) reads $\varphi < +\infty$, which is not restrictive. In (\ref{eq:phi_conds}), the choice of $W_n$, nonlinearity, and identification strength restrict the amount of misspecification allowed in (\ref{eq:phi_conds}). The restrictions (\ref{eq:phi_conds}) are discussed further with Proposition \ref{prop:PLmis} below. The convergence rate $1-\overline{\gamma}$ also depends on $\varphi$, which slows convergence. In the limit, its expression is given by $(1-\overline{\gamma})^2 = 1 - \gamma \underline{\lambda}_P C_1 / 2$, where $C_1 = (\rho\underline{\sigma} - \overline{\lambda}_W^{1/2}L \sqrt{\varphi})^2/[C_3 + C_4 \sqrt{\varphi}]$. The constants $C_3$, $C_4$ appear in Proposition \ref{prop:PLmis} below. The first of the two terms in the upper bound in (\ref{eq:phi_conds}) ensures that $\overline{\gamma}>0$ is feasible. Having $\varphi \neq 0$ makes convergence slower and estimation more challenging.  When $\varphi$ is arbitrarily large, global convergence can fail. This is explained in the next Section, and illustrated with an MA(1) example. Since the magnitude of $\varphi$ depends on the choice of moments $\overline{g}_n$ and weighting matrix $W_n$, a careful selection of these two might mitigate this issue.

\section{Assumption \ref{ass:conds} and its relation to the literature} \label{sec:lit_char}

%\paragraph{Discussion of the rank condition.} \textcolor{red}{[THIS WILL BE REPLACED WITH THE NEW PROPOSITIONS]}
%In the scalar case, where both $\theta$ and $\overline{g}_n$ are one-dimensional, continuity of $G_n$ and the rank condition (\ref{eq:rk}) implies that $\overline{g}_n$ is strictly monotone, i.e. injective. However, this does not imply that $Q_n$ is convex, as illustrated with the MA(1) example above. Under strict monotonicity, univariate methods such as bisection or golden-search converge at a similar rate but do not extend to multivariate estimations. 

%In the multivariate case, (\ref{eq:rk}) implies a unique solution since $0 = \overline{g}_n(\theta) =  G_n(\tilde \theta_n)(\theta-\hat\theta_n) \Leftrightarrow \theta = \hat\theta_n$, for an intermediate value $\tilde \theta_n$. It further implies that $Q_n$ has no local minimum, besides $\hat\theta_n$, since $\partial_\theta Q_n(\theta) = G_n(\theta)^\prime W_n \overline{g}_n(\theta) = 0 \Leftrightarrow \overline{g}_n(\theta) = 0 \Leftrightarrow \theta=\hat\theta_n$, for just-identified models.  

%As discussed in the Introduction, the full rank condition is equivalent to assuming convexity of $\ell_n$ when $\overline{g}_n = \partial_\theta \ell_n$, \footnote{Since Hessian should be positive semi definite at the global minimum, a full rank and continuous Hessian implies that the Hessian is strictly positive definite everywhere. Hence, the original function is global convex.} $\ell_n$ can be a log-likelihood or a sum of squared residuals. 

\paragraph{Convexity, monotonicity and the Polyak-Łojasiewicz condition.}

The following briefly reviews some convexity conditions found in the literature and an important relaxation called the Polyak-Łojasiewicz (PL) condition. The latter has gathered much attention in the machine learning literature in recent years. Because Assumption \ref{ass:conds} is stated on population quantities, the following discussion will focus on $Q$. %It is shown here that Assumption \ref{ass:conds} implies the PL condition in the population. Several conditions that imply Assumption \ref{ass:conds} are given, and conditions are given to ensure Assumption \ref{ass:conds} holds after a non-linear reparameterization.   

For general minimization of an objective $Q$, \textsc{gd}, \textsc{nr} and \textsc{qn} are globally convergent for $\theta^\dagger$ if $Q$ is $\mu$-\textit{strongly convex}, i.e. if for some $\mu >0$: 
\[ Q(\theta_2) \geq Q(\theta_1) + \partial_\theta Q(\theta_1)(\theta_2-\theta_1) + \frac{\mu}{2}\|\theta_1-\theta_2\|^2, \]
for all $\theta_1,\theta_2 \in \mathbb{R}^{d_\theta}$. When $Q$ is twice continuously differentiable it is strongly convex if its Hessian $H(\theta) = \partial^2_{\theta,\theta^\prime} Q(\theta)$ is strictly positive definite everywhere with $0 < \underline{\lambda}_H < \lambda_{\min}[H(\theta)] \leq \lambda_{\max}[H(\theta)] < \overline{\lambda}_H < \infty$.
Under strong convexity, for $\gamma > 0$ sufficiently small and any $\theta_0$:
\[ Q(\theta_{k+1}) - Q(\theta^\dagger) \leq (1-\eta) \left( Q(\theta_{k}) - Q(\theta^\dagger) \right),  \]
for some $\eta \in (0,1)$ which depends on $\gamma$, the choice of algorithm, i.e. $P_{k,n}$, and the eigenvalues of $H$. Iterating on this inequality indicates that the fit improves rapidly from any starting value $\theta_0$: $Q(\theta_{k}) - Q(\theta^\dagger) \leq (1-\eta)^{k} \left( Q(\theta_{0}) - Q(\theta^\dagger) \right)$. Under strong convexity, $Q$ has a unique global minimizer and no local optima.
The literature has considered a number of relaxations of strong convexity under which \textsc{gd} is globally convergent. This includes the so-called \textit{star convexity} condition introduced by \citet{nesterov2006}: 
\[ Q(\theta^\dagger) \geq Q(\theta) + \lambda \partial_\theta Q(\theta)(\theta^\dagger-\theta) + \frac{\mu}{2}\|\theta-\theta^\dagger\|^2\] 
for some $\mu \geq 0$ and $\lambda = 1$. Fast convergence results for $\theta$ require $\mu > 0$. This is similar-looking to strong convexity but only involves the pairs $(\theta_1,\theta_2) = (\theta,\theta^\dagger)$. For these functions, the convexity property only holds on line segments toward $\theta^\dagger$. Star convexity implies that $\theta^\dagger$ is the unique global minimizer of $Q$. This condition can be further weakened to \textit{quasar convexity}, which allows for $\lambda > 1$ in the inequality above. \citet{hinder2020}, Figure 1, plot several functions that satisfy these conditions. %Note that Assumption \ref{ass:conds} is similarly stated for averages $\overline{G}$ of $G$ on line segments between $\theta$ and $\theta^\dagger$.

\citet{karimi2016}, \citet{guminov2017} showed that a number of relaxations of strong convexity imply the so-called \textit{Polyak-{\L}ojasiewicz} (PL) inequality, named after \citet{polyak1963} and \citet{lojasiewicz1963}, which requires that:
\begin{align} \|\partial_\theta Q(\theta)\|^2 \geq \mu \left( Q(\theta) - Q(\theta^\dagger) \right), \tag{PL} \end{align}
for all $\theta \in \mathbb{R}^{d_\theta}$ and some $\mu >0$. When $Q$ satisfies the PL inequality, $\partial_\theta Q(\theta) = 0$ implies $\theta$ is globally optimal, i.e. $Q(\theta) = Q(\theta^\dagger)$. The arg-minimizer may not be unique, however, unlike strong convexity. If the PL inequality holds and $\partial_\theta Q$ is Lipschitz continuous, it can be shown that for $\gamma >0$ small enough: $Q(\theta_{k+1}) - Q(\theta^\dagger) \leq (1-\eta) \left( Q(\theta_{k}) - Q(\theta^\dagger) \right)$ for \textsc{gd} \citep[Th1]{karimi2016}. This does not imply that $\theta_{k+1}$ converges to $\theta^\dagger$, however, unless the arg-minimizer is unique. Because strong convexity implies the PL inequality, \citet{karimi2016} argue that the latter holds locally over a larger area than strong convexity, predicting better optimization performance. They also note that it is difficult to characterize which functions satisfy the PL inequality. They show that $Q(\theta) = h(A\theta)$, with $h$ strongly convex and $A$ a non-zero matrix, satisfies the PL inequality. 

Closely related to the GMM setting, a smaller literature has considered conditions for solving non-linear systems of equations of the form: $g(\theta) = 0$, typically with $g$ and $\theta$ of the same dimension. An important reference is \citet{dennis1996}, who cast the problem as minimizing $Q(\theta) = \|g(\theta)\|^2$, similar to GMM, and derive global convergence results to a local minimum under convexity conditions (Theorems 6.3.3-6.3.4). \citet[Ch3]{deuflhard2005} studies global convergence under alternative conditions. For just and under-determined systems, several authors considered a \textit{strong monotonicity condition}: \[ ( g(\theta_1) - g(\theta_2) )^\prime (\theta_1-\theta_2) \geq \mu \|\theta_1-\theta_2\|^2,\] with $\mu > 0$, e.g. \citet{solodov2000}, \citet{polyak2020}. Note that when $g = \partial_\theta F$, then $g$ is strongly monotone if, and only if, $F$ is strongly convex. Hence, global convergence under strong monotonicity is related to global convergence under strong convexity of $F$. In that case, $g$ is said to be cyclically monotone \citep[p238]{Rockafellar}. %When $G(\theta)+G(\theta)^\prime$ is positive definite, then $g$ is monotone \citep[p240]{Rockafellar}.  % TO RE-ADD LATER? His related global convergence result, Theorem 3.7, implicitly assumes linearity of $g$ in the proof, however.
These results do not consider $g(\theta^\dagger) \neq 0$ which is particularly relevant here. %The functin $g$ is monotone if, and only if, the symmetrized $1/2(G + G^\prime)$ is strictly positive definite.\footnote{When $G$ satisfies this condition, it is said to be quasi-definite, and $g$ is globally injective by the \citet{gale1965} Theorem. This type of result is particularly useful for the analysis of demand functions, see \citet{allen2022} for more details and references.}

A companion paper, \citet{Forneron2022b}, considers correctly specified GMM estimation with non-smooth  sample moments that may not satisfy Assumption \ref{ass:conds}. There are two important differences in that setting: 1) the Jacobian $G_n$ is not defined, and 2) $Q_n$ can have local optima. The methods considered here are not sufficient to find a global optimum, and there is a curse of dimensionality for global convergence. The two papers are complementary.% \citep[e.g.][]{Nemirovsky1983}.
\paragraph{Relation between the different conditions.} Narrowing to the GMM setting specifically, the following shows that the PL inequality holds \textit{in the population for correctly specified models} under Assumption \ref{ass:conds}. A related result is derived under misspecification. %In order to relate all the conditions listed above, the following Assumption is introduced.

%\begin{assumptionbis}{ass:conds} \label{ass:condsbis} There exists $\underline{\sigma} > 0$, $\rho > 0$ such that for all $\theta \in \Theta$ (i) $\|\overline{G}(\theta)(\theta-\theta^\dagger)\| > \underline{\sigma}\|\theta-\theta^\dagger\|$, and (ii) $\|G(\theta)^\prime W \overline{G}(\theta)(\theta-\theta^\dagger)\| > \rho \underline{\sigma} \|\theta-\theta^\dagger\|$.  
%\end{assumptionbis}
%\textcolor{red}{[SIMPLIFY: Condition (ii) implies (i) when $W$ is invertible and $G$ is bounded.]}
As discussed above, Assumption \ref{ass:conds} (a) implies Assumption \ref{ass:conds} (b). The latter confers most of the properties required for minimizing $Q$. It can be useful to re-write the condition in terms of $g$: Assumption \ref{ass:conds} (b) $\|G(\theta)^\prime W [g(\theta)-g(\theta^\dagger)]\| > \rho \underline{\sigma}\|\theta-\theta^\dagger\|$. For correctly specified models, $g(\theta^\dagger) = 0$ implies $G(\theta)^\prime W g(\theta) = \partial_\theta Q(\theta)$. The only critical point is $\theta = \theta^\dagger$. Hence, Assumption \ref{ass:conds} excludes local optima and saddle points when the model is correctly specified.\footnote{A critical point is a $\theta$ such that $\partial_\theta Q(\theta) = 0$. Assuming $Q$ is twice differentiable, it is a local minimum if $\partial^2_{\theta,\theta^\prime} Q(\theta)$ is positive semidefinite, maximum if $\partial^2_{\theta,\theta^\prime} Q(\theta)$ is negative semidefinite, and a saddle point if $\partial^2_{\theta,\theta^\prime} Q(\theta)$ is indefinite, i.e. has both positive and negative eigenvalues.}

\begin{proposition}[Correct Specification] \label{prop:PL} Suppose Assumptions \ref{ass:1prim} (ii), (iii), (vi), \ref{ass:conds} (b) hold and $Q(\theta^\dagger) = 0$, then there exists strictly positive constants $C_1,C_2,C_3$ such that for all $\theta \in \mathbb{R}^{d_\theta}$: 
  \begin{align*} (1) \quad &\|\partial_\theta Q(\theta)\|^2 \geq C_1 \left( Q(\theta) - Q(\theta^\dagger) \right)\\ (2) \quad &C_2 \|\theta - \theta^\dagger\|^2 \leq  Q(\theta) - Q(\theta^\dagger) \leq C_3 \|\theta - \theta^\dagger\|^2.\end{align*}
\end{proposition}

Proposition \ref{prop:PL} shows that Assumption \ref{ass:conds} (b), together with bounds on $W$ and Lipschitz continuity of $G$ imply the PL inequality (1) for $Q$. In addition, (2) implies global identification and is needed to derive the convergence rate of $\theta_k$. Strong convexity also implies (1) and (2). %When $g$ and $\theta$ are scalar, strong convexity implies Assumption \ref{ass:conds}.

\begin{proposition} \label{prop:quasar_convex} Suppose $W$ is invertible, $Q(\theta^\dagger) = 0$. 1) If $Q$ satisfies the PL inequality with $\mu > 0$ and $C_2 \|\theta-\theta^\dagger\|^2 \leq Q(\theta)-Q(\theta^\dagger)$ for $C_2>0$ and all $\theta \in \mathbb{R}^{d_\theta}$, then Assumption \ref{ass:conds} (b) holds. 2) If $Q$ is quasar-convex with $\mu > 0$, then Assumption \ref{ass:conds} (b) holds. %3) If $Q$ is strongly convex and twice differentiable with bounded Hessian, then Assumption \ref{ass:conds} holds.
\end{proposition}

Proposition \ref{prop:quasar_convex} gives a condition under which quasar-convexity and the PL inequality imply  Assumption \ref{ass:conds} (b). On compact sets, Assumption \ref{ass:1prim} (i), (iii), (iv) together imply a $C_2 >0$ exists for correctly specified models. Assumption \ref{ass:conds} (b) does not imply quasar-convexity.\footnote{Quasar-convexity implies $(\theta-\theta^\dagger)^\prime G(\theta)^\prime W \overline{G}(\theta)(\theta-\theta^\dagger) \geq \frac{\mu}{2\lambda} \|\theta-\theta^\dagger\|^2$ for correctly specified models. This is more restrictive than Assumption \ref{ass:conds} (b) when $\theta$ is not scalar.} The following considers strong monotonicity and introduces a \textit{strong injectivity} condition:
\begin{align} \|g(\theta_1) - g(\theta_2)\| \geq \mu \|\theta_1 - \theta_2\|. \tag{SI} \label{eq:SI}\end{align}
It can be shown that the strong injectivity property holds on compact convex sets under the Gale-Nikaidô-Fisher-Rothenberg global identification conditions: $\text{det}(G(\theta)) > 0$ and $G(\theta)$ positive quasi-definite, for all $\theta \in \mathbb{R}^{d_\theta}$, where $\text{det}$ is the determinant.\footnote{$G$ is positive quasi-definite if, and only if, $G+G^\prime$ is positive definite. See \citet{fisher1966}, \citet{rothenberg1971}; and \citet{komunjer2012} for a discussion and alternative conditions.}

\begin{proposition}[Just-Identified] \label{prop:strong_monotone} 1) If $A g$ is strongly monotone for some invertible matrix $A$ and $\mu > 0$, then Assumption \ref{ass:conds} (b) holds. 2) If $g$ is strongly injective with $\mu > 0$, then Assumption \ref{ass:conds} (b) holds.
\end{proposition}

For over-identified models, (\ref{eq:SI}) is not sufficient. As discussed above, the weighting matrix $W$ plays a role in the convergence properties. The following extends (\ref{eq:SI}) appropriately: 
\begin{align} \|G(\theta_1)^\prime W [g(\theta_1) - g(\theta_2)]\| \geq \mu \|\theta_1 - \theta_2\|. \tag{SI'} \label{eq:SIp} \end{align}
Relative to (\ref{eq:SI}), the additional term ensures that $g(\cdot)$ is one-to-one in the row space of $G(\cdot)^\prime W$. 
Taking $(\theta_1,\theta_2) = (\theta,\theta^\dagger)$ yields Assumption \ref{ass:conds} (b). Note that (\ref{eq:SIp}) implies that Assumption \ref{ass:2} holds for \textsc{gn}  (Lemma \ref{lem:SI_ass2}). (\ref{eq:SIp}) is more challenging to verify than (\ref{eq:SI}) as it involves the weighting matrix $W$ and the Jacobian $G$. If (\ref{eq:SI}) holds for a just-identified subset of moments, then it is possible to regularize $W$ so that (\ref{eq:SIp}) holds (Lemma \ref{lem:SI_SIp}).
\begin{figure}[h] \caption{Relationship between conditions for correctly specified models} \label{fig:summary}
  \centering { \small
    \begin{tabular}{|ccccccc|} \hline
      strong convexity & $\Rightarrow$ & star convexity & $\Rightarrow$ & quasar convexity & & \\
       & & & & $\Downarrow$ & & \\
       & & Assumption \ref{ass:conds} (a) & $\Rightarrow$ & Assumption \ref{ass:conds} (b) & $\Leftarrow$ & strong injectivity \\
       & & & & $\Updownarrow$ & & $\Uparrow$\\
       & & & & PL + QLB & & strong monotonicity \\ \hline
    \end{tabular}
    \notes{ \textbf{Legend:} Relations hold when $Q(\theta^\dagger)=0$. QLB = Quadratic Lower Bound, i.e. $C_2 \|\theta-\theta^\dagger\|^2 \leq Q(\theta)-Q(\theta^\dagger)$ for some $C_2>0$. Relation with strong monotonicity is for just-identified models.} 
   }
\end{figure}

Figure \ref{fig:summary} summarizes the results of Propositions \ref{prop:PL}, \ref{prop:quasar_convex}, \ref{prop:strong_monotone}. Since $Q_n(\hat{\theta}_n) = 0$ for just-identified models that are correctly specified, the relationship also applies in the finite samples problems where these conditions are met.  When $g$ and $\theta$ are scalar, Assumption \ref{ass:conds} implies strict monotonicity, $g$ is either increasing or decreasing, but does not imply convexity of $Q$, however, as the MA example below will illustrate. %Also, Assumptions \ref{ass:conds} (a) and (b) are equivalent in the scalar case so that strong, star, and quasar convexity imply Assumption \ref{ass:conds} (a) in that particular setting.

It remains to determine if Assumption \ref{ass:conds} (b) is minimal for global convergence, or if can be weakened further. The following condition is \textit{necessary} for \textsc{gd} and other gradient-based optimizers of the form (\ref{eq:update}) to be globally convergent:
\begin{align} \partial_\theta Q(\theta) = 0 \Leftrightarrow \theta = \theta^\dagger. \tag{N}\end{align}
The following shows, under regularity conditions, that (N) implies Assumption \ref{ass:conds} (b).

\begin{proposition} \label{prop:minimal}
 Suppose condition (N) and Assumption \ref{ass:1prim} (ii)-(iv) and (vi) hold, then Assumption \ref{ass:conds} (b) holds on any compact convex set containing $\theta^\dagger$.
\end{proposition}

%Proposition \ref{prop:minimal} implies that given standard regularity conditions, Assumption \ref{ass:conds} (b) is a necessary condition on compact sets. 

The case of misspecified models is more complicated, as the following shows that the equivalence between the PL inequality and Assumption \ref{ass:conds} (b) is not automatic.

\begin{proposition}[Misspecification] \label{prop:PLmis} Suppose Assumptions \ref{ass:1prim} (ii), (iii), (vi), \ref{ass:conds} (b) hold, then there exists strictly positive constants $C_2,C_3,C_4$ such that for all $\theta \in \mathbb{R}^{d_\theta}$: 
  \begin{align*} (1) \quad &\|\partial_\theta Q(\theta)\| \geq \left( \rho \underline{\sigma} - \sqrt{\varphi} \overline{\lambda}_W^{1/2}L \right)\|\theta-\theta^\dagger\|\\ (2) \quad &(C_2-C_4 \sqrt{\varphi}) \|\theta-\theta^\dagger\|^2 \leq Q(\theta) - Q(\theta^\dagger) \leq  (C_3 + C_4 \sqrt{\varphi}) \|\theta-\theta^\dagger\|^2,\end{align*}
  where $Q(\theta^\dagger) = \varphi >0$, $C_2,C_3$ are the same as in Proposition \ref{prop:PL} and $L$ is the Lipschitz constant of $G$ from in Assumption \ref{ass:1prim} (iii). If in addition $\rho \underline{\sigma} - \sqrt{\varphi \overline{\lambda}_W}L > 0$, then for all $\theta \in \mathbb{R}^{d_\theta}$:
  \begin{align*} (1') \quad &\|\partial_\theta Q(\theta)\|^2 \geq \frac{(\rho \underline{\sigma} - \sqrt{\varphi \overline{\lambda}_W}L)^2}{C_3 + C_4\sqrt{\varphi}} \left( Q(\theta)-Q(\theta^\dagger)\right).%\\ (2') \quad &\left( C_2 - \frac{C_4 \sqrt{\varphi}}{\|\theta-\theta^\dagger\|} \right) \|\theta-\theta^\dagger\|^2 \leq Q(\theta) - Q(\theta^\dagger) \leq  \left( C_3 +\frac{C_4 \sqrt{\varphi}}{\|\theta-\theta^\dagger\|} \right) \|\theta-\theta^\dagger\|^2,
  \end{align*}
\end{proposition}

Proposition \ref{prop:PLmis} (1) is only informative when the amount of misspecification is moderate, i.e. $\varphi < \rho^2 \underline{\sigma}^2/[\overline{\lambda}_W L^2]$. When this holds, there are no local optima besides $\theta^\dagger$. It also implies the PL inequality (1') holds. To recover convergence for $\theta$, the lower bound in (2) should be informative which further requires $\sqrt{\varphi} < C_2 / C_4$.\footnote{The derivations give the following bounds $C_2 = 1/2 \frac{\rho^2\underline{\sigma}^2}{\overline{\sigma}^2 \overline{\lambda}_W}$ and $C_4 = \overline{\lambda}_W^{1/2} L$ so that the condition reads $\sqrt{\varphi} < 1/2 \rho^2\underline{\sigma}^2 [\overline{\sigma}^2\overline{\lambda}_W^{3/2} L]^{-1}$. It is possible to relax this condition at the cost of more complicated derivations using a combination of global and local convergence arguments.} The degree of non-linearity - measured by $L$ - and the choice of weighting matrix - measured by $\overline{\lambda}_W,\underline{\lambda}_W$ and $\varphi$ -  constrain the amount of misspecification permitted to get informative bounds.
For correctly specified models, $Q_n(\hat{\theta}_n) = o_p(1)$ implies that both (1') and (2) hold asymptotically.

%gives some intuition for Theorems \ref{th:global_cv_OI}, \ref{th:global_ms_OI}. It implies that the PL inequality holds for $\|\theta-\theta^\dagger\|$ large relative to $C_4 \sqrt{\varphi}$, but not when $\theta$ is arbitrarily close to $\theta^\dagger$. Theorems \ref{th:global_cv_OI}, \ref{th:global_ms_OI} proceed in two steps. First, a global convergence phase, similar to Theorem \ref{th:global_cv_JI} is derived, away from $\theta^\dagger$. Then, Propositions \ref{prop:local_cv}, \ref{prop:local_cv_ms} are used to predict local convergence around $\theta^\dagger$ using approximate linearity of the moments. These Propositions require the local identification, in Assumption \ref{ass:1prim} (iii), to hold over a sufficiently large neighborhood of $\theta^\dagger$. In Theorem \ref{th:global_ms_OI}, condition (\ref{eq:glotoloc}) ensures that the first step can produce a $\theta_k$ which is local enough to apply Proposition \ref{prop:local_cv_ms}. 

\paragraph{Further characterization of Assumption \ref{ass:conds} (Just-Identified).}

Like star-convexity, Assumption \ref{ass:conds} is stated relative to the unknown $\theta^\dagger$. The following Proposition gives several conditions under which Assumption \ref{ass:conds} (a) holds and properties implied by these conditions.

\begin{proposition} \label{prop:charac} (Sufficient Conditions) Consider the following conditions:\\
    (a) $\sigma_{\min}[\overline{G}(\theta_1,\theta_2)] > \underline{\sigma} > 0$, for all $\theta_1,\theta_2 \in \mathbb{R}^{d_\theta}$,  (b) for all $\theta \in \mathbb{R}^{d_\theta}$, $G(\theta) = U S(\theta) V$ for $U,V$ invertible and $S(\theta)$ symmetric with $0 <  \underline{\lambda}_S < \lambda_{\min}[S(\theta)] < \overline{\lambda}_S < \infty$, for all $\theta \in \mathbb{R}^{d_\theta}$,
    (c) $g(\theta) = \partial_\theta F (\theta)$, for all $\theta \in \mathbb{R}^{d_\theta}$, where $F : \mathbb{R}^{d_\theta} \to \mathbb{R}$ is twice continuously differentiable, strongly convex.\\ %(d) $\sigma_{\min}[G(\theta)] > \underline{\sigma} > 0$ and $\lambda_{\min}[ A G(\theta) + G(\theta)^\prime A^\prime] > \underline{\sigma} > 0$ for $A$ invertible and all $\theta \in \Theta$.\\
    The following holds:
    (1) (c) $\Rightarrow$ (b) $\Rightarrow$ (a) $\Rightarrow$ Assumption \ref{ass:conds} (a) holds; 
    (2) (a) implies $g(\cdot)$ is one-to-one;
    (3) if (a) holds, there exists a reparameterization $h(\cdot) = \psi \circ g \circ \phi (\cdot)$ with $\phi$ one-to-one and $\psi$ affine, such that $1/2 h(\theta)^\prime W h(\theta)$ is strongly convex.%; (4) (d) implies $g(\cdot)$ is one-to-one and Assumption \ref{ass:condsbis} holds.    
\end{proposition}

Condition (a) does not require knowledge of $\theta^\dagger$ and implies that $g(\cdot)$ is one-to-one. The latter is often assumed for indirect inference.\footnote{See e.g. \citet{gourieroux1993}, Assumption (A4).} Condition (a) also implies (\ref{eq:SI}) with $\mu = \underline{\sigma}$.
When the Jacobian can be linearly rearranged into a symmetric positive definite matrix $S(\theta) = U^{-1} G(\theta) V^{-1}$, then condition (a) holds.  These problems can be thought of as \textit{implicitly convex} in the special case where where $S$ is the second derivative of a convex function. For a given $\theta \in \mathbb{R}^{d_\theta}$, the decomposition (b) always exists: the singular value decomposition gives $G(\theta) = U(\theta) S(\theta) V(\theta)$ where $U(\theta),V(\theta)$ are unitary and $S(\theta)$ is diagonal with positive entries. A lesser known result, due to \citet{frobenius1910} shows that any square matrix can be written as the product of two real symmetric matrices; here $G(\theta) = S_1(\theta) S_2(\theta)$. The Jordan normal form of $G(\theta)$ can be used to compute this factorization \citep{bosch1986}. If $G(\theta)$ is invertible, for all $\theta \in \mathbb{R}^{d_\theta}$, and $U,V$ or one of $S_1,S_2$ do not vary with $\theta$, in the singular value or Frobenius decomposition, then (b) holds. Under condition (c), $g$ is cyclically , and thus strongly, monotone.% Condition (d) implies that $A g$ is monotone; this is related to a global identification condition in \citet{fisher1966} and \citet{rothenberg1971} which requires $\text{det}[G(\theta)] > 0$ and $G(\theta) + G(\theta)^\prime$ positive semidefinite for all $\theta$, where $\text{det}$ is the determinant. Their conditions imply $g$ is one-to-one by the \citet{gale1965} Theorem. Similar conditions for overidentified models can be found in Proposition \ref{prop:characOI}. 

\begin{proposition} \label{prop:repar} (Reparameterization)
  Take $h : \mathcal{U} \to \mathbb{R}^{d_\theta}$, one-to-one, continuously differentiable on $\mathcal{U}$, a convex set, with $0 < \underline{\sigma}_h \leq \min_{u \in \mathcal{U}} \sigma_{\min}[\partial_u h(u)] \leq \max_{u \in \mathcal{U}} \sigma_{\max}[\partial_u h(u)] \leq \overline{\sigma}_h < \infty$. Let $u^\dagger=h^{-1}(\theta^\dagger)$, the minimizer of $Q\circ h$.\\ 
  1) Suppose Assumption \ref{ass:conds} (a) holds for $g$, let:
  \begin{align*}
    L_{1,h} = \sup_{u \in \mathcal{U}} \|\partial_u h(u)-\partial_u h(u^\dagger)\|, \,
    L_{2,h} = \sup_{u \in \mathcal{U},\omega \in [0,1]} \|h(\omega u +(1-\omega)u^\dagger) - \omega h(u) - (1-\omega)h(u^\dagger)\|.
  \end{align*}
  If $\underline{\sigma} > [L_{1,h}  \overline{\sigma} + L_{2,h} L \overline{\sigma}_h]/\underline{\sigma}_h$, where $L$ is the Lipschitz constant of $G$,
  then Assumption \ref{ass:conds} (a) holds for $g \circ h$. In particular, if $h = A u + b$ is affine with $A$ invertible then $L_{1,h}=L_{2,h}=0$ and Assumption \ref{ass:conds} (a) holds for $g \circ h$.\\
  2) Suppose Assumption \ref{ass:conds} (b) holds for $g$. If $\|h(u) - h(u^\dagger)\| \geq \mu \|u - u^\dagger\|$, for some $\mu >0 $ and all $u \in \mathcal{U}$, then Assumption \ref{ass:conds} (b) holds for $g \circ h$.  
\end{proposition}

Strong convexity is preserved by affine transformations and reparameterization that satisfy particular component-wise monotonicity constraints on the reparameterization \citep[e.g.][Sec3.2]{boyd2004}. Proposition \ref{prop:repar} shows that Assumption \ref{ass:conds} (a) is also preserved by affine transformations and moderately non-linear reparameterizations $h$. Hence, under Assumption \ref{ass:conds} (a), optimization should be locally robust to the choice of parameterization. Assumption \ref{ass:conds} (b) is preserved if $h$ is strongly injective, a mild requirement. In particular, invertible affine transformations preserve Assumption \ref{ass:conds} (b). Similar statements for overidentified models can be found in Propositions \ref{prop:characOI}, \ref{prop:reparOI}. Propositions \ref{prop:quasar_convex} and \ref{prop:repar} together imply that if $Q$ is strongly convex for a particular parameterization, e.g. reduced-form coefficients, then Assumption \ref{ass:conds} (b) holds for $Q \circ h$ where $h$ satisfies the conditions above, where $h$ is the mapping from reduced form to structural coefficients.

\section{Recommendations for Practice} \label{sec:recs}

\subsection{Checking whether Assumption \ref{ass:conds} holds} \label{sec:check}

The global convergence results hinge on Assumption \ref{ass:conds} (b) as it confers the objective several key properties for optimization. In some cases it may be feasible to verify analytically that one of the conditions in Figure \ref{fig:summary} or Proposition \ref{prop:charac} hold. For some models, it is possible to construct moments that identify the parameters, typically using injectivity arguments. In that case, (\ref{eq:SI}) holds which implies Assumption \ref{ass:conds} (b) holds, under regularity conditions. 

For more complex models, it may only be possible to evaluate numerically over a representative set of points, whether one of these conditions is likely to holds, or not. 
Since Assumption \ref{ass:conds}, and its sample counterpart Assumption \ref{ass:conds_n}, depend on the unknown minimizer or $Q$, resp. $Q_n$, it is not possible to check the conditions numerically before the performing the estimation. It is possible to check a stronger condition which does not take an estimate $\hat{\theta}_n$ as input, however.

In the main results, the constant $\rho \underline{\sigma}$ can be arbitrarily small. In practice, however, when $\rho\underline{\sigma} \to 0$ the convergence rate $(1-\overline{\gamma}) \to 1$ is arbitrarily slow. The following approximates an upper bound for $(1-\overline{\gamma})$, assuming correct specification, and the corresponding number of iterations $\underline{k}$ required to achieve $Q_n(\theta_{\underline{k}}) - Q_n(\hat{\theta}_n) \leq \varepsilon [Q_n(\theta_0) - Q_n(\hat{\theta}_n) ]$ for a user-chosen $\varepsilon \in (0,1)$. In practice, these bounds can be very conservative. The value $\underline{k}$ mainly indicates whether global convergence is practically feasible (e.g. $\underline{k} \leq 10^3$) or not (e.g. $\underline{k} \geq 10^{12}$).

When $\hat{\theta}_n$ is unknown, before the estimation is performed, it is only possible to verify a stronger condition. The following considers a sample analog of (\ref{eq:SIp}), introduced above:
\begin{align} \|G_n(\theta_1)^\prime W_n [\overline{g}_n(\theta_1)-\overline{g}_n(\theta_2)]\| \geq \mu_n \|\theta_1-\theta_2\|, \tag{\ref{eq:SIp}} \end{align}
for some $\mu_n > 0$. In the following, the finite grid of pairs $\Theta_K = \{ (\theta^1_1,\theta^2_1),\dots,(\theta^1_K,\theta^2_K) \}$ will be used for that purpose. It construction is discussed in more detail below. Suppose $\theta^1_k \neq \theta^2_{k}$ for each $k$, compute:
\begin{align*} \mu_k &= \frac{\|P_{k,n} G_n(\theta^1_k)^\prime W_n [\overline{g}_n(\theta^1_k)-\overline{g}_n(\theta^2_{k})]\|}{\|\theta^1_k-\theta^2_{k}\|},\quad
               C_{3,k} = \frac{\|\theta^1_k-\theta^2_{k}\|}{\|\overline{g}_n(\theta^1_k)-\overline{g}_n(\theta^2_{k})\|_{W_n}},\\
               L_{Q,P,k} &= \frac{\|P_{k,n} H_n(\theta^1_k)(\theta^1_k-\theta^2_k)\|}{\|\theta^1_k-\theta^2_{k}\|}, \end{align*}
where $P_{k,n}$ is computed using $\theta^1_k$ and the algorithm of choice.\footnote{Note that $L_{Q,P,k}$ involves a Hessian-vector product, which can be computed using only gradients: $H_n(\theta^1_k)(\theta^1_k-\theta^2_{k}) \simeq [\partial_\theta Q_n(\theta^1_k + \epsilon[\theta^2_{k}-\theta^1_k])-\partial_\theta Q_n(\theta^1_k)]/\epsilon$, for $\epsilon$ small.}  Then compute: \[ (1-\overline{\gamma})^2 = \max\left(0,1 - [\hat{\mu}_n \hat{C}_{3,n}]^2 /[4 \hat{L}_{Q,P,n}]\right), \quad \underline{k} \geq \frac{\log(\varepsilon)}{\log(1-\overline{\gamma})}, \]
where $\hat{\mu}_n = \min_k \mu_k$, $\hat{C}_{3,n} = \min_k C_{3,k}$, and $\hat{L}_{Q,P,n} = \max_k L_{Q,P,k}$. The normalization using $P_{k,n}$ ensures that these values are invariant to linear reparameterizations of the parameters and/or moments for \textsc{gn} or \textsc{nr}. As a reference, with the normalization linear models have $\hat{\mu}_n=1$ under the standard rank condition. To ensure the product $P_{k,n} G_n(\theta_k^1)$ is well behaved, it is recommended to compute a pseudo-inverse of $G_n(\theta_k^1)^\prime W_n G_n(\theta_k^1)$ in the case of \textsc{gn}. This yields $\mu_k = 0$ and $\overline{\gamma}=0$ when $G_n(\theta_k^1)$ is numerically close to singular in the relevant direction. If the conditions fail or the bounds indicate that convergence is not practically feasible, typically when $\hat{\mu}_n < 10^{-2}$ for \textsc{gn},\footnote{Dividing $\mu_n$ by 10 approximately reduces $\overline{\gamma}$ by a factor of 100, by a local expansion argument. Convergence becomes significantly slower when $\hat{\mu}_n$ approaches $0$.} gradient-based methods need to be modified to ensure global convergence, using multiple starting values or a hybrid approach with theoretical guarantees, see \citet{Forneron2022b} for an explicit algorithm in that setting. 

\paragraph{Constructing $\Theta_K$.} Take a set $\Theta$ large enough that $\hat{\theta}_n \in \Theta$ is likely. The grid $\Theta_K$ should be dense in $\Theta$ so that, as the number of points $K$ increases, any $\theta \in \Theta$ is arbitrarily close to some value in the grid.  For $\Theta = [0,1]^{d_\theta}$, the Sobol and Halton sequences have this property, and are readily available in statistical software (R, Matlab, Python, Julia). In general, when $\Theta = [\underline{\theta}_1,\overline{\theta}_1] \times \dots \times [\underline{\theta}_{d_\theta},\overline{\theta}_{d_\theta}]$, where $\underline{\theta}_1,\overline{\theta}_1$ denote lower (resp. upper) bounds on each coefficient, a sequence can be constructed from the Sobol or Halton sequence, denoted $(\vartheta_{i,k})$, $i=1,\dots,d_\theta$, $k=1,\dots,K$, by setting $\theta_{i,k} = \underline{\theta}_i + (\overline{\theta}_i-\underline{\theta}_i)\vartheta_{i,k}$. 

\subsection{Iteration dependent choice of learning rate $\gamma_k$} 
The results are stated for a fixed learning rate. In practice, adaptive choices of $\gamma_k$ are common, using a line search for instance. If the adaptive algorithm is tuned to satisfy the requirements for global convergence, then it is also globally convergent. To preserve convergence properties, additional tuning parameters are typically involved \citep[][Ch3.1]{nocedal-wright:06}. A backtracking line search, a simple and popular way to set the learning rate \citep[Ch3.1]{nocedal-wright:06}, is used as a benchmark comparison for the fixed learning rate used in the applications.% It is described below.

\begin{algorithm}[H]
  \small
  % Define some things:
  \SetKwInOut{Input}{Inputs}
  \SetKwInOut{Tuning}{Tuning Parameters}
  \SetKwInOut{Output}{Output}
  \SetKwInOut{Compute}{Compute}
  \SetKwInOut{Transport}{Transport}
  \SetKwInOut{Set}{Set}
  \SetKwInOut{Update}{Update}
  \SetKwInOut{Estimates}{Estimates}
  \Tuning{Initial $\gamma_{\text{init}}$, $\rho \in (0,1)$, $c \in (0,1)$. }
  \Input{Previous iterate $\theta_k$, moments $\overline{g}_n(\theta_k)$, Jacobian $G_n(\theta_k)$} 
  \Compute{Search direction: $p_k = (G_n(\theta_k)^\prime W_n G_n(\theta_k))^{-1} G_n(\theta_k)^\prime W_n \overline{g}_n(\theta_k)$,\\ $J_k = G_n(\theta_k)^\prime W_n \overline{g}_n(\theta_k)$.}
  \Set{$\gamma_k = \gamma_{\text{init}}$ and $\theta_{k+1} = \theta_k - \gamma_k p_k$}
  \While{ $Q_n(\theta_{k+1}) > Q_n(\theta_{k}) - c \gamma_k J_k^\prime p_k$ }{
    \Set{$\gamma_k = \rho \gamma_{k}$ and $\theta_{k+1} = \theta_k - \gamma_k p_k$}
  }
  \Output{New iterate $\theta_{k+1}$, Learning Rate $\gamma_k$.}
  \caption{Backtracking Line Search for Gauss-Newton} \label{algo:gnback}
 \end{algorithm}
By construction, $J_k^\prime p_k \geq 0$ so that the final $\gamma_k$ decreases the value of the objective function. The while loop terminates once the so-called \textit{Armijo condition} is met:\footnote{See \citet[p33]{nocedal-wright:06}, \citet[pp28-29]{Nesterov2018} for discussions.} $Q_n(\theta_{k+1}) \leq Q_n(\theta_{k}) - c \gamma_k J_k^\prime p_k$. For just-identified models, the termination criterion is feasible if $c$ is sufficiently small.\footnote{A sample analog of Proposition \ref{prop:PL} implies that $Q_n(\theta_{k+1}) \leq (1 - \overline{\gamma})^2 Q_n(\theta_k)$, for any $\theta_k \in \Theta$, when $\gamma \in (0,1)$ small enough for some $\overline{\gamma} \in (0,\gamma)$. Proposition \ref{prop:PL} (1)-(2) further imply for just-identified models that $Q_n(\theta_k) - Q_n(\hat{\theta}_n) \leq c_n J_k^\prime p_k$ for some $c_n>0$. The Armijo condition is feasible if $c$ is small enough.} Having $\theta_k = \hat{\theta}_n$ implies $p_k = 0$;  the condition holds for any $\gamma_k \in (0,1]$. %If $\gamma_k$ is sufficiently small, a linear approximation yields $Q_n(\theta_{k+1}) \simeq Q_n(\theta_k) - \gamma_k J_k^\prime p_k$. The constant $c \in (0,1)$ in the Armijo condition accounts for the inaccuracy in the approximation. 
A common choice is $c = 10^{-4}$, $\gamma_{\text{init}} = 1$, $\rho = 0.8$. These were used in all examples.\footnote{When there are bounds for parameters values, one can set $Q_n(\theta_{k+1}) = +\infty$ if $\theta_{k+1}$ is outside the bounds. Another approach is to project $\theta_{k+1}$ inside the bounds when $\gamma_k$ is too large.}

\section{Numerical and Empirical Applications} \label{sec:empirics}

\subsection{A pen and pencil example: the MA(1) model} 
\begin{figure}[h]
  \caption{MA(1): illustration of non-convexity and the rank condition} \label{fig:MA1}
\begin{center}
  \includegraphics[scale=0.5]{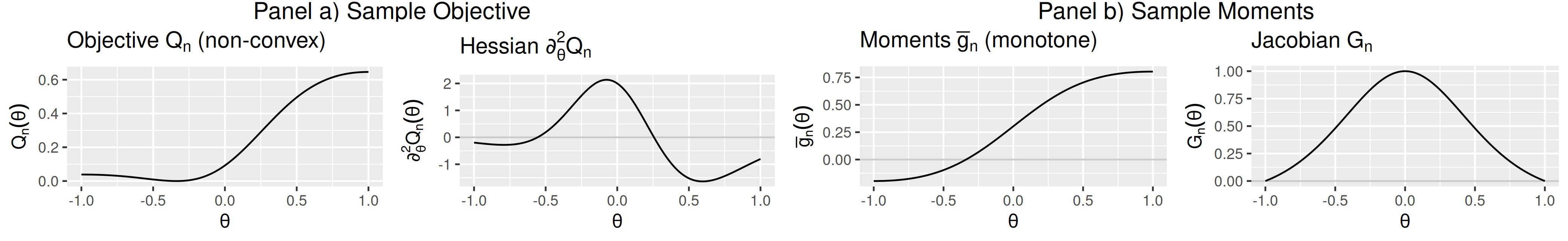}\\
  \notes{ \textbf{Legend:} simulated sample of size $n = 200$, $\theta^\dagger = -1/2$, $\overline{g}_n(\theta) = \hat\beta_1 - \theta/(1+\theta^2)$, $W_n = I_d$. The GMM objective (panel a) is non-convex but the sample moments (panel b) satisfy the rank condition.}
\end{center}
\end{figure}

The first example illustrates the main results using a simple MA(1) process:
\[ y_t = e_t - \theta^\dagger e_{t-1}, \quad e_{t} \overset{iid}{\sim} \mathcal{N}(0,1), \quad \theta^\dagger \in (-1,1), \]
for $t=1,\dots,n$. $\theta^\dagger$ is the parameter of interest. Set $p \geq 1$, following \citet[Ch4.3]{gourieroux1996}, $\theta^\dagger$ is estimated by matching coefficients from an auxiliary AR(p) model: $y_t = \beta_1 y_{t-1} + \dots + \beta_p y_{t-p} + u_t$. For $p = 1$, $\hat\beta_1 \overset{p}{\to} -\theta^\dagger/(1+\theta^{^\dagger 2})$ defines the moment condition: \[ \overline{g}_n(\theta) = \hat\beta_1 + \frac{\theta}{1+\theta^2},\] with Jacobian $G_n(\theta) =(1-\theta^2)/(1+\theta^2)^2 > 0$ for any $\theta \in (-1,1)$ and  $G_n(\theta) = 0$ for $\theta \in \{-1,1\}$. It has full rank on any interval of the form $[-1+\varepsilon,1-\varepsilon]$, $\varepsilon \in (0,1)$. However, Figure \ref{fig:MA1} shows that the Hessian $\partial^2_{\theta,\theta} Q_n(\theta)$ can be positive, negative, or equal to zero depending on the value of $\theta$ -- $Q_n$ is non-convex, especially when $\overline{g}_n(\theta)$ is large. 
Now notice that:  $\overline{g}_n(\theta) = \partial_\theta F_n(\theta) \text{ where } F_n(\theta) = \hat\beta_1 \theta + \frac{1}{2}\log(1+\theta^2),$ which not a GMM objective but is nevertheless convex on $[-1,1]$, strongly convex on any $[-1+\varepsilon,1-\varepsilon]$, $\varepsilon \in (0,1)$. Hence, $\overline{g}_n$ is cyclically monotone and statisfies Assumption \ref{ass:conds} (a). Note that implicitly, \textsc{gn} minimizes the convex $F_n$ -- whereas \textsc{nr} explicitly minimizes the non-convex $Q_n$. This is specific to the just-identified case ($p=1$), since an $F_n$ cannot be defined in the over-identified case ($p>1$).

Table \ref{tab:MA1} shows the search paths for \textsc{nr} and \textsc{gn} with a fixed $\gamma=0.1$ as well as R's built-in \textit{optim}'s \textsc{bfgs} implementation and the bound-constrained \textsc{l-bfgs-b}. \textsc{nr} diverges, because the objective is locally concave at $\theta_0 = -0.6$. This is surprising given how close $\theta_0$ is to $\theta^\dagger$. Although $Q_n$ is locally convex around $\hat\theta_n$, which is useful for local optimization, the corresponding neighborhood can be fairly small from a practical standpoint.

\begin{table}[ht]
  \centering \caption{MA(1): search paths for \textsc{nr}, \textsc{gn}, \textsc{bfgs}, and \textsc{l-bfgs-b}} \label{tab:MA1}
  \setlength\tabcolsep{4.0pt}
    \renewcommand{\arraystretch}{0.9}
    {\footnotesize \begin{tabular}{r|cccccccccc|c}
    \hline \hline
  $k$ & 0 & 1 & 2 & 3 & 4 & 5 & 6 & 7 & \dots & 99 & $Q_n(\theta_{99})$ \\ 
    \hline 
    & \multicolumn{11}{c}{$p = 1$}\\ \hline
  \textsc{nr} & -0.600 & -0.689 & -0.722 & -0.749 & -0.772 & -0.793 & -0.811 & -0.828 & \dots & -0.993 & 0.038\\  \rowcolor{gray}
  \textsc{gn} & -0.600 & -0.560 & -0.529 & -0.504 & -0.484 & -0.466 & -0.451 & -0.438 & \dots & -0.338 & $7 \cdot 10^{-8}$ \\ \rowcolor{gray}
  \textsc{gn}-\textsc{back} & -0.600 & -0.202 & -0.326 & -0.338 & -0.338 & -0.338 & -0.338 & -0.338 & \dots & -0.338 & $7 \cdot 10^{-8}$\\
  \textsc{bfgs} & -0.600 & -0.505 & 4.425 & -0.307 & -0.359 & -0.338 & -0.337 & -0.337 & \dots & -0.337 & $7 \cdot 10^{-8}$\\ 
  \textsc{l-bfgs-b} & -0.600 & -0.505 & 1.000 & -0.455 & -0.375 & -0.318 & -0.341 & -0.339 & \dots & -0.338 & $7 \cdot 10^{-8}$\\ 
  \textsc{bfgs}$^\star$ & -0.600 & -0.462 & -0.286 & -0.345 & -0.340 & -0.338 & -0.338 & -0.338 & \dots & -0.338 & $7 \cdot 10^{-8}$\\ 
  \textsc{l-bfgs-b}$^\star$ & -0.600 & -0.462 & -0.286 & -0.345 & -0.339 & -0.338 & -0.338 & -0.338 & \dots & -0.338 & $7 \cdot 10^{-8}$\\
  \hline
  & \multicolumn{11}{c}{$p = 12$}\\ \hline
  \textsc{nr} & 0.950 & 0.956 & 0.961 & 0.965 & 0.969 & 0.972 & 0.975 & 0.978 &  \dots & 1.000 & 4.786 \\  \rowcolor{gray}
      \textsc{gn} & 0.950 & 0.890 & 0.860 & 0.834 & 0.810 & 0.787 & 0.763 & 0.740 &  \dots & -0.623 & 0.101 \\  \rowcolor{gray}
      \textsc{gn}-\textsc{back} & 0.950 & 0.350 & -0.089 & -0.478 & -0.591 & -0.616 & -0.616 & -0.623 &  \dots & -0.626 & 0.101 \\ 
      \textsc{bfgs} & 0.950 & -8.290 & -8.279 & -8.267 & -8.256 & -8.244 & -8.233 & -8.221 & \dots & -6.979 & 0.397 \\ 
      \textsc{l-bfgs-b} & 0.950 & -1.000 & -1.000 & -1.000 & -1.000 & -1.000 & -1.000 & -1.000 & \dots & -1.000 & 1.7 \\ 
     \hline \hline
  \end{tabular} } 
  \notes{ \textbf{Legend:} simulated data with sample size $n = 200$, $\theta^\dagger = -1/2$. For $p = 1$, $\overline{g}_n(\theta) = \hat\beta_1 - \theta/(1+\theta^2)$. For $p =12$, $\overline{g}_n(\theta) = \hat\beta_n - \beta(\theta)$ where $\beta(\theta)$ is the p-limit of the AR(p) coefficients, evaluated at $\theta$. $W_n = I_d$. The solutions are $\hat\theta_n = -0.339$ ($p=1$) and $\hat\theta_n = -0.626$ ($p=12$). \textsc{nr} = Newton-Raphson, \textsc{gn} = Gauss-Newton, \textsc{gn}-\textsc{back} = Gauss-Newton with backtracking line search (Algorithm \ref{algo:gnback}). The learning rate is $\gamma=0.1$ for \textsc{nr} and \textsc{gn}. \textsc{bfgs} = R's \textit{optim}, \textsc{l-bfgs-b} = R's \textit{optim} with bound constraints $\theta \in [-1,1]$. \textsc{bfgs}$^\star$ and \textsc{l-bfgs-b}$^\star$ apply the same optimizers to $F_n$ instead of $Q_n$. } 
\end{table}

\textsc{gn} converges steadily from the same $\theta_0$.  \textsc{bfgs} is more erratic, especially when $\theta_k \simeq -0.5$, i.e. $k = 1$, leading to a search outside the unit circle ($k=2$), before reaching an area where the iterations are better behaved ($k=3$ onwards). While here this is not too problematic, the objective function is well defined outside the bounds, this is more concerning in applications where the model cannot be solved outside the bounds -- this is illustrated in Section \ref{sec:blp}. A natural solution is to introduce bounds using \textsc{l-bfgs-b}. The search, however, remains somewhat erratic as seen in the Table. Compare these to \textsc{bfgs}$^\star$ and \textsc{l-bfgs-b}$^\star$ which minimize $F_n$, instead of $Q_n$, using the same \textit{optim}. Like \textsc{gn}, they steadily converge to $\hat\theta_n$.

\begin{figure}[ht] \caption{MA(1): illustration of the strong injectivity condition} \label{fig:MA1_p12_SI}
  \includegraphics[scale=0.5]{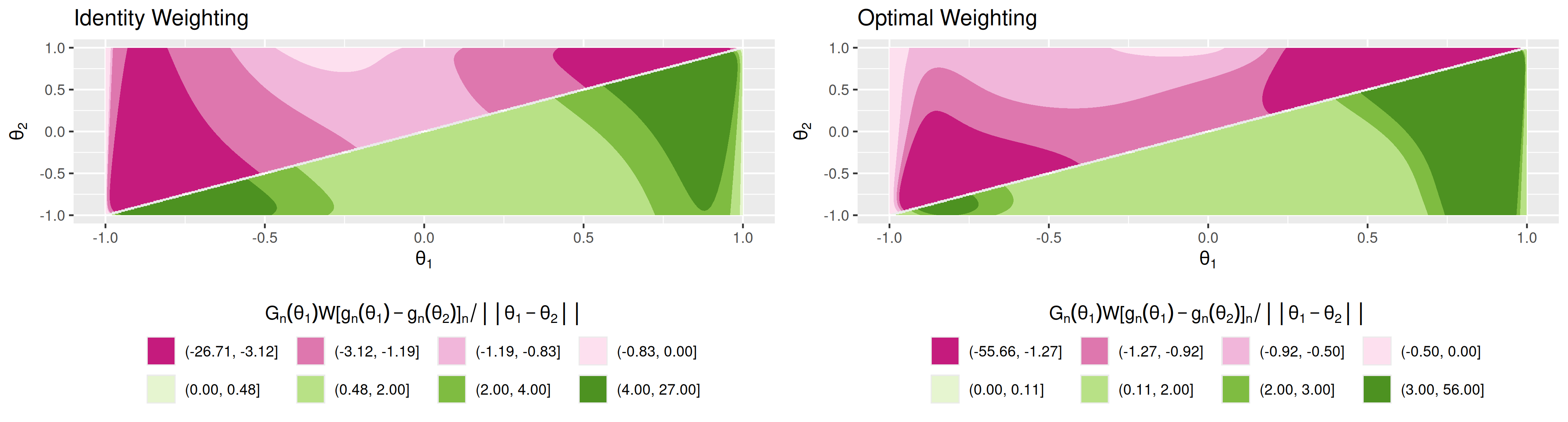}\\
  \notes{ \textbf{Legend:} the color groups are given by quantiles on the positive and negative values, so that the colors represent the same fraction of values on the left and right panels.}
\end{figure}

For $p=12$, the model is over-identified and the conditions are more challenging to check analytically. Figure \ref{fig:MA1_p12_SI} indicates that the strong injectivity condition (\ref{eq:SIp}) appear to hold, the value is bounded away from zero, except at the boundary. The choice of weighing matrix affects the constant $\mu$ in (\ref{eq:SIp}), as it appears to be smaller with optimal weighting.

Table \ref{tab:MA1} shows that \textsc{nr}, \textsc{bfgs} and \textsc{l-bfgs-b} all fail to converge from $\theta_0 = 0.95$, a starting value with negative curvature, with identity weighting.\footnote{\textsc{l-bfgs-b} relies on projection descent which maps search directions outside the unit circle to $-1$ or $1$ where $\partial_\theta Q_n(-1) = \partial_\theta Q_n(1) = 0$, a stationary point for (\ref{eq:update}).} Compare with \textsc{gn}, which steadily converges to $\hat\theta_n$. Starting closer to the solution, \textsc{bfgs} and \textsc{l-bfgs-b} also fail to converge using $\theta_0 = 0.6$; \textsc{gn} remains accurate (not reported). R codes can be found in Appendix \ref{apx:Rcode}.

%, and the condition for global convergence requires $G_n(\theta_1)^\prime W_n G_n(\theta_2)$ to be non-singular for all pairs $(\theta_1,\theta_2) \in \Theta \times \Theta$. For just-identified models, this amounts to $G_n(\theta)$ non-singular for all $\theta \in \Theta$. Figure \ref{fig:MA1over} in Appendix \ref{apx:addise} illustrates, similar to Figure \ref{fig:MA1}, that $Q_n$ is non-convex and that the rank condition holds for $\Theta = [-1+\varepsilon,1-\varepsilon]$. $F_n$ is no longer defined because of over-identification. 
%In this example, the erratic behaviour of \textsc{bfgs} and \textsc{l-bfgs-b} does not prevent the algorithms from eventually converging. However, in multivariate settings this result in failed convergence as illustrated in Example [non-invertible MA(1)] and the empirical application(s).

\paragraph{The impact of misspecification on estimation.} As discussed in Section \ref{sec:lit_char}, when the degree of misspecification $\varphi$ becomes large, local optima may appear and making gradient-based optimizers non-globally convergent. To illustrate this issue, consider the true DGP:
\[ y_t = e_t - \theta_1 e_{t-1} - \theta_2 e_{t-2},  \quad e_{t} \overset{iid}{\sim} \mathcal{N}(0,1), \quad \theta^\dagger \in (-1,1),\]
where $\theta_2 \in \{0,0.4,0.8\}$ determines the degree of misspecification. The following sets $\theta_1 = - 0.1$ to ensure invertibility as $\theta_2$ varies. The moments are the same as above with $p=12$. 
\begin{figure}[ht]
  \caption{MA(1): effect of misspecification on $Q_n$} \label{fig:MA1_miss}
\begin{center}
  \includegraphics[scale=0.5]{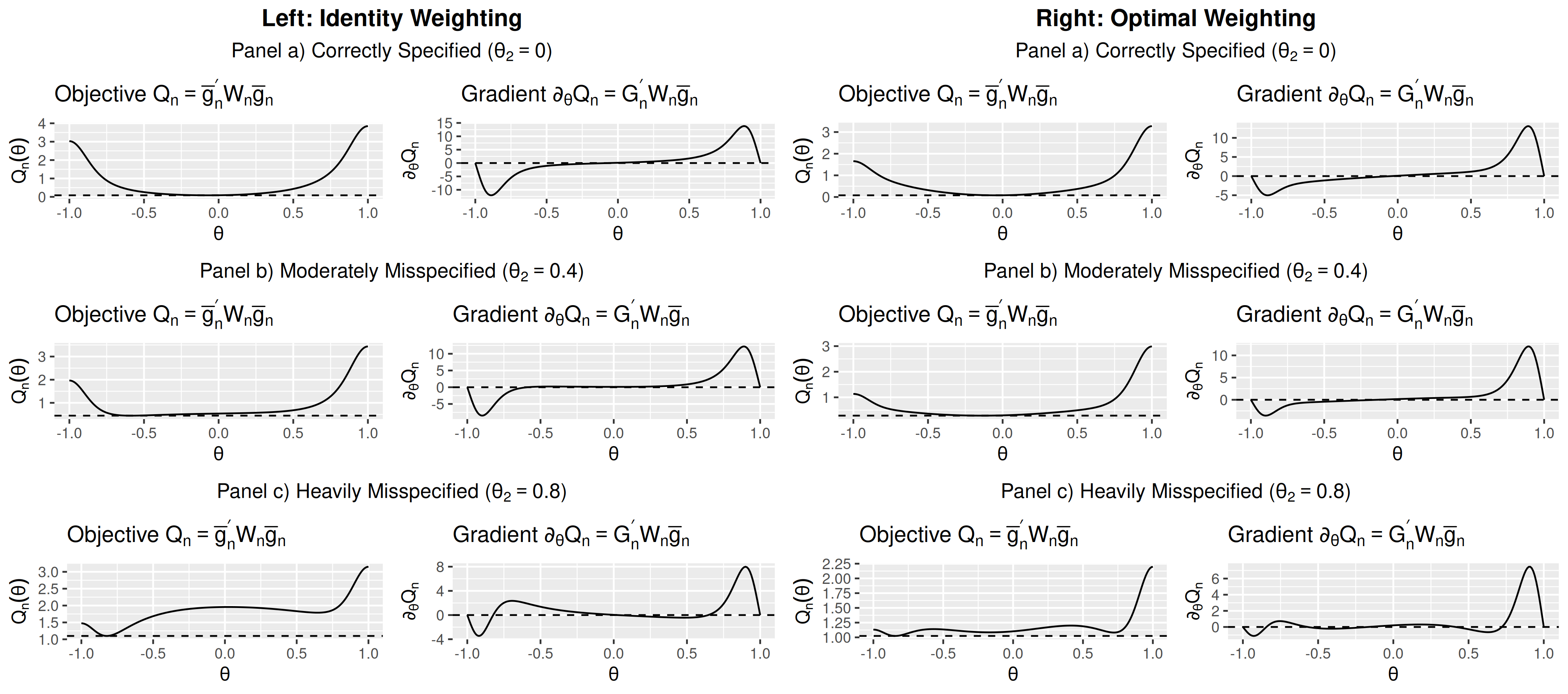}\\
  \notes{ \textbf{Legend:} simulated sample of size $n = 200$, $\theta_1 = -0.1$, $p=12$. }
\end{center}
\end{figure}

\begin{table}[ht]
  \centering \caption{MA(1):  estimates under misspecification} \label{tab:MA1miss}
  \setlength\tabcolsep{4.0pt}
    \renewcommand{\arraystretch}{0.9}
    {\footnotesize \begin{tabular}{r|cc|cc|cc|cc|cc|cc}
    \hline \hline 
    & \multicolumn{4}{c|}{Correctly Specified} & \multicolumn{4}{c|}{Moderately Misspecified} & \multicolumn{4}{c}{Heavily Misspecified}\\ \hline
    $W_n$ & \multicolumn{2}{c|}{Identity} & \multicolumn{2}{c|}{Optimal} & \multicolumn{2}{c|}{Identity} & \multicolumn{2}{c|}{Optimal} & \multicolumn{2}{c|}{Identity} & \multicolumn{2}{c}{Optimal}\\ \hline
     & $\hat{\theta}_n$ & $Q_n$ & $\hat{\theta}_n$ & $Q_n$ & $\hat{\theta}_n$ & $Q_n$ & $\hat{\theta}_n$ & $Q_n$ & $\hat{\theta}_n$ & $Q_n$ & $\hat{\theta}_n$ & $Q_n$\\ \hline
     \textsc{true} & -0.070 & 0.084 & -0.040 & 0.078 & -0.590 & 0.447 & -0.140 & 0.285 & -0.82 & 1.10 & -0.84 & 1.02\\ \rowcolor{gray}
     \textsc{gn} & -0.070 & 0.084 & -0.043 & 0.078 & -0.588 & 0.447 & -0.139 & 0.285 & 0.645 & 1.789 & 0.722 & 1.081\\
     \textsc{bfgs} & -14.4 & 0.084 & -22.8 & 0.078 & -6.95 & 0.52 & -7.35 & 0.285 & -1.21 & 1.10 & -5.92 & 1.09\\
     \textsc{l-bfgs-b} & -1.00 & 3.03 & -1.00 & 1.65 & -1.00 & 1.97 & -1.00 & 1.13 & -1.00 & 1.47 & -1.00 & 1.13\\ \hline \hline
    \end{tabular} }\notes{ \textbf{Legend:} simulated sample of size $n = 200$, $\theta_1 = -0.1$, $p=12$, $W_n = I_d$. \textsc{true} is the actual sample estimator. Starting value $\theta_0 = 0.9$. $\hat{\theta}_n$: estimates returned by optimizer, $Q_n$: minimized objective.}
\end{table}

Figure \ref{fig:MA1_miss} compares $Q_n$ and its gradient $\partial_\theta Q_n$, with identity and optimal weighting, at different degrees of misspecification. On intervals $[-1+\varepsilon,1-\varepsilon]$, $Q_n$ has no local optima for $\theta_2 \in \{0,0.4\}$, in line with Proposition \ref{prop:PLmis}. For the larger $\theta_2=0.8$, there are local optima: equal weighting has two (one maximum and minimum), optimal weighting has four (two maxima and minima). Also, the objective becomes flatter as $\theta_2$ increases, making optimization more challenging. Table \ref{tab:MA1miss} shows how this translates into estimation properties. As predicted, \textsc{gn} is robust to moderate misspecification but only converges to a local minimum under heavier misspecification. Other methods (\textsc{l-bfgs-b}, \textsc{bfgs}) systematically fail to converge.

%The MA(1) example above gave some analytical insights for the main results. The following illustrates the properties of \textsc{gn} in empirical settings.
\subsection{Estimation of a Random Coefficient Demand Model Revisited} \label{sec:blp}
The following revisits the results for random coefficient demand estimation in \citet{knittel2014} with the `fake' cereal data generated by \citet{nevo2001}.\footnote{It available in the R package BLPestimatoR \citep{brunner2017}. The data consists of 2,256 observations for 24 products (brands) in 47 cities over two quarters in 94 markets. The specification is identical to Nevo's, with cereal brand dummies, price, sugar content (sugar), a mushy dummy indicating whether the cereal gets soggy in milk (mushy), and 20 IV variables. } This is a non-linear instrumental variable regression with sample moment conditions: $\overline{g}_n(\theta,\beta) = \frac{1}{n} \sum_{j,t} z_{jt}[ \delta_{jt}(\theta) - x_{tj}^\prime \beta ]$,
where $z_{jt}$ are the instruments, $x_{jt}$ the linear regressors in market $j$ at time period $t$. The $8$ parameters of interest are the random coefficients $\theta$,\footnote{8 parameters are the unobserved standard deviation and the income coefficient on the constant term, price, sugar, and mushy.} which enter $\delta_{jt}$, recovered from market shares $s_{jt}$ using the fixed point algorithm of \citet{Berry1995}. The $25$ linear coefficients $\beta$ are nuisance parameters concentrated out by two-stage least squares for each $\theta$. The replication sets the maximum number of iterations for the contraction mapping to $20000$ and the tolerance level for convergence to $10^{-12}$. This is important for the optimization to be well-behaved; see e.g. \citet{brunner2017}, \citet{conlon2020}. The range of starting values used here is much wider than in these papers,\footnote{\citet[p25]{conlon2020} draw ``starting values from a uniform distribution with support 50\%
above and below the true parameter value.''} which explains why optimizers are more prone to crashing here than in their replications. Initial values are constructed as follows: the Sobol sequence generates values in $[0,1]^8$, the coefficients for standard deviations are adjusted to lie in $[0,10]$, those for income in $[-10,10]$. Values for which the contraction mapping produces an error are discarded until $50$ valid starting values are available.% for estimation.

%The regressors $x_{jt}$ includes the endegeous price term $p_{jt}$.

  %Since the goal here is to illustrate the complicated interactions between numerical optimizers and the inner loop which solves the model for each $\theta$, a common feature of structural estimation, error handling is not applied. In comparison, \textsc{gn} converges all the time to an accurate estimate. This indicates that the interactions between \textsc{gn} iterations and the fixed-point iterations are more numerically stable. 

% in $\Theta = [-10,10] \times \dots \times [-10,10]$. \textsc{gn} is run for $150$ iterations for all starting values with the same tuning parameter $\gamma=0.1$. 

\begin{table}[ht]
  \centering \caption{Demand for Cereal: performance comparison}  \label{tab:BLP}
  \setlength\tabcolsep{4.5pt}
    \renewcommand{\arraystretch}{0.9}
{\footnotesize \begin{tabular}{rr|cccc|cccc|c|c|c}
  \hline \hline
  & & \multicolumn{4}{c|}{\textsc{stdev}} & \multicolumn{4}{c|}{\textsc{income}}  & \multirow{2}{*}{objs} &  & \multirow{2}{*}{time}\\
    &  & const. & price & sugar & mushy  & const. & price & sugar & mushy &   &  \multirow{-2}{*}{crash}  & \\ 
  \hline
%  \multicolumn{4}{l}{\textsc{True}} &  &  &  &  &  &  &   \\
\multirow{2}{*}{\textsc{true}} & est & 0.28 & 2.03 & -0.01 & -0.08 & 3.58 & 0.47 & -0.17 & 0.69 & 33.84 & \multirow{2}{*}{-} &  \\ 
& se & 0.11 & 0.76 & 0.01 & 0.15 & 0.56 & 3.06 & 0.02 & 0.26 & - &  & \multirow{-2}{*}{-}\\
   \hline
%\multicolumn{4}{l}{\textsc{gn}} &  &  &  &  &  &  &\\ 
\rowcolor{gray}
  & avg & 0.28 & 2.03 & -0.01 & -0.08 & 3.58 & 0.47 & -0.17 & 0.69 & 33.84 &   &    \\  
\rowcolor{gray}
  \multirow{-2}{*}{\textsc{gn}}  &   std  & 0.00 & 0.00 & 0.00 & 0.00 & 0.00 & 0.00 & 0.00 & 0.00 & 0.00 & \multirow{-2}{*}{0}   & \multirow{-2}{*}{00:03:51} \\ 
      \hline \rowcolor{gray}
%\multicolumn{4}{l}{\textsc{bfgs}} &  &  &  &  &  &  &   \\
 & avg & 0.28 & 2.03 & -0.01 & -0.08 & 3.58 & 0.47 & -0.17 & 0.69 & 33.84 &   &  \\ \rowcolor{gray}
 \multirow{-2}{*}{\textsc{gn}-\textsc{b}} &  std & 0.00 & 0.00 & 0.00 & 0.00 & 0.00 & 0.00 & 0.00 & 0.00 & 0.00 & \multirow{-2}{*}{0} & \multirow{-2}{*}{00:00:20}  \\ 
   \hline
& avg & 0.29 & 2.18 & -0.01 & -0.08 & 3.59 & 0.43 & -0.17 & 0.71 & 35.17 &   &  \\ 
 \multirow{-2}{*}{\textsc{gd}-\textsc{b}} &  std & 0.01 & 0.23 & 0.00 & 0.01 & 0.70 & 4.58 & 0.01 & 0.06 & 1.14 & \multirow{-2}{*}{0} & \multirow{-2}{*}{09:59:04}\\ 
   \hline
\multirow{2}{*}{\textsc{bfgs}} & avg & 0.53 & 1.90 & -0.29 & -1.72 & 5.03 & 0.97 & -0.22 & 0.21 & 4555.97 &  \multirow{2}{*}{23} & \\ 
 &  std & 1.26 & 0.69 & 1.45 & 8.54 & 7.55 & 2.63 & 0.24 & 2.51 & $2.35 \cdot 10^4$ &  & \multirow{-2}{*}{00:01:46}  \\ 
   \hline
%\multicolumn{4}{l}{\textsc{nm}} &  &  &  &  &  &  &\\
\multirow{2}{*}{\textsc{nm}} & avg & 1.10 & 5.28 & -0.09 & 0.78 & 4.99 & 3.68 & -0.28 & 3.29 & 543.20 &  \multirow{2}{*}{3} & \\ 
&  std & 1.44 & 7.74 & 0.11 & 1.86 & 4.43 & 8.99 & 0.26 & 3.47 &  700.29 &  & \multirow{-2}{*}{00:01:19} \\ 
   \hline
\multirow{2}{*}{\textsc{sa}} & avg & 7.66 & 9.52 & -0.94 & 10.45 & -0.27 & 2.01 & 3.73 & 3.07 & $8.27 \cdot 10^4$ & \multirow{2}{*}{2} & \\ 
&   std & 3.25 & 3.73 & 0.58 & 4.09 & 5.78 & 6.66 & 3.94 & 6.35 & $8.60 \cdot 10^4$ &  & \multirow{-2}{*}{01:45:59} \\ 
   \hline
%\multicolumn{4}{l}{\textsc{sa+nm}} &  &  &  &  &  &  &\\
\multirow{2}{*}{\textsc{sa+nm}} & avg & 1.02 & 8.90 & -0.13 & 1.00 & 4.75 & 7.64 & -0.29 & 4.65 & 613.70 & \multirow{2}{*}{2} & \\ 
&   std & 1.26 & 8.95 & 0.15 & 1.63 & 4.19 & 11.08 & 0.26 & 5.68 &  558.19 &  & \multirow{-2}{*}{01:47:33} \\ 
   \hline \hline
\end{tabular}  } 
  \notes{ \textbf{Legend:} Comparison for 50 starting values where $[0,10] \times \dots \times [0,10]$ for standard deviations and $[-10,10] \times \dots \times [-10,10]$ for income coefficients. Avg, Std: sample average and standard deviation of optimizer outputs. \textsc{true}: full sample estimate (est)  and standard errors (se). Objs: avg and std of minimized objective value. crash: optimization terminated by an error. time: average run time for optimizers in hours:minutes:seconds. \textsc{gn} uses $\gamma=0.1$, $k=150$ iterations. \textsc{gn}-\textsc{b} and \textsc{gd}-\textsc{b} use a backtracking line search, terminates once $Q_n(\theta_{k}) - Q_n(\theta_{k+1}) \leq 10^{-8}$. Additional results can be found in Appendix \ref{apx:cereal_additional}.  }
\end{table}

\begin{figure}[H]
  \caption{Demand for Cereal: distribution of minimized objective values} \label{fig:BLPobjs}
  \centering
  \includegraphics[scale=0.5]{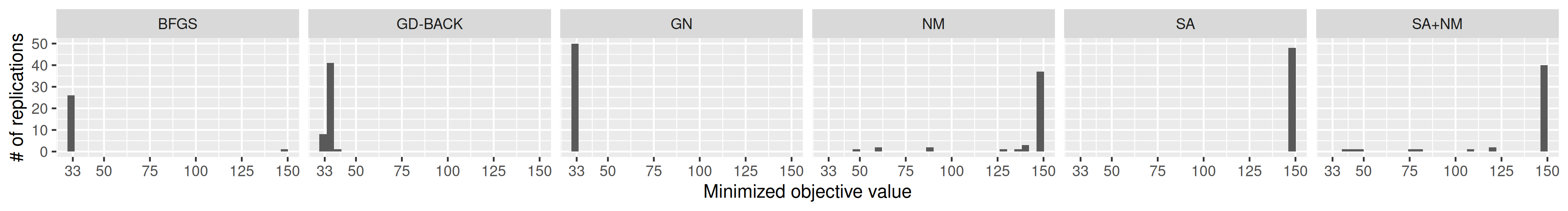}\\
  \notes{ \textbf{Legend:} Comparison for 50 starting values. Minimized objective values for non-crashed optimizations. Objective values are truncated from above at $Q_n(\theta) = 150$. }
%\end{center}
\end{figure}

Table \ref{tab:BLP} and Figure \ref{fig:BLPobjs} compare the performance of quasi-Newton (\textsc{bfgs}), Nelder-Mead (\textsc{nm}), Simulated-Annealing (\textsc{sa}), and Nelder-Mead after Simulated-Annealing (\textsc{sa}+\textsc{nm}), using R's default optimizer \textit{optim}, with Gauss-Newton (\textsc{gn}) and Gradient-Descent (\textsc{gd}) for 50 different starting values.\footnote{The solution of the contraction mapping is not well defined for all values in $\Theta$, so we use the first $50$ values produced by the Sobol sequence such that $\delta_{jt}$ is finite for all $j,t$. }  As reported in \citet{knittel2014}, optimization can crash often.\footnote{The optimizers will crash when the fixed point algorithms fail to return finite values. This is typically the case when the search direction was poorly chosen at the previous iteration.} Crashes could be avoided using error handling (try-catch statements). However, this may not be enough to produce accurate estimates as the next application will illustrate.\footnote{\citet{conlon2020} illustrate that modifications to the fixed-point algorithm and specific optimizer implementations to handle near-singularity of the Hessian can also improve performance for \textsc{bfgs}.} Only \textsc{gn} systematically produces accurate estimates; \textsc{bfgs} crashes 46\% of the time and has one highly inaccurate estimate. Derivative-free optimizers (\textsc{nm}, \textsc{sa}, \textsc{sa+nm}) can produce inaccurate estimates. \textsc{gd} can be very slow to converge. Using a backtracking line search, \textsc{gn} converges in 11 iterations on average, compared to 8816 for \textsc{gd} -- which has a higher maximum number of iterations set at 10000, compared to 150 for \textsc{gn}. Increasing the maximum number of iterations for \textsc{gd} would improve the estimates at the expense of further computation time. 

\begin{figure}[H]
  \caption{Demand for Cereal: Gauss-Newton iterations for 5 starting values} \label{fig:BLPconv}
  \centering %\begin{center}
  \includegraphics[scale=0.5]{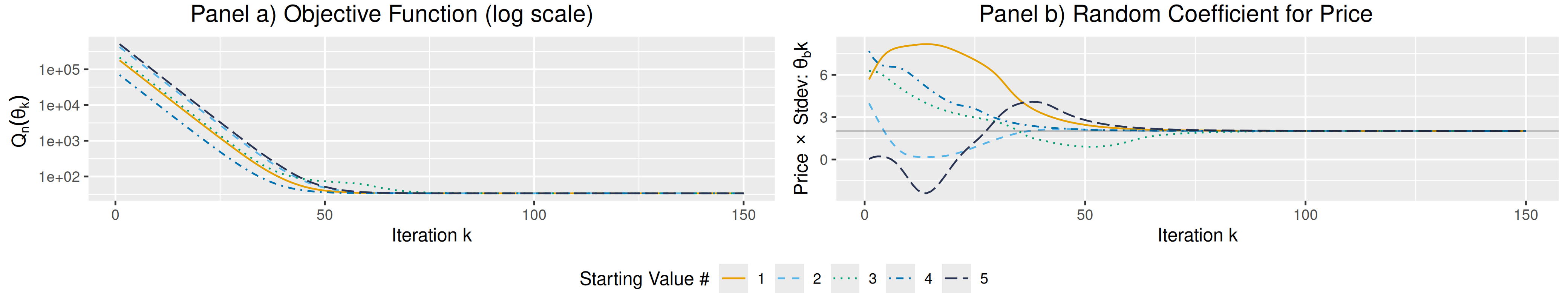}\\
  \notes{ \textbf{Legend:} 150 \textsc{gn} iterations for 5 starting values in $[0,10] \times \dots \times [0,10]$ for standard deviations and $[-10,10] \times \dots \times [-10,10]$ for income coefficients. Panel b) horizontal grey line = full sample estimate.}
%\end{center}
\end{figure}

Figure \ref{fig:BLPconv}, illustrates the convergence of \textsc{gn} for the first 5 starting values. In line with the predictions of Theorem \ref{th:global_cv_cs}, though $Q_n$ is non-convex, \textsc{gn} iterations steadily converge to the solution. This type of ``Gauss-Newton regression'' is related to \citet{salanie2022} who compute two-stage least-squares for linearized BLP.

\subsection{Innovation, Productivity, and Monetary Policy} \label{sec:impulse}
The second application revisits \citet{moran2018}'s estimation of a model with endogenous total factor productivity (TFP) growth \citep[see][Sec2, for details about the model]{moran2018}. They estimate parameters related to Research and Development (R\&D) by matching the impulse response function (IRF) of an identified R\&D shock to R\&D and TFP in a small-scale Vector Auto-Regression (VAR) estimated on U.S. data. 

The parameters of interest are $\theta = (\eta,\nu,\rho_s,\sigma_s)$ which measure, respectively, the elasticity of technology creation to R\&D, R\&D spillover to adoption, the persistence coefficient and size of impulse to the R\&D wedge. The sample moments are $\overline{g}_n(\theta) = \hat \psi_n - \psi(\theta)$, $\hat \psi_n$ and $\psi(\theta)$ are the sample and predicted IRFs, respectively. The latter is computed using Dynare in Matlab. To minimize $Q_n$, the authors use Sims's \textsc{csminwel} (\textsc{sims} in the Table, Figures)\footnote{Details about \textsc{csminwel} and code can be found at: \url{http://sims.princeton.edu/yftp/optimize/}.} algorithm with a reparameterization which bounds the coefficients.\footnote{The replication uses the mapping $\theta_j = \underline{\theta}_j + \frac{\overline{\theta}_j-\underline{\theta}_j}{1 + \exp(-\vartheta_j)}$, where each $\vartheta_j$ is unconstrained. The original study relied on $\theta_j = 1/2(\overline{\theta}_j + \underline{\theta}_j) + 1/2(\overline{\theta}_j - \underline{\theta}_j) \frac{\vartheta_j}{\sqrt{1 + \vartheta_j^2}}$, which we found to make optimizers very unstable.} Although this type of reparameterization is commonly used, the Jacobian is singular at the boundary; this matters for both local and global convergence, according to the results. As in the demand estimation, initial values are constructed using the Sobol sequence and adjusted to match the bounds used in the original study, reported in the last two rows of Table \ref{tab:imp_no}.

%project iterations outside the bounds inside, this is the approach used by \textsc{l-bfgs-b} in R.\footnote{The projection replaces the $j$-th coefficient $\theta_{k j}$ at iteration $k$ with $\underline{\theta}_j$ when $\theta_{k j}< \underline{\theta}_j$, or $\overline{\theta}_j$ when $\theta_{k j}> \overline{\theta}_j$. This is how box constraints are implemented when using \textsc{l-bfgs-b} in R with \textit{optim}. See Table \ref{tab:imp_proj} in Appendix \ref{apx:IRF_additional} for results using \textsc{gn} with projection.}

\begin{table}[h]
  \centering \caption{Impulse Response Matching: performance comparison}  \label{tab:imp_no}
  \setlength\tabcolsep{3.25pt}
    \renewcommand{\arraystretch}{0.9}
{\footnotesize \begin{tabular}{rr|cccc|c|c|c||cccc|c|c|c}
  \hline \hline
  &  & $\eta$ & $\nu$ & $\rho_s$ & $\sigma_s$  & objs  & crash & time & $\eta$ & $\nu$ & $\rho_s$ & $\sigma_s$  & objs  & crash & time \\  \hline
  \multirow{1}{*}{\textsc{true}} & est & 0.30 & 0.29 & 0.39 & 0.17 & 4.65 &   - &  - &0.30 & 0.29 & 0.39 & 0.17 & 4.65 &   - & - \\ 
  \hline
  & & \multicolumn{7}{c||}{\textsc{without reparameterization}} & \multicolumn{7}{c}{\textsc{with reparameterization}} \\
   \hline
   %\multicolumn{4}{l}{\textsc{gn}} &    &  &\\
   \rowcolor{gray}
   & avg &  0.30 & 0.29 & 0.39 & 0.17 & 4.65 &  &  & 0.30 & 0.29 & 0.39 & 0.17 & 4.65  &  &   \\  \rowcolor{gray}
   \multirow{-2}{*}{\textsc{gn}}  & std  &  0.00 & 0.00 & 0.00  & 0.00 & 0.00 & \multirow{-2}{*}{1} &\multirow{-2}{*}{00:00:56} & 0.00 & 0.00 & 0.00 & 0.00 & 0.00  &  \multirow{-2}{*}{9}  & \multirow{-2}{*}{00:00:55} \\ 
     \hline \rowcolor{gray}
   & avg &  0.30 & 0.29 & 0.39 & 0.17 & 4.65 &  &  & 0.30 & 0.29 & 0.39 & 0.17 & 4.65 &  & \\  \rowcolor{gray}
   \multirow{-2}{*}{\textsc{gn}-\textsc{b}}   & std  &  0.00 & 0.00 & 0.00  & 0.00 & 0.00 &  \multirow{-2}{*}{0} &\multirow{-2}{*}{00:00:04} & 0.00 & 0.00 & 0.00  & 0.00 & 0.00 & \multirow{-2}{*}{1} &\multirow{-2}{*}{00:00:06} \\ 
     \hline
%\multicolumn{4}{l}{\textsc{bfgs}} &    &  &   \\
\multirow{2}{*}{\textsc{gd}-\textsc{b}} & avg & 0.30 & 0.29 & 0.39 & 0.17 & 4.65 &  \multirow{2}{*}{0} & \multirow{2}{*}{04:25:16} & 0.31 & 0.29 & 0.39 & 0.17 & 4.65 &  \multirow{2}{*}{27} & \multirow{2}{*}{08:58:30} \\ 
& std & 0.00 & 0.00 & 0.00 & 0.00 & 0.00 &  &  & 0.00 & 0.00 & 0.00 & 0.00 & 0.00 &  &   \\ 
\hline
\multirow{2}{*}{\textsc{bfgs}} & avg & -0.04 & -0.11 & -0.38 & 4.87 & $2 \cdot 10^4$ &  \multirow{2}{*}{0} & \multirow{2}{*}{00:00:12} & 0.44 & 0.27 & 0.29 & 0.15 & 65.1 &  \multirow{2}{*}{0} & \multirow{2}{*}{00:00:08}\\ 
& std & 0.25 & 0.93 & 0.45 & 3.79 & $2 \cdot 10^4$ & & & 0.32 & 0.16 & 0.50 & 0.07 & 101 &  &  \\ 
\hline
   %\multicolumn{4}{l}{\textsc{sims}} &    &  &\\
   \multirow{2}{*}{\textsc{sims}} & avg & 0.23 & -0.23 & 0.31 & 0.18 & 42.2   &  \multirow{2}{*}{0} & \multirow{2}{*}{00:00:40}  & 0.61 & 0.25 & 0.09 & 0.14 & 118  &  \multirow{2}{*}{0} & \multirow{2}{*}{00:00:38} \\ 
  & std & 0.42 & 2.00 & 0.38 & 0.12 & 105 &  &  & 0.36 & 0.26 & 0.73 & 0.07 & 123 &  &  \\ 
   \hline
%\multicolumn{4}{l}{\textsc{nm}} &    &  &\\
\multirow{2}{*}{\textsc{nm}} & avg & 0.43 & -4.98 & 0.38 & 0.17 & 16.96 &  \multirow{2}{*}{0} & \multirow{2}{*}{00:00:17}  & 0.56 & 0.25 & 0.41 & 0.15 & 21.6 &  \multirow{2}{*}{0} &  \multirow{2}{*}{00:00:16}  \\ 
&  std & 0.44 & 37.3 & 0.22 & 0.05 & 39.9 & & & 0.34 & 0.16 & 0.29 & 0.05 & 31.5 &  & \\ 
\hline
%\multicolumn{4}{l}{\textsc{sa+nm}} &    &  &\\
\multirow{2}{*}{\textsc{sa}} &  avg & 1.55 & -1.45 & 0.50 & 0.09 & 74.8  &  \multirow{2}{*}{0} & \multirow{2}{*}{00:04:45} & 0.66 & 0.19 & 0.63 & 0.05 & 194   &  \multirow{2}{*}{0} & \multirow{2}{*}{00:02:32} \\ 
& std & 2.14 & 2.71 & 0.25 & 0.09 & 92.0 & & & 0.45 & 0.28 & 0.66 & 0.07 & 87.2  & & \\
   \hline
%\multicolumn{4}{l}{\textsc{sa+nm}} &    &  &\\
\textsc{sa} &  avg & 0.96 & -79.0 & 0.44 & 0.10 & 63.2  &  \multirow{2}{*}{0} & \multirow{2}{*}{00:04:52} & 0.66 & 0.24 & 0.59 & 0.06 & 168   &  \multirow{2}{*}{0} & \multirow{2}{*}{00:02:49}  \\ 
\textsc{+nm}& std & 2.03 & 122 & 0.15 & 0.09 & 78.7 & & & 0.43 & 0.27 & 0.66 & 0.07 & 98.5  &  & \\ 
 \hline
  %\multicolumn{4}{l}{\textsc{True}} &    &  &\\
\multicolumn{2}{c|}{lower b.} & 0.05 & 0.01 & -0.95 & 0.01 & - & - & - & 0.05 & 0.01 & -0.95 & 0.01 & - & - & -\\
\multicolumn{2}{c|}{upper b.} & 0.99 & 0.90 &  0.95 & 12  & - & - & - & 0.99 & 0.90 &  0.95 & 12  & - & - & -\\ \hline \hline
\end{tabular}  } 
  \notes{ \textbf{Legend:} Comparison for 50 starting values.  \textsc{true}: full sample estimate (est). Objs: avg and std of minimized objective value. crash: optimization terminated because objective returned error. time: average run time for optimizers in hours:minutes:seconds. Lower/upper bound used for the reparameterization. \textsc{gn} run with $\gamma=0.1$ for $k=150$ iterations for all starting values. Standard errors were not computed in the original study. \textsc{gn}-\textsc{b} and \textsc{gd}-\textsc{b} use a backtracking line search, terminates once $Q_n(\theta_{k}) - Q_n(\theta_{k+1}) \leq 10^{-8}$. Additional results for \textsc{gn}, using a range of values $\gamma \in (0,1]$ can be found in Appendix \ref{apx:IRF_additional}.}
\end{table}

In the original paper, the authors initialize the estimation at $\theta_0 = (\eta_0,\nu_0,\rho_{s0},\sigma_{s0}) = (0.20,0.20,0.30,0.10)$, very close to $\hat\theta_n$. Here, 50 starting values are generated within the bounds in Table \ref{tab:imp_no}. The model is estimated using \textsc{csminwel} and the same set of optimizers used in the previous replication. Table \ref{tab:imp_no} reports the results with and without the non-linear reparameterization. Similar to the MA(1) model with $p=12$, without the reparameterization, several optimizers return values outside the parameter bounds, which motivates the constraints in these cases. \textsc{gn} correctly estimates the parameters for all starting values but crashes twice for starting values for which both $\eta$ and $\nu$ are close to their lower bounds where the Jacobian is nearly singular. With the reparameterization, \textsc{gn} crashed more often, nines times in total, but is otherwise accurate. With backtracking line search, crashes are fewer for \textsc{gn}, and converges in 14 iterations, on average, with or without reparameterization, compared to 832 for \textsc{gd} without reparameterization and 2912 with reparameterization (both with cap of 10000). The crashes might also occur at values strictly within the parameter bounds for which Dynare cannot solve the model and returns an error. There is no obvious way to modify \textsc{gn} or \textsc{gd} to avoid this problem.

\begin{figure}[H]
  \caption{Impulse Response Matching: distribution of minimized objective values} \label{fig:IRFobjs}
  \centering%\begin{center}
  \includegraphics[scale=0.5]{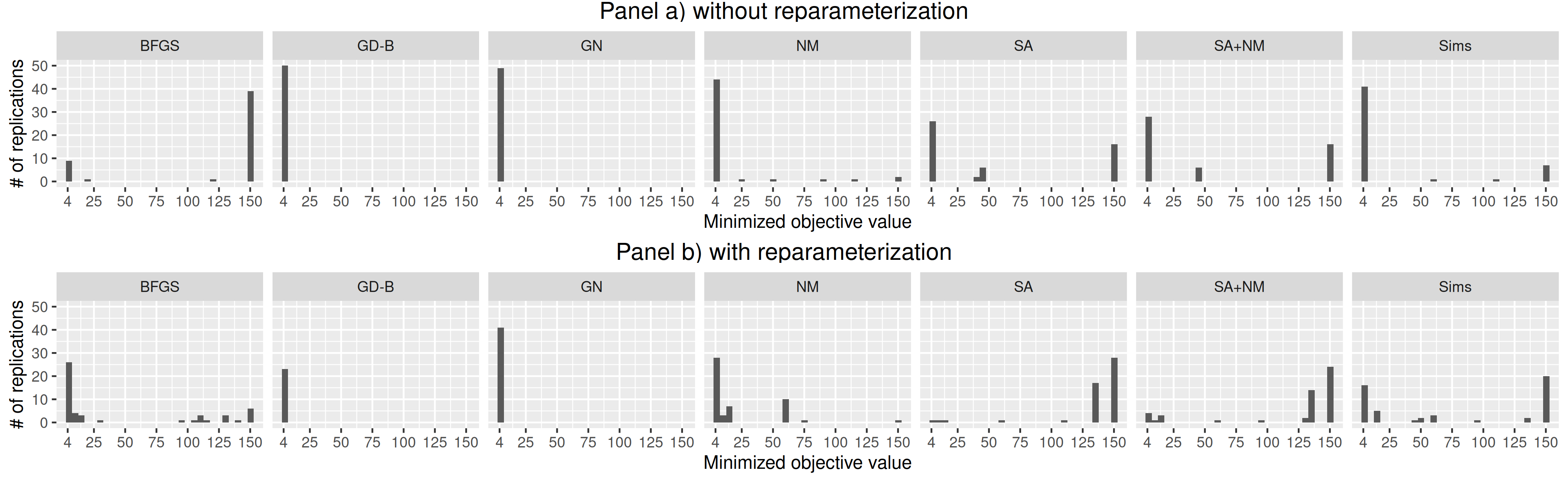}\\
  \notes{ \textbf{Legend:} Comparison for 50 starting values. Minimized objective values for non-crashed optimizations. Objective values are truncated from above at $Q_n(\theta) = 150$.}
%\end{center}
\end{figure}

\begin{figure}[H]
  \caption{Impulse Response Matching: Gauss-Newton iterations for 5 starting values} \label{fig:impconv}
  \centering%\begin{center}
  \includegraphics[scale=0.5]{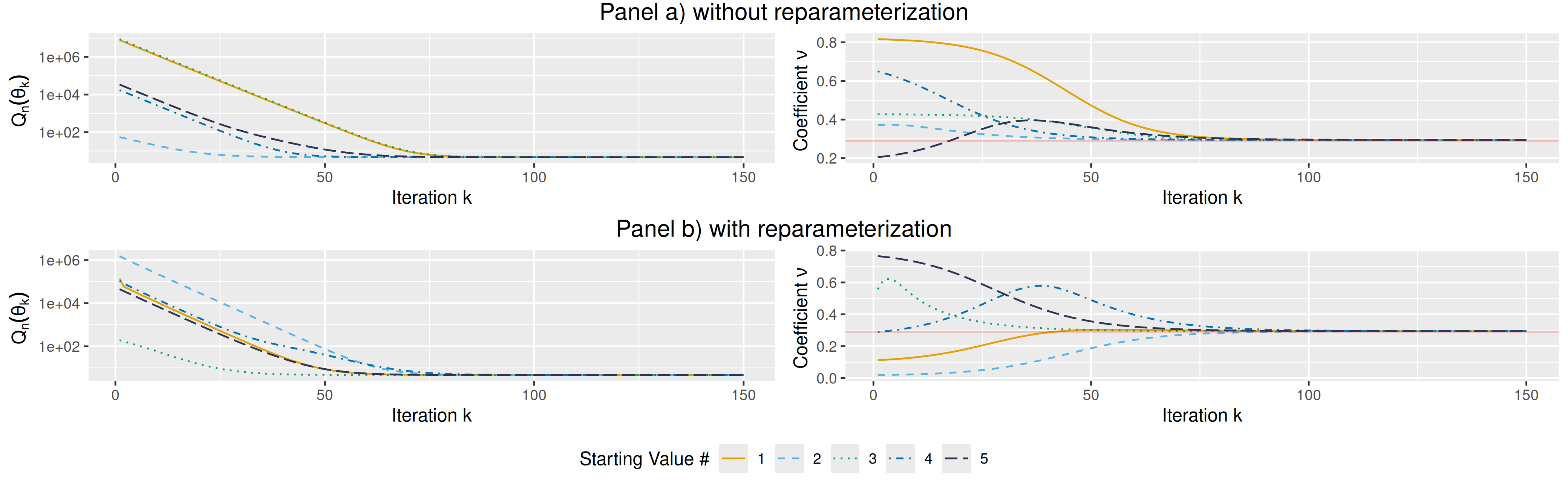}\\
  \notes{ \textbf{Legend:} 150 \textsc{gn} iterations for 5 non-crashing starting values. Left: value of the objective function at each iteration; Right: coefficient $\eta$ at each iteration; horizontal light red line = full sample estimate.}
%\end{center}
\end{figure}

The other two gradient-based optimizers, \textsc{bfgs} and \textsc{sims}(\textsc{csminwel}), never crash because of better error handling in Matlab. They produce valid estimates less often than \textsc{gn}. Figure \ref{fig:IRFobjs} illustrates that \textsc{csminwel} is sensitive to reparameterization. Likewise, derivative-free methods can be inaccurate, as illustrated in Table \ref{tab:imp_no} and Figure \ref{fig:IRFobjs}; some crashes occur despite Matlab's error handling. Finally, Figure \ref{fig:impconv} shows 5 optimization paths for which \textsc{gn} does not crash with and without the reparameterization. Appendix \ref{apx:IRF_additional} gives additional results for larger values of $\gamma \in (0,1]$ and error handling. %It systematically converges using the same number of iterations ($k=150$), for all starting values, regardless of $\gamma \in (0,1]$. %As one would expect from Theorem \ref{th:global_cv_OI}, the convergence properties degrade as $\gamma$ becomes larger. Surprisingly, this is not the case in the BLP example (Table \ref{tab:BLPgamma}, Appendix \ref{apx:cereal_additional}).

\subsection{Convexity, Strong Injectivity, and Assumption \ref{ass:conds} (b)} \label{sec:comp_rk}

Table \ref{tab:compare_rank} illustrates the strong injectivity conditions, Assumption \ref{ass:conds} (b), and convexity for the MA(1) model and the two empirical applications. A grid of $100$ Sobol points was used to construct values within parameter bounds, respectively \textsc{sb} and \textsc{lb}, as described in Section \ref{sec:check}; the first grid value is enforced to take only values from the bounds. The objective $Q_n$ is locally convex at $\theta$ if the Hessian is positive definite, i.e. $H_n(\theta)>0$. The Table reports an estimate for $\mu_n$, $\rho\underline{\sigma}$ and the proportion of grid values where $Q_n$ is locally convex. To evaluate $\rho\underline{\sigma}$, the same step from Section \ref{sec:check} were used setting $\theta_k^2 = \hat{\theta}_n$ for all $k$. The $P_{k,n}$ for \textsc{gn} was used so that the $\mu_n$, $\rho\underline{\sigma}$ reported here are invariant to linear reparameterizations of both the parameters and the moments.

\begin{table}[h]
  \centering \caption{Empirical and Illustrative Examples: Conditions, Convexity}  \label{tab:compare_rank}
  \setlength\tabcolsep{4.5pt}
    \renewcommand{\arraystretch}{0.9}
{\footnotesize \begin{tabular}{cr|ccc|ccc||c}
  \hline \hline
   & & \multicolumn{3}{c|}{ Strong Injectivity (\ref{eq:SIp}) } & \multicolumn{3}{c||}{ Assumption \ref{ass:conds} (b) }    & Convexity  \\ \hline
   & & $\hat{\mu}_n$ & $\overline{\gamma}$ & $\underline{k}$ &  $\widehat{\rho \underline{\sigma}}_n$ & $\overline{\gamma}$ & $\underline{k}$ & $H_n > 0$ (\%) \\ \hline
  \hline
%  \multicolumn{4}{l}{\textsc{True}} &  &  &  &  &  &  &   \\
\textsc{ma}(1), $p=1$ & \textsc{sb} & 0.5 & $3.1 \cdot 10^{-3}$ & $2.2 \cdot 10^{3}$ & 0.9 & $3.1 \cdot 10^{-3}$ & $2.1 \cdot 10^{3}$ & 46 \\ 
 & \textsc{lb} & 0.0 & 0.0 & $\infty$ & 0.0 & 0.0 & $\infty$  & 40 \\
   \hline
%\multicolumn{4}{l}{\textsc{gn}} &  &  &  &  &  &  &\\ 
%\rowcolor{gray}
\textsc{ma}(1), $p=12$ & \textsc{sb} & 0.15 & $4 \cdot 10^{-3}$ & $1.6 \cdot 10^{3}$ & 0.17 & $6 \cdot 10^{-3}$ & $1.2 \cdot 10^{3}$ & 98   \\  %\rowcolor{gray}
$W_n = I_d$   &   \textsc{lb}  & 0.0 & 0.0 & $\infty$ & 0.0 & 0.0 & $\infty$ & 90 \\  \hline
 \textsc{ma}(1), $p=12$ & \textsc{sb} & 0.12 & $3.5 \cdot 10^{-3}$ & $1.9 \cdot 10^{3}$ & 0.13 & $3.4 \cdot 10^{-3}$ & $2.0 \cdot 10^{3}$ & 98   \\  %\rowcolor{gray}
  $W_n = \hat{V}_n^{-1}$   &   \textsc{lb}  & 0.0 & 0.0 & $\infty$ & 0.0 & 0.0 & $\infty$ & 90 \\ 
      \hline %\rowcolor{gray}
%\multicolumn{4}{l}{\textsc{bfgs}} &  &  &  &  &  &  &   \\
 & \textsc{sb} & 0.38 & $1.0$ & $1$ & 0.75 & $1.0$ & $1$ & 95    \\ %\rowcolor{gray}
 \multirow{-2}{*}{\textsc{blp}} &  \textsc{lb} & 0.62 & $1.6 \cdot 10^{-5}$ & $4.3 \cdot 10^{5}$ & 0.68 & $2.3 \cdot 10^{-7}$ & $3.0 \cdot 10^{7}$ & 1  \\ 
   \hline
   \multirow{2}{*}{\textsc{dsge}} & \textsc{sb} & 0.16 & $4.0 \cdot 10^{-10}$ & $1.7 \cdot 10^{10}$ & 0.37 & $1.3 \cdot 10^{-9}$ & $5.2 \cdot 10^{9}$  & 3  \\ 
 &  \textsc{lb} & 0.13 & $1.3 \cdot 10^{-11}$ & $5.5 \cdot 10^{11}$  & 0.33 & $2.7 \cdot 10^{-10}$ & $2.5 \cdot 10^{10}$ & 5  \\ 
   \hline
%\multicolumn{4}{l}{\textsc{nm}} &  &  &  &  &  &  &\\
\textsc{dsge} & \textsc{sb} & 0.13 & $3.5 \cdot 10^{-12}$ & $2.0 \cdot 10^{12}$ & 0.23 & $3.5 \cdot 10^{-13}$ & $1.98 \cdot 10^{13}$ & 0  \\ 
(\textsc{re}) &  \textsc{lb} & $1.2 \cdot 10^{-14}$ & $0$ & $\infty$ &  0.24 & $5.0 \cdot 10^{-13}$ & $1.4 \cdot 10^{13}$  & 0  \\ \hline \hline
\end{tabular}  } 
  \notes{ \textbf{Legend:} Results for 100 sobol grid points, adjusted to match the bounds (Smaller Bounds \textsc{sb}, or Larger Bounds \textsc{lb}), for which the moments are well defined. \textsc{dsge}, \textsc{dsge} \textsc{(re)} with/without reparameterization. \textbf{Bounds:} MA(1): \textsc{sb}  $\Theta =  [-0.9,0.9]$ (rank conditions hold); \textsc{lb} $\Theta = [-1.0,1.0]$ (rank conditions fail). \textsc{blp}: \textsc{sb} $\Theta = $ values $50\%$ above/below the true value \citep[p25]{conlon2020}, \textsc{lb} $\Theta = [-10,10] \times \dots \times [-10,10]$. \textsc{dsge}: \textsc{sb} same as original paper plus/minus  $0.1$ for lower/upper bounds; \textsc{lb} same as original paper $\Theta = [0.05,0.99] \times [ 0.01,0.90] \times [ -0.95,0.95] \times [ 0.01, 12]$. \textbf{Convexity:} percentage (\%) of points for which $H_n$ is strictly positive definite. \textbf{Sample sizes:} MA(1) $n = 200$, \textsc{blp} $n = 2256$, \textsc{dsge} $n = 63$.} 
\end{table}

For the MA(1) model, strong injectivity and Assumption \ref{ass:conds} (b) fail at the boundary where $\theta = \pm 1$. This is visible in the results for \textsc{lb}. For \textsc{sb}, both conditions hold as illustrated in the Table. Optimal weighting has some effect on the conditions and the predicted convergence properties.  Convexity fails more often with a single moment condition ($p=1$).

For BLP, the conditions appear to hold and predict fast convergence for \textsc{sb}, used in \citet{conlon2020}, where $Q_n$ is almost everywhere locally convex. With wider bounds (\textsc{lb}), convexity almost always fails, but the estimates for  $\mu_n$, $\rho\underline{\sigma}$ are very close to \textsc{sb}. This confirm the good optimization properties for \textsc{gn} reported above.

For the DSGE model,  $\mu_n$, $\rho\underline{\sigma}$ are of the same order of magnitude as the other applications without reparameterization. With reparameterization, the conditions can fail at the boundary which is visible in the Table under \textsc{lb}. With and without reparameterization, $Q_n$ is rarely locally convex, which confirms the challenges \textsc{bfgs} and \textsc{csminwell} can have. 

In both empirical applications, the estimates for $\overline{\gamma}$ and $\underline{k}$ tend to be very small and large, respectively, despite $\mu_n$, $\rho\underline{\sigma}$ being away from zero. This reflect the large amount of non-linearity, measured by $C_{3,K}$ and $L_{Q,P,K}$. As discussed in Section \ref{sec:check}, these estimates can be fairly conservative which is clearly the case here. Also, because the moments are evaluated numerically, using a fixed-point algorithm for BLP, and the derivatives are computed by finite differences, the second-order derivatives can be fairly inaccurate. This issue is explained in Appendix \ref{apx:sensitivity}. Innacurate second-order derivatives can make optimizers like \textsc{bfgs} and \textsc{csminwell} numerically unstable, but will also affect the value of $L_{Q,P}$, which tend to be very large in the empirical applications resulting in a very conservative bound for $\overline{\gamma}$.\footnote{The estimate is $L_{Q,P,K} = 2 \cdot 10^4$ for BLP with large bounds, and $L_{Q,P,K} = 9$ with small bounds.}

%Generally, Table \ref{tab:compare_rank} indicates that the rank conditions are more likely to hold than convexity with wider bounds. This is useful since parameters magnitudes are \textit{a priori} unknown when the estimation is carried out. Now, reading Table \ref{tab:survey} in light of Table \ref{tab:compare_rank}, it becomes clear why convex optimizers are generally absent from the survey. They might perform poorly given that convexity often fails in the empirical examples. %The rank conditions can also fail, this is the case for the estimation in \citet{kelly2016}. Because the moments are fairly costly to evaluate, this is illustrated on a one-dimensional sub-problem in Appendix \ref{apx:rank_convexity}.

%Avg: Sample mean of the convergence points; Std: Sample standard error of the convergence points; $\eta$: Elasticity of technology creation to R\&D Statistics; $\nu$: R\&D spillover to adoption; $\rho_s$: Persistence coefficient of $\Delta^s_t$; $\sigma_s$: Size of impulse to$\Delta^s_t$. Avg, Std: sample average and standard deviation of optimizer outputs.

% CSMINWEL: \footnote{Matlab and R implementations can be found at: \url{http://sims.princeton.edu/yftp/optimize/}

\section{Conclusion}
Non-convexity of the GMM objective function is considered to be an important challenge for structural estimation. This paper considers alternative conditions under which there are globally convergent algorithms. The results are robust to non-convexity, moderately non-linear one-to-one reparameterizations, and moderate misspecification. Though off-the-shelf methods might fail to converge due to the non-convexity of the optimization problem, the paper has shown that this does not necessarily imply that it will be difficult in practice.  
Econometric theory emphasizes the role of the weighting matrix $W_n$ on the statistical efficiency of the estimator $\hat\theta_n$. Here, Assumption \ref{ass:conds} may or may not hold, depending on the choice of weighting matrix $W_n$. Its condition number $\kappa_W$ also affects local convergence which highlights an important role for the weighting matrix: it may facilitate or hinder the estimation itself. % -- it systematically outperforms commonly used methods. 
\newpage
\bibliographystyle{ecta}
\bibliography{refs}

%-----------------------------------------------------------
%         Appendices
%-----------------------------------------------------------
\begin{appendices}
  \renewcommand\thetable{\thesection\arabic{table}}
  \renewcommand\thefigure{\thesection\arabic{figure}}
  \renewcommand{\theequation}{\thesection.\arabic{equation}}
  \renewcommand\thelemma{\thesection\arabic{lemma}}
  \renewcommand\thetheorem{\thesection\arabic{theorem}}
  \renewcommand\thedefinition{\thesection\arabic{definition}}
    \renewcommand\theassumption{\thesection\arabic{assumption}}
  \renewcommand\theproposition{\thesection\arabic{proposition}}
    \renewcommand\theremark{\thesection\arabic{remark}}
    \renewcommand\thecorollary{\thesection\arabic{corollary}}
\setcounter{equation}{0}
\setcounter{lemma}{0}
\clearpage \baselineskip=18.0pt
\appendix
%\section{Preliminary Results for Section \ref{sec:main}} \label{apx:prelim}

\makeatletter
\@addtoreset{assumption}{section}
\makeatother

\section{Proofs for the Main Results} \label{apx:proofs}

The proofs will make repeated use of the following mean value identity.
\begin{lemma}[Mean Value Identity] \label{lem:OMV} For any $g(\cdot)$ continuous differentiable on $\mathbb{R}^{d_\theta}$ with Jacobian $G(\cdot)$, let $\overline{G}(\theta_1,\theta_2) = \int_0^1 G(\omega \theta_1 + (1-\omega)\theta_2)d\omega$. For any $\theta_1,\theta_2 \in \mathbb{R}^{d_\theta}$: \[ g(\theta_1) - g(\theta_2) = \overline{G}(\theta_1,\theta_2)(\theta_1-\theta_2).\]
\end{lemma}

\paragraph{Proof of Lemma \ref{lem:OMV}:} Let $h:[0,1] \to \mathbb{R}^{d_g}$ be defined as $h(\omega) = g(\omega \theta_1 + (1-\omega)\theta_2)$, so that $g(\theta_1) - g(\theta_2) = h(1) - h(0) = \int_0^1 \partial_\omega h (\omega) d\omega$. By composition and the chain rule: $\partial_\omega h (\omega) = \partial_\theta g( \omega \theta_1 + (1-\omega)\theta_2)(\theta_1-\theta_2) = G(\omega\theta_1 + (1-\omega)\theta_2)(\theta_1-\theta_2)$. Plug this into the integral to find: $g(\theta_1) - g(\theta_2) = \overline{G}(\theta_1,\theta_2)(\theta_1-\theta_2)$, as desired. \qed

\subsection{Implications of Assumptions \ref{ass:1prim}, \ref{ass:conds}} \label{apx:pop}
In the following we will use the notation: $\overline{g}_n(\theta) = 1/n \sum_{i=1}^n g(\theta ;x_i)$, $g(\theta) = \mathbb{E}[\overline{g}_n(\theta)]$, $G(\theta ;x_i) = \partial_\theta g(\theta ;x_i)$, $G_n(\theta) = 1/n\sum_{i=1}^n G(\theta ;x_i)$, $G(\theta) = \mathbb{E}[G_n(\theta)]$, $Q_n(\theta) = 1/2\overline{g}_n(\theta)^\prime W_n \overline{g}_n(\theta)$, and $Q(\theta) = 1/2g(\theta)^\prime W g(\theta)$. $W_n$ and $W$ are symmetric. With probability approaching 1 will be abbreviated as wpa1. $\mathcal{B}_{R}(\theta^\dagger)$ is a closed ball of radius $R$, centered around $\theta^\dagger$. In the following, $\tilde{\Theta}$ generically denotes a compact convex subset of $\mathbb{R}^{d_\theta}$ such that $\theta^\dagger \in \text{interior}(\tilde{\Theta})$.

\begin{assumption} \label{ass:1} With probability approaching 1: i. $Q_n$ has a global minimizer on $\tilde{\Theta}$, $\hat\theta_n \in \text{interior}(\tilde{\Theta})$, ii. $\overline{g}_n$ is twice continuously differentiable on $\tilde{\Theta}$, iii. $G_n$ is Lipschitz continuous with constant $L \geq 0$ on $\tilde{\Theta}$, and for some $R_G>0$ such that, $\sigma_{\min}[G_n(\theta)] \geq \underline{\sigma}>0$ for all $\|\theta-\hat\theta_n\| \leq R_G$, iv. $W_n$ is such that $0 < \underline{\lambda}_W \leq \lambda_{\min}(W_n) \leq \lambda_{\max}(W_n) \leq \overline{\lambda}_W  < \infty$.
\end{assumption}

\paragraph{Remarks.} The condition that $x_i$ are iid can also be weakened to allow for non-identically distributed dependent observations by appropriately adjusting the moment conditions in \ref{ass:1prim}i, iii which are used to derive uniform laws of large numbers for $\overline{g}_n$ and $G_n$.

\begin{lemma} \label{lemma:pop}
Assumption \ref{ass:1prim} implies Assumption \ref{ass:1}.
\end{lemma}

\begin{lemma} \label{lem:unif_cv} Suppose Assumption \ref{ass:1prim} holds, then $\sup_{\theta \in \mathbb{R}^{d_\theta}}\|G_n(\theta)-G(\theta)\| = o_p(1)$. This implies that $\sup_{\theta_1,\theta_2 \in \mathbb{R}^{d_\theta}}\|\overline{G}_n(\theta_1,\theta_2)-\overline{G}(\theta_1,\theta_2)\| = o_p(1)$ and $\sigma_{\max}[\overline{G}_n(\theta_1,\theta_2)] \leq \overline{\sigma}$, wpa1, uniformly in $\theta_1,\theta_2 \in \mathbb{R}^{d_\theta}$.
\end{lemma}

\begin{lemma} \label{lem:ass1_2} Suppose Assumption \ref{ass:1prim} holds. Then, for some $r > 0$, Assumption \ref{ass:conds} (a) holds for all $\theta \in \mathcal{B}_r(\theta^\dagger)$ with the same choice of $\rho$, $\underline{\sigma}$.
\end{lemma}

\begin{lemma} \label{lem:ass2_1} Suppose Assumption \ref{ass:1prim} (iii), (v), (vi) and \ref{ass:conds} (b) hold. Then, Assumption \ref{ass:1prim} (iv) holds for some strictly positive $\tilde{\underline{\sigma}},\tilde{R}$.
\end{lemma}

The following results are stated in terms of $\overline{G}_n(\theta) = \int_0^1 \{ G_n(\omega \theta + (1-\omega) \hat{\theta}_n) \} d\omega$.

\begin{assumption} \label{ass:conds_n} With probability approaching 1, for all $\theta \in \mathbb{R}^{d_\theta}$: (a) $\sigma_{\min}[G_n(\theta)^\prime W_n \overline{G}_n(\theta)] \geq \rho \underline{\sigma}$, (b) $\| G_n(\theta)^\prime W_n \overline{G}_n(\theta)(\theta - \theta^\dagger) \| \geq \rho \underline{\sigma} \|\theta - \theta^\dagger\|$.
\end{assumption}

\begin{lemma} \label{lemma:conds_n} 
  Suppose Assumptions \ref{ass:1prim} holds. 1) If Assumption \ref{ass:conds} (a) holds, Assumption \ref{ass:conds_n} (a)  holds. 2) If Assumption \ref{ass:conds} (b), Assumption \ref{ass:conds_n} (b) holds. 
\end{lemma}

\paragraph{Proof of Lemma \ref{lem:consistency}.} Lemma \ref{lem:unif_cv} implies that $\overline{G}_n(\theta_1,\theta_2)$ is uniformly consistent in $\theta_1,\theta_2 \in \mathbb{R}^{d_\theta}$. With this in mind, Lemma \ref{lem:OMV} implies:
\begin{align*}
  \overline{g}_n(\theta)-\overline{g}_n(\theta^\dagger) = \overline{G}_n(\theta,\theta^\dagger)(\theta-\theta^\dagger) &= [\overline{G}(\theta) + o_p(1)](\theta-\theta^\dagger),
\end{align*}
uniformly in $\theta \in \mathbb{R}^{d_\theta}$. Now Assumption \ref{ass:conds} (b) implies:
\[ \|\overline{g}_n(\theta)- \overline{g}_n(\theta^\dagger)\| \geq ( \rho\underline{\sigma}/[\overline{\lambda}_W \overline{\sigma}]- o_p(1))\|\theta-\theta^\dagger\|. \] 
Using the triangular inequality: $\|\overline{g}_n(\theta)\|_{W_n} \geq \underline{\lambda}_W^{1/2}( \rho\underline{\sigma}/[\overline{\lambda}_W \overline{\sigma}]- o_p(1))\|\theta-\theta^\dagger\| - \|\overline{g}_n(\theta^\dagger)\|_{W_n}$, uniformly in $\theta \in \mathbb{R}^{d_\theta}$. For any $\|\theta - \theta^\dagger\| \geq [2 \|g(\theta^\dagger)\|_W + 1] \overline{\lambda}_W \overline{\sigma}/[\underline{\lambda}_W^{1/2}\rho\underline{\sigma}]$, this implies:
\[ \|\overline{g}_n(\theta)\|_{W_n} \geq \|g(\theta^\dagger)\|_{W} + 1 - o_p(1). \]
Now, given that $\|\overline{g}_n(\theta^\dagger)\|_{W_n} \leq \|g(\theta^\dagger)\|_{W} + 1$ wpa1, this implies that $\|\hat{\theta}_n-\theta^\dagger\| \leq [2 \|g(\theta^\dagger)\|_W + 1] \overline{\lambda}_W \overline{\sigma}/[\underline{\lambda}_W^{1/2}\rho\underline{\sigma}],$ wpa1. Then, uniform convergence on compact sets (Lemma \ref{lemma:pop}), and the identification conditions imply that $\hat{\theta}_n \overset{p}{\to} \theta^\dagger$, using standard arguments \citep[e.g.][Th2.1]{newey-mcfadden-handbook}. Again, $Q_n$ is uniformly consistent on compact sets, so $Q_n(\hat{\theta}_n) \overset{p}{\to} Q(\theta^\dagger)$. This concludes the proof.
\qed

\paragraph{Proof of Lemma \ref{lemma:pop}.} In the following, all the strict inequalities are replaced by weak inequalities with some slackness $\delta >0$, e.g. $\sigma_{\min}(G(\theta)) \geq (1+\delta)\underline{\sigma}>0$ instead of $\sigma_{\min}(G(\theta)) > \underline{\sigma}>0$, and $\lambda_{\max}(W) \leq  (1-\delta)\overline{\lambda}_W < \infty$ instead of $\lambda_{\max}(W) < \overline{\lambda}_W < \infty$. Assumption \ref{ass:1}ii, iv follow from \ref{ass:1prim}ii, iv. Use Weyl's perturbation inequality for singular values \citep[][Problem III.6.5]{bhatia2013} to find $\lambda_{\min}(W_n) \geq \lambda_{\min}(W) - \sigma_{\max}(W_n-W) \geq (1+\delta)\underline{\lambda}_W - o_p(1) \geq \underline{\lambda}_W$, wpa 1. Likewise, $\lambda_{\max}(W_n) \leq \overline{\lambda}_W$, wpa1. This yields Assumption \ref{ass:1}v.

Assumption \ref{ass:1prim}iii and compactness imply uniform convergence of the sample Jacobian $\sup_{\theta\in\tilde{\Theta}} \|G_n(\theta)-G(\theta)\| = o_p(1)$, see \citet{jennrich1969}. We also have uniform convergence for the same moments. Condition ii implies $\overline{g}_n(\theta) - g(\theta) = o_p(1)$, for all $\theta \in \tilde{\Theta}$. Notice that $\|[\overline{g}_n(\theta_1) - g(\theta_1)]-[\overline{g}_n(\theta_2) - g(\theta_2)]\| = \| [\overline{G}_n(\theta_1,\theta_2)-\overline{G}(\theta_1,\theta_2)](\theta_1-\theta_2) \| \leq [\sup_{\theta \in \tilde{\Theta}} \| G_n(\theta)-G(\theta)\|]\|\theta_1-\theta_2\|$, where the $\sup$ is a $o_p(1)$ by uniform convergence of $G_n$, and $\overline{G}(\theta_1,\theta_2) = \int_0^1 G( \omega \theta_1 + (1-\omega)\theta_2 ) d\omega$.  Using a finite cover and arguments similar to \citet{jennrich1969}, this implies uniform convergence: $\sup_{\theta \in \tilde{\Theta}}\|\overline{g}_n(\theta) - g(\theta)\| = o_p(1)$. 

Then, uniform convergence of $\overline{g}_n$ and $W_n \overset{p}{\to} W$ imply uniform converge of $Q_n$ to $Q$. Continuity and the global identification condition \ref{ass:1prim}i. imply $\hat\theta_n \overset{p}{\to} \theta^\dagger$ \citep[][Th2.1]{newey-mcfadden-handbook}. This implies that $\|\theta-\hat\theta_n\| \leq R_G \Rightarrow \|\theta-\theta^\dagger\| \leq R_G +o_p(1) \leq (1+\delta)R_G$, wpa 1, i.e. $\mathcal{B}_{R_G}(\hat\theta_n) \subseteq \mathcal{B}_{(1+\delta)R_G}(\theta^\dagger) \subseteq \tilde{\Theta}$. This implies $\hat\theta_n \in \text{interior}(\tilde{\Theta})$, wpa1. Then, for the same $\theta$, $\sigma_{\min}[G(\theta)] \geq (1+\delta)\underline{\sigma}$, wpa1. Apply Weyl's inequality for singular values  to find that, uniformly in $\theta$: $\sigma_{\min}[G_n(\theta)] \geq \sigma_{\min}[G_n(\theta)] - \sigma_{\max}[G(\theta)-G_n(\theta)] \geq (1+\delta)\underline{\sigma} - o_p(1) \geq \underline{\sigma} > 0$, wpa 1. Take any two $\theta_1,\theta_2$ in $\tilde{\Theta}$, $\|G_n(\theta_1)-G_n(\theta_2)\| \leq 1/n \sum_{i=1}^n \bar{L}(x_i)\|\theta_1-\theta_2\| \leq [(1-\delta)L + o_p(1)]\|\theta_1-\theta_2\| \leq L\|\theta_1-\theta_2\|$, wpa1, using a law of large numbers for $\bar{L}(x_i)$. This yields all the conditions in Assumption \ref{ass:1}iii. \qed

\paragraph{Proof of Lemma \ref{lem:unif_cv}.} Pick $\delta >0$, set $(1+R) \geq \frac{3 M }{\delta}$ so that $\|G(\theta)-G(\theta_R)\| \leq \delta/3$ for any $\theta \in \mathbb{R}^{d_\theta}$. Since $\Theta_R = \{ \theta \in \mathbb{R}^{d_\theta}, \|\theta\| \leq R\}$ is compact, $\sup_{\theta \in \Theta_R} \|{G}_n(\theta)-G(\theta)\| = o_p(1)$, using Lemma \ref{lemma:pop}. Likewise,
\[ \|{G}_n(\theta)-{G}_n(\theta_R)\| \leq \left[\frac{1}{n} \sum_{i=1}^n \bar{M}(x_i)\right]/(1+R) \leq [M+o_p(1)]/(1+R) \leq \delta/3 +o_p(1). \]
Then, combine these results to find:
\begin{align*}
  \|{G}_n(\theta)-G(\theta)\| &\leq \|{G}_n(\theta)-{G}_n(\theta_R)\| + \|{G}_n(\theta_R)-G(\theta_R)\| + \|G(\theta)-G(\theta_R)\|\\
  &\leq 2/3 \delta + o_p(1),
\end{align*}
uniformly in $\theta \in \mathbb{R}^{d_\theta}$. This implies uniform consistency: $\lim_{n\to\infty}\mathbb{P}(\sup_{\theta \in \mathbb{R}^{d_\theta}}\|{G}_n(\theta)-G(\theta)\| > \delta) = 0$. Also,$\|\overline{G}_n(\theta_1,\theta_2)-\overline{G}(\theta_1,\theta_2)\| \leq \sup_{\theta \in \mathbb{R}^{d_\theta}}\|{G}_n(\theta)-G(\theta)\| = o_p(1)$ and $\sigma_{\max}[\overline{G}_n(\theta_1,\theta_2)] \leq \sup_{\theta \in \mathbb{R}^{d_\theta}}\sigma_{\max}[{G}_n(\theta)] \leq \overline{\sigma}$, wpa1, which is the desired result. \qed

\paragraph{Proof of Lemma \ref{lem:ass1_2}:} Under Assumption \ref{ass:1prim}, $\sigma_{\min}[G(\theta)] \geq (1+\delta)\underline{\sigma}$ for all $\theta \in \mathcal{B}_{R_G}(\theta^\dagger)$ and some $\delta >0$. Also, $G$ is Lipschitz continuous with constant $L$ since $\|G(\theta_1)-G(\theta_2)\| \leq \mathbb{E}[\|G(\theta_1;x_i)-G(\theta_2;x_i)\|] \leq L\|\theta_1-\theta_2\|$. As a result, $\|\overline{G}(\theta)-G(\theta^\dagger)\| \leq L \|\theta-\theta^\dagger\|$. Then,
\[ \|G(\theta)^\prime W \overline{G}(\theta) - G(\theta^\dagger)^\prime W G(\theta^\dagger)\| \leq 2 \overline{\sigma}\overline{\lambda}_W L \|\theta-\theta^\dagger\|. \]
Apply Weyl's inequality to find:
\[ \sigma_{\min}[G(\theta)^\prime W \overline{G}(\theta)] \geq \big\{ (1+\delta)[\underline{\lambda}_W \underline{\sigma}]  - 2 \frac{\overline{\sigma}\overline{\lambda}_W L}{\underline{\sigma}} \|\theta-\theta^\dagger\| \big\} \underline{\sigma}. \]
Pick $\|\theta-\theta^\dagger\| \leq r$ with $r$ such that $\delta > 2 \overline{\sigma}L \overline{\lambda}_W/[\underline{\lambda}_W\underline{\sigma}^2]  r$ to find:
$\sigma_{\min}[G(\theta)^\prime W \overline{G}(\theta)] > [\underline{\lambda}_W \underline{\sigma}] \underline{\sigma}$,
given that $0 < \rho \leq \underline{\lambda}_W \underline{\sigma}$ in Assumption \ref{ass:conds} (a), this yields the result.\qed

\paragraph{Proof of Lemma \ref{lem:ass2_1}:} Take $\theta = \theta^\dagger + \varepsilon v$, with $v$ unitary. Assumption \ref{ass:conds} (b) and the regularity conditions (Assumption \ref{ass:1prim} (iii), (v), (vi)) imply:
$\|G(\theta^\dagger)^\prime W G(\theta^\dagger) v\| > \rho \underline{\sigma} - 2\overline{\sigma} \overline{\lambda}_W L \varepsilon$,
take $\varepsilon \to 0$, to find $\lambda_{\min}[G(\theta^\dagger)^\prime W G(\theta^\dagger)] > \rho \underline{\sigma}$ so that $\sigma_{\min}[G(\theta^\dagger)] > \rho \underline{\sigma} / [\overline{\sigma} \overline{\lambda}_W]$. Pick $\tilde{\underline{\sigma}} = 1/2 \rho \underline{\sigma} / [\overline{\sigma} \overline{\lambda}_W]$. The Lipschitz continuity of $G$ implies $\sigma_{\min}[G(\theta)] \geq \sigma_{\min}[G(\theta^\dagger)] - L\|\theta-\theta^\dagger\| > 1/2 \rho \underline{\sigma} / [\overline{\sigma} \overline{\lambda}_W]$ for $\|\theta-\theta^\dagger\| \leq \tilde{R} < 1/2 \rho \underline{\sigma} / [L \overline{\sigma} \overline{\lambda}_W]$.
\qed

\paragraph{Proof of Lemma \ref{lemma:conds_n}.} Lemmas \ref{lem:consistency} and \ref{lem:unif_cv} apply so that $G_n$ is uniformly convergent and Lipschitz continuous, $\hat{\theta}_n$ is consistent. 1) This implies that:
\begin{align*}
  \|\overline{G}_n(\theta) - \overline{G}(\theta)\| &= \|\int_0^1 \{ G_n(\omega \theta + (1-\omega)\hat{\theta}_n) - G(\omega \theta + (1-\omega)\theta^\dagger) \} d\omega\|\\
  &\leq L \|\hat{\theta}_n-\theta^\dagger\| + \sup_{\theta \in \Theta} \|G_n(\theta)-G(\theta)\| = o_p(1).
\end{align*}
Then apply Weyl's inequality to find that, uniformly in $\theta$ and wpa1: $\sigma_{\min}[G_n(\theta)] \geq \sigma_{\min}[G(\theta)] - o_p(1)$,  $\sigma_{\min}[\overline{G}_n(\theta)] \geq \sigma_{\min}[\overline{G}(\theta)] - o_p(1)$, and $\sigma_{\min}[G_n(\theta)^\prime W_n \overline{G}_n(\theta)] \geq \sigma_{\min}[G(\theta)^\prime W \overline{G}(\theta)] - o_p(1)$, which yields the result.

2) Lemma \ref{lem:ass1_2} implies Assumption \ref{ass:conds} (a) holds locally, i.e. for $\|\theta-\theta^\dagger\| \leq r$, with $r > 0$. With the derivations above, this implies that Assumption \ref{ass:conds_n} (a) holds locally as well, i.e. for $\|\theta-\hat{\theta}_n\| \leq r/2$, wpa1. Recall that Assumption \ref{ass:conds_n} (a) implies Assumption \ref{ass:conds_n} (b).

Take $\|\theta - \hat{\theta}_n\| \geq r/2$. By uniform consistency and boundedness of $G_n$ and $\overline{G}_n$, we have: $G_n(\theta)^\prime W_n \overline{G}_n(\theta) = G(\theta)^\prime W \overline{G}(\theta) + o_p(1)$, uniformly in $\theta$ using $\sigma_{\max}[G_n(\theta)] \leq \overline{\sigma}$ wpa1. Since $\hat{\theta}_n$ is consistent, we have uniformly in $\|\theta - \hat{\theta}_n\| \geq r/2$:
\begin{align*}
  \|G_n(\theta)^\prime W_n \overline{G}_n(\theta)(\theta-\hat{\theta}_n)\| &\geq \|G(\theta)^\prime W \overline{G}(\theta)(\theta-\hat{\theta}_n)\| - o_p(1)\|\theta-\hat{\theta}_n\|\\
  &\geq \|G(\theta)^\prime W \overline{G}(\theta)(\theta-\theta^\dagger)\| - o_p(1)\|\theta-\hat{\theta}_n\| - \overline{\sigma}^2 \overline{\lambda}_W o_p(1)\\
  &\geq (1+\delta)\rho\underline{\sigma}\|\theta-\theta^\dagger\| - o_p(1)\|\theta-\hat{\theta}_n\| - \overline{\sigma}^2 \overline{\lambda}_W o_p(1)\\
  &\geq [(1+\delta)\rho\underline{\sigma}-o_p(1)]\|\theta-\hat{\theta}_n\| - [\overline{\sigma}^2 \overline{\lambda}_W + (1+\delta)\rho\underline{\sigma}] o_p(1)\\
  &\geq \left[(1+\delta)\rho\underline{\sigma}-o_p(1) -  o_p(1) 2\frac{\overline{\sigma}^2 \overline{\lambda}_W + (1+\delta)\rho\underline{\sigma}}{r}\right]\|\theta-\hat{\theta}_n\|,
\end{align*}
using $\|\theta-\hat{\theta}_n\| / (r/2) \geq 1$ for the last inequality. The leading term is greater or equal than $\rho\underline{\sigma}$ wpa1 which yields the result.
\qed

\subsection{Proofs for Section \ref{sec:global_cv}} \label{apx:global_cv}

\paragraph{Proof of Theorem \ref{th:global_cv_cs}:} 

Take $\theta_0 \in \Theta$, let  $\Theta_n = \Big\{ \tilde{\theta}\in \mathbb{R}^{d_\theta}, Q_n(\tilde{\theta}) \leq Q_n(\theta_0) \Big\}$. 
From the proof of Lemma \ref{lem:consistency}, we have:
\[ \sqrt{2 Q_n(\tilde{\theta})} \geq ( \rho\underline{\sigma}/[\overline{\lambda}_W \overline{\sigma}] -o_p(1) )\|\tilde{\theta}-\theta^\dagger\| - \|g(\theta^\dagger)\|_W - o_p(1), \]
uniformly in $\tilde{\theta} \in \mathbb{R}^{d_\theta}$. Now take $Q_0 = \sup_{\theta \in \Theta} Q(\theta)$ and let:
\[ \Theta_0 = \Big\{ \tilde{\theta} \in \mathbb{R}^{d_\theta}, \|\tilde{\theta}-\theta^\dagger\| \leq 2\frac{ \sqrt{2 Q_0}+\sqrt{2 Q(\theta^\dagger)}+1 }{\rho\underline{\sigma}/[\overline{\lambda}_W]\overline{\sigma}} \Big\}, \]
a compact subset of $\mathbb{R}^{d_\theta}$. We have $\Theta \subseteq \Theta_0$ and $\Theta_n \subseteq \Theta_0$, wpa1. Now, let:
$R_\Theta = 4 \overline{\lambda}_P \overline{\sigma}^2 \overline{\lambda}_W \text{diam}(\Theta_0)$,
which bounds $\|\theta_{k+1}-\theta_k\|$ wpa1, uniformly in $\theta_k \in \Theta_0$, for any choice of $\gamma \in [0,1]$. Uniformly in $\theta_k \in \Theta_0$, $\theta_{k+1}$ computed in (\ref{eq:update}) satisfies $\theta_{k+1} \in \Theta_{R} = \cup_{\theta \in \Theta_0} B_{R_\Theta}(\theta)$ wpa1. The sample moments and Jacobian are uniformly consistent on $\Theta_{R} \supseteq \Theta_0$. Then, by recursion over $k \geq 0$, the following establishes that uniformly in $\theta_k \in \Theta_0$, $Q_n(\theta_{k+1}) \leq Q_n(\theta_k)$. Hence, $\theta_{k} \in \Theta_0$ for all $k\geq 0$ wpa1, uniformly in $\theta_0 \in \Theta$. So the derivations below can proceed under Assumption \ref{ass:1}, with $\tilde{\Theta} = \Theta_R$, since the path is compact-valued wpa1.

\textbf{Case 1) Just-identifed:} Since Assumptions \ref{ass:1} and \ref{ass:conds_n} hold (using Lemmas \ref{lemma:pop}, \ref{lemma:conds_n}), Proposition \ref{prop:PL} (1)-(2) holds, with probability approaching $1$, for the sample moments with the same choice of strictly positive constants $C_1,C_2,C_3$. Denote by $L_Q$ the Lipschitz constant of $\partial_\theta Q_n$. The mean value value theorem implies that for some $\tilde{\theta}_k$ between $\theta_k$ and $\theta_{k+1}$:
\begin{align*}
  Q_n(\theta_{k+1}) &= Q_n(\theta_k) - \gamma \partial_\theta Q_n(\theta_k) P_{k,n} \partial_\theta Q_n(\theta_k) - \gamma \{ \partial_\theta Q_n(\tilde{\theta}_k)- \partial_\theta Q_n(\theta_k) \} P_{k,n} \partial_\theta Q_n(\theta_k)\\
  &\leq Q_n(\theta_k) - \gamma \underline{\lambda}_P \|\partial_\theta Q_n(\theta_k)\|^2 + \gamma^2 L_Q \overline{\lambda}_P^2 \|\partial_\theta Q_n(\theta_k)\|^2\\
  &\leq Q_n(\theta_k) + \gamma\{ -  \underline{\lambda}_P  + \gamma L_Q \overline{\lambda}_P^2\} \|\partial_\theta Q_n(\theta_k)\|^2\\
  &\leq Q_n(\theta_k) - \gamma \underline{\lambda}_P / 2 \|\partial_\theta Q_n(\theta_k)\|^2\\
  &\leq Q_n(\theta_k) - \gamma \underline{\lambda}_P  C_1 / 2 (Q_n(\theta_k)  - Q_n(\hat{\theta}_n)),
\end{align*}
if $0 <  \gamma \leq \underline{\lambda}_P / [2 L_Q \overline{\lambda}_P^2]$. Substract $Q_n(\hat{\theta}_n)$ on both sides to find: 
\begin{align*}
  Q_n(\theta_{k+1}) - Q_n(\hat{\theta}_n) &=  \{ 1 - \gamma C_1  \underline{\lambda}_P/2 \}(Q_n(\theta_k)  - Q_n(\hat{\theta}_n)).
\end{align*}
Set $(1-\overline{\gamma})^2 = 1 - \gamma \underline{\lambda}_P C_1 / 2$ and iterate over $k = 0,\dots$ to find:
\[ \|\theta_{k+1} - \hat{\theta}_n \| \leq (1-\overline{\gamma})^{k+1} \sqrt{C_3/C_2} \|\theta_{0} - \hat{\theta}_n \|, \]
which is the desired result.

\textbf{Case 2) Over-identifed:} Since Assumptions \ref{ass:1} and \ref{ass:conds_n} hold (using Lemmas \ref{lemma:pop}, \ref{lemma:conds_n}), and $Q_n(\hat{\theta}_n) = o_p(1)$, Proposition \ref{prop:PLmis} (1')-(2) holds, with probability approaching $1$, for the sample moments with the same choice of strictly positive constants $C_2,C_3,C_4$. Let 
\[ C_{1n} = \frac{(\rho \underline{\sigma} - \overline{\lambda}_W^{1/2}L \|\overline{g}_n(\hat{\theta}_n)\|_{W_n})^2}{C_3 + C_4\|\overline{g}_n(\hat{\theta}_n)\|_{W_n} } = C_1 + o_p(1),\]
for the same $C_1$ found in Proposition \ref{prop:PL} (1). Denote by $L_Q$ the Lipschitz constant of $\partial_\theta Q_n$. The mean value theorem implies that for some $\tilde{\theta}_k$ between $\theta_k$ and $\theta_{k+1}$:
\begin{align*}
  Q_n(\theta_{k+1}) &= Q_n(\theta_k) - \gamma \partial_\theta Q_n(\theta_k) P_{k,n} \partial_\theta Q_n(\theta_k) - \gamma \{ \partial_\theta Q_n(\tilde{\theta}_k)- \partial_\theta Q_n(\theta_k) \} P_{k,n} \partial_\theta Q_n(\theta_k)\\
  &\leq Q_n(\theta_k) + \gamma \{ - \underline{\lambda}_P + \gamma L_Q \overline{\lambda}_P^2 \} \|\partial_\theta Q_n(\theta_k)\|^2\\
  &\leq Q_n(\theta_k) - \gamma  \underline{\lambda}_P C_{1n} /2 (Q_n(\theta_k)  - Q_n(\hat{\theta}_n)),
\end{align*}
if $0 <  \gamma \leq \underline{\lambda}_P / [2 L_Q \overline{\lambda}_P^2]$. Substract $Q_n(\hat{\theta}_n)$ on both sides to find: 
\begin{align*}
  Q_n(\theta_{k+1}) - Q_n(\hat{\theta}_n) &=  \{ 1 - \underline{\lambda}_P \gamma C_{1n}/2 \}(Q_n(\theta_k)  - Q_n(\hat{\theta}_n)).
\end{align*}
Set $(1-\overline{\gamma})^2 = 1 - \gamma \underline{\lambda}_P C_{1n} /2$ and iterate over $k = 0,\dots$ to find:
\[ \|\theta_{k+1} - \hat{\theta}_n \| \leq (1-\overline{\gamma})^{k+1} \frac{\sqrt{C_3 + C_4 \|\overline{g}_n(\hat{\theta}_n)\|_{W_n}}}{\sqrt{C_2 - C_4 \|\overline{g}_n(\hat{\theta}_n)\|_{W_n}}} \|\theta_{0} - \hat{\theta}_n \|, \]
which is the desired result.
\qed

\paragraph{Proof of Theorem \ref{th:global_cv_ms}:} The proof is similar to Theorem \ref{th:global_cv_cs}, the condition on $\varphi$ ensures that inequalities (1')-(2) in Proposition \ref{prop:PLmis} hold with strictly positive constants, with probability approaching $1$, for the sample moments.
\qed

%Let $\overline{G}(\theta_1,\theta_2) = \int_0^1 G(\omega \theta_1 + (1-\omega) \theta_2)$

\section{Proofs and additional results for Section \ref{sec:lit_char}} \label{apx:lit_char}

\subsection{Properties related to Strong Injectivity}

\begin{lemma}[From (\ref{eq:SIp}) to (\ref{eq:SI}), Assumption \ref{ass:2}] \label{lem:SI_ass2} Suppose Assumption \ref{ass:1prim} (iii), (vi) hold, then: 1)  (\ref{eq:SIp}) implies (\ref{eq:SI}), and 2) (\ref{eq:SIp}) implies Assumption \ref{ass:2} holds for \textsc{gn}.
\end{lemma}

\begin{lemma}[From (\ref{eq:SI}) to (\ref{eq:SIp})] \label{lem:SI_SIp} Suppose Assumption \ref{ass:1prim} (iii), (vi) hold and $g = (g_1^\prime,g_2^\prime)^\prime$ where $g_1$ is just-identified and satisfies (\ref{eq:SI}) for some $\mu_1 > 0$. Let $\tilde{W}(\lambda) = \lambda W + (1-\lambda) \text{blockdiag}(W_1,0)$, where $W_1$ is the upper block of $W$ corresponding to $g_1$. If $W_1$ is invertible, then there exists $\lambda^\star \in (0,1]$ such that (\ref{eq:SIp}) holds using $\tilde{W}(\lambda)$ for any $0 \leq \lambda \leq \lambda^\star$.
\end{lemma}

\subsection{Additional Results for Over-Identified Models}

\begin{proposition} \label{prop:characOI} (Sufficient Conditions: Over-Identified) Consider the following three conditions: (a) $\sigma_{\min}[G(\theta)^\prime W \overline{G}(\theta_1,\theta_2)] > \underline{\sigma} > 0$, for all $\theta,\theta_1,\theta_2 \in \mathbb{R}^{d_\theta}$, (b) for all $\theta \in \mathbb{R}^{d_\theta}$, $G(\theta) = U S(\theta) V$ for $U,V$ full rank, $S(\theta)$ symmetric with $0 < \underline{\lambda}_S < \lambda_{\min}[S(\theta)] < \lambda_{\max}[S(\theta)] <\overline{\lambda}_S < \infty$, and $U^\prime W U$ invertible.\\
  %\item[(c)] $g(\theta) = \partial_\theta F (\theta)$, for all $\theta \in \Theta$, where $F : \Theta \to \mathbb{R}$ is twice continuously differentiable, strongly convex.
The following holds: (1) (b) $\Rightarrow$ (a) $\Rightarrow$ Assumption \ref{ass:conds} (a), 
(2) (a) implies $G(\theta_1)^\prime W g(\cdot)$ is one-to-one, for any $\theta_1 \in \mathbb{R}^{d_\theta}$.
\end{proposition}

\begin{proposition} \label{prop:reparOI} (Reparameterization: Over-Identified)
  Take $h$ as in Proposition \ref{prop:repar}. 1) If Assumption \ref{ass:conds} (a) holds for $g$ and $\underline{\sigma} > \overline{\lambda}_W[C_1 \overline{\sigma}_h \overline{\sigma}^2 + C_2 L \overline{\sigma}_h^2 \overline{\sigma} ]/\underline{\sigma}_h^2$,
  then Assumption \ref{ass:conds} (a) holds for $g \circ h$. In particular, if $h = A u + b$ is affine with $A$ invertible then $C_1=C_2=0$ and Assumption \ref{ass:conds} (a) holds for $g \circ h$. 2) Suppose Assumption \ref{ass:conds} (b) holds for $g$. If $\|h(u) - h(u^\dagger)\| \geq \mu \|u - u^\dagger\|$, for some $\mu >0 $ and all $u \in \mathcal{U}$, then Assumption \ref{ass:conds} (b) holds for $g \circ h$.  
\end{proposition}

\subsection{Proofs for Section \ref{sec:lit_char} and the additional results}

\paragraph{Proof of Proposition \ref{prop:PL}:} We first prove (2). For any $\theta \in \mathbb{R}^{d_\theta}$, $g(\theta) = g(\theta) - g(\theta^\dagger) = \overline{G}(\theta)(\theta-\theta^\dagger)$, for correctly specified models. 
This implies that $Q(\theta) = 1/2 (\theta-\theta^\dagger)^\prime \overline{G}(\theta)^\prime W \overline{G}(\theta) (\theta-\theta^\dagger)$. Assumption \ref{ass:1prim} (iii) implies $\sigma_{\max}[\overline{G}(\theta)] \leq \max_{\theta \in \mathbb{R}^{d_\theta}}\sigma_{\max}[G(\theta)] \leq \overline{\sigma} < +\infty$. Assumption \ref{ass:conds} (b) implies $\overline{\sigma}\overline{\lambda}_W^{1/2} \|W^{1/2}\overline{G}(\theta)(\theta-\theta^\dagger)\| \geq \rho \underline{\sigma} \|\theta-\theta^\dagger\|$ and $\|W^{1/2}\overline{G}(\theta)(\theta-\theta^\dagger)\| = \sqrt{2[Q(\theta)-Q(\theta^\dagger)]}$. Putting these together yields:
\[ 1/2 \frac{\rho^2 \underline{\sigma}^2}{\overline{\sigma}^2 \overline{\lambda}_W} \|\theta-\theta^\dagger\|^2 \leq Q(\theta) - Q(\theta^\dagger) \leq 1/2 \overline{\sigma}^2 \overline{\lambda}_W \|\theta-\theta^\dagger\|^2. \]
Now, we prove (1). We have $\partial_\theta Q(\theta) = G(\theta)^\prime W g(\theta) = G(\theta)^\prime W \overline{G}(\theta)(\theta - \theta^\dagger)$. Assumption \ref{ass:conds} (b) implies:
\begin{align*} \|\partial_\theta Q(\theta)\|^2 &\geq \rho^2 \underline{\sigma}^2 \|\theta-\theta^\dagger\|^2 \geq \frac{\rho^2 \underline{\sigma}^2}{1/2 \overline{\sigma}^2 \overline{\lambda}_W} [Q(\theta) - Q(\theta^\dagger)], \end{align*}
using (2). This is the desired result. \qed

\paragraph{Proof of Proposition \ref{prop:quasar_convex}:} For correctly specified models, $\partial_{\theta} Q(\theta) = G(\theta)^\prime W \overline{G}(\theta)(\theta-\theta^\dagger)$. 1) If the PL inequality holds, the quadratic lower bound implies $\|G(\theta)^\prime W \overline{G}(\theta)(\theta-\theta^\dagger)\|^2 \geq \mu C_2 \|\theta-\theta^\dagger\|^2$, i.e. Assumption \ref{ass:conds} (b) holds.\\
2) By definition, $Q$ is quasar-convex if, and only if, there are $\lambda \geq 1$ and $\mu \geq 0$ such that:
\[ \partial_\theta Q(\theta)(\theta-\theta^\dagger) \geq \frac{1}{\lambda}\{ Q(\theta) - Q(\theta^\dagger) \} + \frac{\mu}{2\lambda}\|\theta-\theta^\dagger\|^2, \]
where $\partial_\theta Q(\theta)(\theta-\theta^\dagger) = (\theta-\theta^\dagger)^\prime G(\theta)^\prime W \overline{G}(\theta)(\theta-\theta^\dagger)$. Since $Q(\theta)-Q(\theta^\dagger) \geq 0$ we have:
\[ (\theta-\theta^\dagger)^\prime G(\theta)^\prime W \overline{G}(\theta)(\theta-\theta^\dagger) \geq \frac{\mu}{2 \lambda}\|\theta-\theta^\dagger\|^2. \]
Now apply the Cauchy-Schwarz inequality to find:
\[ \|\theta-\theta^\dagger\|\|G(\theta)^\prime W \overline{G}(\theta)(\theta-\theta^\dagger)\| \geq (\theta-\theta^\dagger)^\prime G(\theta)^\prime W \overline{G}(\theta)(\theta-\theta^\dagger) \geq \frac{\mu}{2 \lambda}\|\theta-\theta^\dagger\|^2, \]
which implies Assumption \ref{ass:conds} (b).
\qed

\paragraph{Proof of Proposition \ref{prop:strong_monotone}:} 1) Strong monotonicity of $A g$ implies $(\theta_1-\theta_2)^\prime A \overline{G}(\theta_1,\theta_2) (\theta_1-\theta_2) \geq \mu\|\theta_1-\theta_2\|^2$ since $g(\theta_1)-g(\theta_2)=\overline{G}(\theta_1,\theta_2)(\theta_1-\theta_2)$. For any unit vector $v$, take $\theta_2 = \theta_1 + \varepsilon v$ and let $\varepsilon \to 0$ to find $v^\prime A G(\theta_1)v = \frac{1}{2} v^\prime [AG(\theta_1)+ G(\theta_1)^\prime A^\prime]v\geq \mu$ so that $G(\theta_1)$ has full rank and $AG(\theta_1)+G(\theta_1)^\prime A^\prime$ is positive definite. We have $\sigma_{\min}[G(\theta)] \geq \mu \sigma_{\min}(A)^{-1} := \underline{\sigma} > 0$, as a normalization. Pick $\theta_2 = \theta^\dagger$, use the Cauchy-Schwarz inequality to find $ \|A^\prime (\theta-\theta^\dagger)\| \|\overline{G}(\theta,\theta^\dagger) (\theta-\theta^\dagger)\| \geq (\theta-\theta^\dagger)^\prime A \overline{G}(\theta,\theta^\dagger) (\theta-\theta^\dagger) \geq \mu\|\theta-\theta^\dagger\|^2$. 

Because $G(\theta)^\prime W$ is invertible, we can write $\|G(\theta)^\prime W \overline{G}(\theta,\theta^\dagger) (\theta-\theta^\dagger)\| \geq \underline{\sigma}\underline{\lambda}_W \|\overline{G}(\theta,\theta^\dagger) (\theta-\theta^\dagger)\| \geq  \underline{\sigma} \mu \underline{\lambda}_W  \sigma_{\max}(A)^{-1}\|\theta-\theta^\dagger\|$; Assumption \ref{ass:conds} (b) holds for any appropriate choice of $0 < \rho \leq \mu \underline{\lambda}_W \sigma_{\max}(A)^{-1}$.\\
2) Strong injectivity of $g$ implies $\|\overline{G}(\theta_1,\theta_2) (\theta_1-\theta_2)\| \geq \mu\|\theta_1-\theta_2\|$, for any pair $\theta_1,\theta_2$. Using the same arguments as above: $G(\theta)$ has full rank for all $\theta$ and $\|G(\theta)^\prime W \overline{G}(\theta,\theta^\dagger) (\theta-\theta^\dagger)\| \geq \underline{\sigma} \underline{\lambda}_W \mu \|\theta-\theta^\dagger\|$; Assumption \ref{ass:conds} (b) holds for any appropriate choice of $0 < \rho \leq \mu \underline{\lambda}_W $. \qed

\paragraph{Proof of Proposition \ref{prop:minimal}.} Assumption \ref{ass:1prim} (ii)-(vi) implies Assumption \ref{ass:conds} (a) holds locally (Lemma \ref{lem:ass1_2}). Hence, for $\|\theta-\theta^\dagger\|\leq r$, we have $\|G(\theta)^\prime W \overline{G}(\theta-\theta^\dagger)\| \geq \rho \underline{\sigma} \|\theta-\theta^\dagger\|$. Condition (N) implies that for $R \geq \|\theta-\theta^\dagger\| \geq r$ we have:
\[ \inf_{\theta, R \geq \|\theta-\theta^\dagger\| \geq r} \|\partial_\theta Q(\theta)\| \geq \delta(r,R) \geq \frac{\delta(r,R)}{R}\|\theta-\theta^\dagger\|, \]
by continuity, compactness and the Weierstrass Theorem. We can pick $\rho < \frac{\delta(r,R)}{R \underline{\sigma}}$. \qed

\paragraph{Proof of Proposition \ref{prop:PLmis}:} 
For any $\theta \in \mathbb{R}^{d_\theta}$, we have:
\begin{align*}
  Q(\theta) - Q(\theta^\dagger) &= \frac{1}{2}\left( g(\theta)^\prime W g(\theta) - g(\theta^\dagger)^\prime W g(\theta^\dagger) \right)\\
  &= \frac{1}{2}\left( g(\theta) + g(\theta^\dagger) \right)^\prime W \left( g(\theta) - g(\theta^\dagger) \right)\\
  &= \frac{1}{2}\left( g(\theta) + g(\theta^\dagger) \right)^\prime W \overline{G}(\theta)(\theta-\theta^\dagger)\\
  &= \frac{1}{2} (\theta-\theta^\dagger)^\prime \overline{G}(\theta)^\prime W \overline{G}(\theta)(\theta-\theta^\dagger) - g(\theta^\dagger)^\prime W \overline{G}(\theta)(\theta-\theta^\dagger),
\end{align*}
the first term in the last display matches the one in the proof of Proposition \ref{prop:PL}. Note that $g(\theta^\dagger)^\prime W G(\theta^\dagger) = 0$ and $\|G(\theta^\dagger) - \overline{G}(\theta)\| \leq L \|\theta-\theta^\dagger\|$, together these allow to bound the second term:
\[ \| g(\theta^\dagger)^\prime W \overline{G}(\theta)(\theta-\theta^\dagger) \| = \| g(\theta^\dagger)^\prime W [\overline{G}(\theta)-G(\theta^\dagger)](\theta-\theta^\dagger) \| \leq \overline{\lambda}_W^{1/2} L \sqrt{\varphi} \|\theta-\theta^\dagger\|^2.\]
Let $C_2 = 1/2 \frac{\rho^2 \underline{\sigma}^2}{\overline{\sigma}^2 \overline{\lambda}_W}$ and $C_3 = 1/2 \overline{\sigma}^2 \overline{\lambda}_W$, as in the proof of Proposition \ref{prop:PL}. Take $C_4 = \overline{\lambda}_W^{1/2} L$, this yields (2):
\[ (C_2 - C_4 \sqrt{\varphi} )\|\theta-\theta^\dagger\|^2 \leq Q(\theta) - Q(\theta^\dagger) \leq  (C_3 + C_4 \sqrt{\varphi}) \|\theta-\theta^\dagger\|^2. \]
For (1), we have $\partial_\theta Q(\theta) = G(\theta)^\prime W g(\theta)$ and $G(\theta^\dagger)^\prime W g(\theta^\dagger) = 0$, so that:
\[ \partial_\theta Q(\theta) = G(\theta)^\prime W \overline{G}(\theta)(\theta-\theta^\dagger) + \{ G(\theta)-G(\theta^\dagger)\}^\prime W g(\theta^\dagger). \]
Apply the reverse triangular inequality to find:
\begin{align*}
  \|\partial_\theta Q(\theta)\| &\geq \rho \underline{\sigma}\|\theta-\theta^\dagger\| - \sqrt{\varphi \overline{\lambda}_W}L \|\theta-\theta^\dagger\|\\
  &= \left( \rho \underline{\sigma} - \sqrt{\varphi \overline{\lambda}_W}L \right) \|\theta-\theta^\dagger\|,
\end{align*}
where $L$ is the Lipschitz constant of $G$. Finally, (1') can be derived from (1) and (2) assuming $(\rho \underline{\sigma} - \sqrt{\varphi \overline{\lambda}_W}L) > 0$.
\qed

\paragraph{Proof of Proposition \ref{prop:charac}:} We first prove (1). (a) $\Rightarrow$ Assumption \ref{ass:conds} (a) is immediate. Under (c), $G(\theta) = \partial^2_{\theta,\theta^\prime} F(\theta)$ is symmetric and strictly positive definite so (b) holds. Suppose (b) holds, then $\overline{G}(\theta_1,\theta_2) = U \{ \int_0^1 S( \omega \theta_1 + (1-\omega)\theta_2 )d\omega \} V$ where $\int_0^1 S( \omega \theta_1 + (1-\omega)\theta_2 )d\omega$ is symmetric. Concavity of the smallest positive eigenvalue on the set of positive definite matrices, and Jensen's inequality imply: $\lambda_{\min}[ \int_0^1 S( \omega \theta_1 + (1-\omega)\theta_2 )d\omega ] \geq \int_0^1 \lambda_{\min}[S( \omega \theta_1 + (1-\omega)\theta_2 )] d\omega \geq \min_{\theta \in \Theta} \lambda_{\min}[S(\theta)] > 0$, by positive definiteness and continuity of $S(\cdot)$.
Finally, \begin{align*}\sigma_{\min}[\overline{G}(\theta_1,\theta_2)] \geq \sigma_{\min}(U) \sigma_{\min}(V) \min_{\theta \in \Theta} \lambda_{\min}[S(\theta)] > \underline{\lambda}_S \underline{\sigma}_U \underline{\sigma}_V > 0,\end{align*} 
taking $\underline{\sigma}_U \underline{\sigma}_V$ to be smallest singular values of $U,V$. Hence (a) holds. 

For (2), note that $g(\theta_1)-g(\theta_2) = \overline{G}(\theta_1,\theta_2)(\theta_1-\theta_2)$, using Lemma \ref{lem:OMV}. With condition (a), we have $g(\theta_1)-g(\theta_2) = 0 \Leftrightarrow \theta_1=\theta_2$, i.e. $g(\cdot)$ is one-to-one. 

For (3), $g(\cdot)$ is one-to-one, take $\phi(\cdot) = g^{-1}(\cdot)$, one-to-one, and $\psi = I_d - \theta^\dagger$, we get that $h(\theta) = \theta - \theta^\dagger$ is linear, the associated GMM loss is strictly quadratic; i.e. strongly convex.\qed

\paragraph{Proof of Proposition \ref{prop:repar}:} 1) Under Assumption \ref{ass:conds} (a), $G$ has full rank for all $\theta \in \mathbb{R}^{d_\theta}$.
Take $u \in \mathcal{U}$, let $\theta = h(u)$, the chain rule implies that $\partial_u g\circ h (u) = \partial_\theta g \circ h (u) \partial_u h(u)$ has full rank for all $u \in \mathcal{U}$. Then, we have:
\begin{align*}
  &\int_0^1  G\circ h(\omega u + (1-\omega) u ) \partial_u h(\omega u + (1-\omega) u^\dagger ) d\omega \\ &= \int_0^1  G(\omega \theta + (1-\omega)\theta^\dagger) d\omega \partial_u h(u^\dagger)  \\ &+  \int_0^1  G(\omega \theta + (1-\omega)\theta^\dagger) [\partial_u h(\omega u + (1-\omega) u^\dagger ) - \partial_u h(u^\dagger)] d\omega\\ &+  \int_0^1  [G\circ h(\omega u + (1-\omega) u )-G(\omega \theta + (1-\omega)\theta^\dagger)]\partial_u h(\omega u + (1-\omega) u^\dagger ) d\omega,
\end{align*}
using Weyl's inequality and a minoration of the singular value for a matrix product, we get:
\begin{align*}
  &\sigma_{\min}[\int_0^1 \partial_u h(\omega u + (1-\omega) u^\dagger ) G\circ h(\omega u + (1-\omega) u ) d\omega] \geq \underline{\sigma}_h \underline{\sigma} - C_1 \overline{\sigma} - C_2 L \overline{\sigma}_h,
\end{align*}
which is strictly positive under the stated condition. After the change of variable, the Assumption \ref{ass:conds} (a) holds if:
\[ \partial_u h(u)^\prime G( g(u) )^\prime W \Big\{ \int_0^1 \partial_u h(\omega u + (1-\omega) u^\dagger ) G\circ h(\omega u + (1-\omega) u ) d\omega \Big\}, \]
has singular values bounded below by a strictly positive term, which is the case for $C_1,C_2$ bounded as in the Proposition statement.  
In particular, when $h$ is affine, $C_1=C_2 = 0$ and $0<\underline{\sigma}_h = \sigma_{\min}[A] \leq \sigma_{\max}[A] \leq \overline{\sigma}_h < \infty$, so that the condition is automatically satisfied.\\
2) Take $u$, let $\theta = h(u)$, since Assumption \ref{ass:conds} (b) holds, we have:
\[ \|G\circ h (u)^\prime W [g\circ h (u) - g\circ h (u^\dagger)]\| \geq \rho \underline{\sigma}\|h(u)-h(u^\dagger)\| \geq  \rho \underline{\sigma} \mu \|u-u^\dagger\|.\]
Using the bounds on the Jacobian of $\partial_u h$, we get the desired result:
\[ \| \partial_u h(u)^\prime G\circ h (u)^\prime W [g\circ h (u) - g\circ h (u^\dagger)]\| \geq  \underline{\sigma}_h  \rho \underline{\sigma} \mu \|u-u^\dagger\|.\]
\qed

\paragraph{Proof of Lemma \ref{lem:SI_ass2}} 1) (\ref{eq:SIp}) implies $\|g(\theta_1)-g(\theta_2))\| \geq \mu/[\overline{\sigma} \overline{\lambda}_W] \|\theta_1-\theta_2\|$, where $\mu/[\overline{\sigma} \overline{\lambda}_W]>0$. 2) Take $\|v\|=1$, $\theta_2 = \theta_1 + \varepsilon v$ and let $\varepsilon \to 0$ in (\ref{eq:SIp}) to find: $\|G(\theta_1)^\prime W G(\theta_1) v\| \geq \mu \|v\|$. Apply the min-max theorem for singular values \citep[p75]{bhatia2013} to find that $\sigma_{\min}[G(\theta)^\prime W G(\theta)] \geq \mu$ for all $\theta \in \mathbb{R}^{d_\theta}$. Since $G(\theta)^\prime W G(\theta)$ is positive semidefinite, singular and eigenvalues coincide, which implies that $\underline{\lambda}_P = \mu > 0$. Then Assumption \ref{ass:1prim} (iii), (vi) implies $\overline{\lambda}_P \leq \overline{\sigma}^2 \overline{\lambda}_W$. Uniform consistency then yield the desired result (Lemma \ref{lem:unif_cv}). \qed

\paragraph{Proof of Lemma \ref{lem:SI_SIp}:} For just-identified models, (\ref{eq:SI}) implies (\ref{eq:SIp}) as long as the weighting matrix is finite and invertible. Indeed, (\ref{eq:SI}) implies $\sigma_{\min}[G_1(\theta)] \geq \mu_1 > 0$ so that $\|G_1(\theta_1)^\prime W [g_1(\theta_1)-g_1(\theta_2)]\| \geq \mu_1^2 \lambda_{\min}(W_1)\|\theta_1-\theta_2\|$ so that (\ref{eq:SIp}) holds using $W_1$ as weighting matrix. Some calculations imply that, by construction of $\tilde{W}$:
\begin{align*} \|G(\theta_1)^\prime \tilde{W}(\lambda)[g(\theta_1)-g(\theta_2)]\| &\geq \mu_1^2 \lambda_{\min}(W_1)\|\theta_1-\theta_2\| - \lambda \Big[ \|G_2(\theta_1)^\prime W_{21}[g_1(\theta_1)-g_1(\theta_2)]\| \\ &+ \|G_1(\theta_1)^\prime W_{12}[g_2(\theta_1)-g_2(\theta_2)]\| + \|G_2(\theta_1)^\prime W_{22}[g_2(\theta_1)-g_2(\theta_2)]\| \Big], \end{align*} 
where $W_2$ is the lower block of $W$ corresponding to $g_2$, $W_{12}$ and $W_{21}$ are the top right and bottom left corners of $W$, respectively. Let $L_1,L_2$ be the Lipschitz constants of $g_1,g_2$ respectively we can conservatively bound the last terms with:
\begin{align*} \lambda \|G\|_{\infty} \|W\|_{\infty}[L_1+2 L_2] \|\theta_1-\theta_2\| < \mu_1^2 \lambda_{\min}(W_1)\|\theta_1-\theta_2\|, \end{align*} 
for any $0 \leq \lambda \leq \lambda^\star < \mu_1^2 \lambda_{\min}(W_1) /(\|G\|_{\infty} \|W\|_{\infty}[L_1+2 L_2])$, where $\|\cdot\|_{\infty}$ denotes the $\ell_\infty$ norm.
\qed

\paragraph{Proof of Proposition \ref{prop:characOI}:} First, we prove (1). (a) $\Rightarrow$ Assumption \ref{ass:conds} (a) is immediate. Suppose (b) holds, take any $\theta,\theta_1,\theta_2 \in \mathbb{R}^{d_\theta}$, then $G(\theta)^\prime W \overline{G}(\theta_1,\theta_2) = V^\prime S(\theta) U^\prime W U \int_0^1 \{ S(\omega \theta_1 + (1-\omega)\theta_2) \} d\omega V$. By assumption, $V^\prime S(\theta)$ and  $U^\prime W U$ have full rank. As in the proof of Proposition \ref{prop:charac}, $\int_0^1 \{ S(\omega \theta_1 + (1-\omega)\theta_2) \} d\omega$ has full rank for any $\theta_1,\theta_2$, and $V$ is invertible. Hence, $S(\theta) U^\prime W U \int_0^1 \{ S(\omega \theta_1 + (1-\omega)\theta_2) \} d\omega V$ is invertible, $U$ has full rank  so that $G(\theta)^\prime W \overline{G}(\theta_1,\theta_2)$ has full rank for all $\theta,\theta_1,\theta_2$.

For part (2), take any $\theta_1,\theta_2,\theta_3$. Suppose $G(\theta_1)^\prime W g(\theta_2) = G(\theta_1)^\prime W g(\theta_3)$, apply Lemma \ref{lem:OMV} to find $G(\theta_1)^\prime W \overline{G}(\theta_2,\theta_3)(\theta_2-\theta_3)=0 \Rightarrow \theta_2= \theta_3$ under condition (a). \qed

\paragraph{Proof of Proposition \ref{prop:reparOI}:} 1) We'll proceed similarly to the proof of Proposition \ref{prop:repar}:
\begin{align*}
  &\int_0^1 \partial_u^\prime h(\omega u + (1-\omega)u^\dagger) G\circ h (\omega u + (1-\omega)u^\dagger)^\prime d\omega W G\circ h (u) \partial_u h(u)\\
  &= \partial_u^\prime h( u^\dagger)\int_0^1  G(\omega \theta + (1-\omega)\theta^\dagger)^\prime d\omega W G(\theta) \partial_u h(u)\\
  &+ \int_0^1 [\partial_u h(\omega u + (1-\omega)u^\dagger)  - \partial_u h( u^\dagger)]^\prime G(\omega \theta + (1-\omega)\theta^\dagger)^\prime d\omega W G(\theta) \partial_u h(u)\\
  &+ \int_0^1 \partial^\prime_u h(\omega u + (1-\omega)u^\dagger)[G\circ h (\omega u + (1-\omega)u^\dagger)-G(\omega \theta + (1-\omega)\theta^\dagger)]^\prime d\omega W G(\theta) \partial_u h(u).
\end{align*}
As before, we get: $\sigma_{\min}[\int_0^1 \partial_u^\prime h(\omega u + (1-\omega)u^\dagger) G\circ h (\omega u + (1-\omega)u^\dagger)^\prime d\omega W G\circ h (u) \partial_u h(u)] \geq \underline{\sigma} \underline{\sigma}_h^2 - C_1 \overline{\sigma}_h \overline{\sigma}^2\overline{\lambda}_W - C_2 L \overline{\sigma}_h^2 \overline{\sigma} \overline{\lambda}_W$ which is positive under the stated condition. As before, for $h$ affine we have $C_1=C_2 = 0$ so that the condition holds for $A$ finite and invertible. 2) The proof is the same as in the just-identified case.
\qed

\newpage

\begin{titlingpage} 
  \emptythanks
  \title{ {Supplement to\\ \lQ {\bf Convexity Not Required:
  Estimation of Smooth Moment Condition Models}''}}
  \author{ Jean-Jacques Forneron\thanks{Department of Economics, Boston University, 270 Bay State Road, Boston, MA 02215 USA.\newline Email: \href{mailto:jjmf@bu.edu}{jjmf@bu.edu}, Website: \href{http://jjforneron.com}{http://jjforneron.com}. } \and Liang Zhong\thanks{Department of Economics, Boston University, 270 Bay State Road, Boston, MA 02215 USA.\newline Email: \href{samzl@bu.edu}{samzl@bu.edu}, Website: \href{https://samzl1.github.io/}{https://samzl1.github.io/}.} }
  \setcounter{footnote}{0}
  \setcounter{page}{0}

  \clearpage 
  \maketitle 
  \thispagestyle{empty} 
  \begin{center}
  This Supplemental Material consists of Appendices \ref{apx:local}, \ref{sec:survey_properties}, \ref{apx:Rcode}, \ref{apx:addise}, and \ref{apx:optizs} to the main text.
  \end{center}
\end{titlingpage}

\setcounter{page}{1}

\section{Local Convergence Results} \label{apx:local}

%\subsection{Local Convergence} \label{sec:local_cv}
The following considers local convergence under correct specification, where $g(\theta^\dagger)=0$, and misspecification, where $g(\theta^\dagger)\neq 0$. These results highlight how several quantities affect the estimation. Here, Assumptions \ref{ass:1prim}, \ref{ass:2} are sufficient to study local convergence. Throughout it is assumed that $\hat{\theta}_n$ and $Q_n(\hat{\theta}_n)$ are consistent for $\theta^\dagger$ and $Q(\theta^\dagger)$.

\begin{proposition}[Correctly Specified] \label{prop:local_cv}
If Assumptions \ref{ass:1prim}, \ref{ass:2} hold, then for $\gamma \in (0,1)$ small enough, with probability approaching $1$, there exist $0 < R_n \leq R_G$ and $\tilde{\gamma} \in (0,1)$ such that :
\begin{align}
  \|\theta_{k+1}-\hat\theta_n\| \leq (1-\tilde{\gamma})\|\theta_{k}-\hat\theta_n\| \leq \dots \leq (1-\tilde{\gamma})^{k+1}\|\theta_{0}-\hat\theta_n\|\label{eq:local_cv}
\end{align}
for any $\|\theta_0 - \hat\theta_n\| \leq R_n$. For just-identified models, $\overline{g}_n(\hat\theta_n)=0$ implies $R_n >0$ with probability $1$. For over-identified models, $\overline{g}_n(\hat\theta_n) = o_p(1)$ implies $R_n >0$ with probability approaching $1$.
\end{proposition}

This result is comparable to those found for non-linear systems of equations \citep[e.g.][Ch11]{dennis1996,nocedal-wright:06}, with some notable differences. First, if the model is over-identified, $\overline{g}_n(\theta)=0$ does not have a solution, and standard results do not apply. Second, the area of local convergence $R_n$ is tied to a) the choice of tuning parameter $\gamma$, b) the size of the moments at the solution $\overline{g}_n(\hat\theta_n)$, c) the choice of weighting matrix. %This will be important for global convergence with over-identified and misspecified models. %The radius $R_n$ also depends on $\underline{\sigma}$, as shown below. 
For \textsc{gn}, the area of local convergence $R_n = \min(R_G,\tilde{R}_n)$ is the smallest of $R_G$ and:
\[ \tilde{R}_n = (1-\tilde{\gamma}/\gamma)\frac{\underline{\sigma}}{L\sqrt{\kappa_W}} - \frac{1}{\underline{\sigma}\sqrt{\underline{\lambda}_W}}\|\overline{g}_n(\hat\theta_n)\|_{W_n}, \]
where $\kappa_W = \overline{\lambda}_W/\underline{\lambda}_W$ bounds the condition number of the weighting matrix $W_n$. Having $\|\overline{g}_n(\hat\theta_n)\|_{W_n} \neq 0$ reduces the area of local convergence in finite samples. For correctly specified models $\overline{g}_n(\hat\theta_n) = o_p(1)$ implies $\tilde{R}_n \overset{p}{\to} \tilde{R} = (1-\tilde{\gamma}/\gamma)\underline{\sigma}/(\sqrt{\kappa_W}L) > 0$. Note that for \textsc{gn}, Proposition \ref{prop:local_cv} holds for any choice of $\gamma \in (0,1)$. This is typically not the case for other choices of $P_{k,n}$: \textsc{gd} requires $0 < \gamma < [\overline{\lambda}_W \overline{\sigma}^2]^{-1}$ to be sufficiently small. \textsc{gd} and \textsc{gn} iterations use the same inputs $G_n$ and $\overline{g}_n$, but the latter converges more quickly. %As \textsc{nr} and \textsc{qn} iterations require an exact or approximate Hessian, they are more costly than \textsc{gd}, \textsc{gn}.

The expression for $\tilde{R}_n$ illustrates that the choice of weighting matrix $W_n$ matters. Equal weighting, $W = I_d$, has $\kappa_W = 1$ whereas an ill-conditioned matrix has $\kappa_W \gg 1$. %This can make local optimization challenging. When the sample moments are highly correlated, the optimal weighting matrix can be ill-conditioned. Using a diagonal weighting matrix, or regularizing the optimal weighting matrix using $W_n = ( \hat{V}_n + \lambda I_d )^{-1}$, where  $\hat{V}_n$ estimates the variance of $\sqrt{n}\overline{g}_n(\theta^\dagger)$, may improve numerical stability.

In applications, $\|\overline{g}_n(\hat\theta_n)\|_{W_n}$ can be relatively large so that misspecification becomes a concern. Understanding the robustness of Proposition \ref{prop:local_cv} to non-negligible deviations from $Q(\theta^\dagger)=0$ is then empirically relevant. The following considers models where the quantity:
\[ Q_n(\hat{\theta}_n) \overset{p}{\to} Q(\theta^\dagger) := \varphi/2 >0 \] 
does not vanish asymptotically which implies that $\|\overline{g}_n(\hat{\theta}_n)\|_{W_n}$ matters for local convergence, even in large samples. 
Since $G_n$ cannot be full rank at $\theta=\hat\theta_n$ when the model is both just-identified and misspecified, the results presented here solely consider over-identified models.\footnote{The solution $\hat\theta_n$ is s.t. $G_n(\hat\theta_n)^\prime W_n \overline{g}_n(\hat\theta_n)=0$, misspecification implies $\overline{g}_n(\hat\theta_n) \neq 0$, and since $W_n$ has full rank, it must be that $G_n(\hat\theta_n)$ is singular for just-identified models. For over-identified models, $\overline{g}_n(\hat\theta_n)$ is in the null space of $G_n(\hat\theta_n)^\prime W_n$, which allows $G_n(\hat\theta_n)$ to be full rank.}

%The following considers ``moderate'' amounts of misspecification in the sense that: 
%\[ \text{plim}_{n\to \infty} Q_n(\hat\theta_n) = Q(\theta^\dagger) := \varphi^2 \geq 0\] exists and can be non-zero in the limit. 

%When $\varphi > 0$, the degree of misspecification is non-negligible asymptotically and the statistic $n \|\overline{g}_n(\hat\theta_n)\|_{W_n}^2 \to \infty$ can diverge. However, $\varphi$ cannot be too large for the local and global convergence results to hold as shown below. For simplicity, only Gauss-Newton will be considered in the results. Also, s 

%For correctly specified models, a test for over-identifying restrictions can diagnose global convergence \citep[][Sec3.3]{andrews1997}. For misspecified models, such test would frequently reject in large samples. Then the issue is that, when the test rejects, either 1) the optimizer has not found valid estimates, or 2) the model fits the data poorly in some dimension(s). When $Q_n$ is globally convex, a given value is the global solution if, and only if, it satisfies the first and second-order optimality conditions.\footnote{The first is $\partial_\theta Q_n(\hat\theta_n)=0$ and the second $\partial^2_{\theta,\theta^\prime} Q_n(\hat\theta_n)$ positive semidefinite.} Without convexity, this only guarantees a local optimum. For moderately misspecified models, Assumption \ref{ass:conds} provides an alternative to convexity in these settings.

\begin{proposition}[Misspecified] \label{prop:local_cv_ms}
  Suppose Assumptions \ref{ass:1prim}, \ref{ass:2} hold, and $\varphi$ is such that:
  \begin{align}
    \sqrt{\varphi} < \frac{\underline{\sigma}^2 \sqrt{\overline{\lambda}_W}}{L \kappa_W \kappa_P},  \label{eq:phi_mis}
  \end{align}
  where $\kappa_P =  \overline{\lambda}_P / \underline{\lambda}_P$.  For $\gamma \in (0,1)$ small enough, there exists $\tilde{\gamma} \in (0,\gamma)$, such that, with probability approaching $1$, for any $\|\theta_0-\hat\theta_n\| \leq R_n$, and all $k \geq 0$:
  \begin{align}
    \|\theta_{k+1}-\hat\theta_n\| \leq (1-\tilde{\gamma})\| \theta_{k}-\hat\theta_n\| \leq \dots \leq (1-\tilde{\gamma})^{k+1} \|\theta_0 - \hat\theta_n\|, \tag{\ref{eq:local_cv}} \label{eq:local_cvp}
  \end{align}
  with the same $R_n$ found in Proposition \ref{prop:local_cv}; such that $\text{plim}_{n \to \infty} R_n = R > 0$ when (\ref{eq:phi_mis}) holds.
  \end{proposition}
  
The result shows that under `moderate' amounts of misspecification, the area of local convergence is asymptotically non-empty. For \textsc{gn}, the condition simplifies to:
  $\sqrt{\varphi} < \frac{\underline{\sigma}^2 \sqrt{\underline{\lambda}_W}}{L \sqrt{\kappa_W}}$.
 Several terms restrict the amount of misspecification in (\ref{eq:phi_mis}): $\underline{\sigma}$, $L$, and the pair $\underline{\lambda}_W,\kappa_W$. The first measures local identification strength, the second non-linearity, and the latter comes from the weighting matrix.   For linear models, $L=0$, the conditions reads $\sqrt{\varphi} < +\infty$; misspecification only matters in nonlinear problems with $L>0$. Note that the area of local convergence is asymptotically smaller than in Proposition \ref{prop:local_cv}.

%\subsection{Proofs for Section \ref{sec:local_cv}} \label{apx:local_cv}

\paragraph{Proof of Proposition \ref{prop:local_cv} (Gauss-Newton).}
 Take $\theta_k \in \mathbb{R}^{d_\theta}$, the update (\ref{eq:update}) can be re-written as:
\begin{equation}
  \begin{aligned}
    \theta_{k+1} - \hat\theta_n = &\Big( I_d - \gamma P_{k,n} G_n(\theta_k)^\prime W_n G_n(\theta_k)\Big)(\theta_k-\hat\theta_n)\\ &- \gamma P_{k,n} G_n(\theta_k)^\prime W_n [\overline{g}_n(\theta_k) - G_n(\theta_k)(\theta_k - \hat\theta_n)].
  \end{aligned}  \label{eq:update2}
\end{equation}
For \textsc{gn}, $P_{k,n} G_n(\theta_k)^\prime W_n G_n(\theta_k)\ = I_d$ so that we have:
\begin{equation}
  \begin{aligned}
    \theta_{k+1} - \hat\theta_n = & (1-\gamma)(\theta_k-\hat\theta_n)\\ &- \gamma P_{k,n} G_n(\theta_k)^\prime W_n [\overline{g}_n(\theta_k) - \overline{g}_n(\hat\theta_n) - G_n(\theta_k)(\theta_k - \hat\theta_n)]\\
    &- \gamma P_{k,n} [G_n(\theta_k)-G_n(\hat\theta_n)]^\prime W_n \overline{g}_n(\hat\theta_n),
  \end{aligned}  \tag{\ref{eq:update2}'}
\end{equation}
using the first-order condition $G_n(\hat\theta_n)^\prime W_n \overline{g}_n(\hat\theta_n) = 0$. From Assumption \ref{ass:1}, there exists $R_G > 0$ such that: $\underline{\sigma} \leq \sigma_{\min}[G_n(\theta_k)]$ for any $\|\theta_k-\hat\theta_n\| \leq R_G$, which implies that $P_{k,n}$ is well defined and bounded. Since $G_n$ is Lipschitz continuous with constant $L \geq 0$:
\[ \|P_{k,n} G_n(\theta_k)^\prime W_n [\overline{g}_n(\theta_k) - \overline{g}_n(\hat\theta_n) - G_n(\theta_k)(\theta_k - \hat\theta_n)]\| \leq \underline{\sigma}^{-1} \sqrt{\overline{\lambda}_W/\underline{\lambda}_W} L \|\theta_k - \hat\theta_n\|^2, \]
 We also have:
\[ \|P_{k,n} [G_n(\theta_k)-G_n(\hat\theta_n)]^\prime W_n \overline{g}_n(\hat\theta_n)\| \leq \underline{\sigma}^{-2} (\sqrt{\overline{\lambda}_W}/\underline{\lambda}_W) L \|\overline{g}_n(\hat\theta_n)\|_{W_n} \|\theta_k - \hat\theta_n\|.   \]
Combine these two inequalities into (\ref{eq:update2}') to find:
\begin{equation}
  \begin{aligned}
    &\|\theta_{k+1} - \hat\theta_n\|\\ &\leq \left(1-\gamma + \gamma \left[\underline{\sigma}^{-1}\sqrt{\overline{\lambda}_W/\underline{\lambda}_W} L \|\theta_k - \hat\theta_n\| +\underline{\sigma}^{-2} (\sqrt{\overline{\lambda}_W}/\underline{\lambda}_W) L \|\overline{g}_n(\hat\theta_n)\|_{W_n} \right] \right)\|\theta_k-\hat\theta_n\|.
  \end{aligned}  \tag{\ref{eq:update2}''}
\end{equation}

Now take any $\tilde{\gamma} \in (0,\gamma)$, let:
\[ \tilde{R}_n = \frac{\gamma - \tilde{\gamma}}{\gamma }\left[L^{-1} \underline{\sigma}\sqrt{\underline{\lambda}_W/\overline{\lambda}_W}\right] - (\underline{\sigma}^{-1}/\sqrt{\underline{\lambda}_W}) \|\overline{g}_n(\hat\theta_n)\|_{W_n}.\]% \overset{p}{\to} R = \frac{\gamma - \tilde{\gamma}}{\gamma}[L \underline{\sigma}^{-1}\sqrt{\overline{\lambda}_W/\underline{\lambda}_W} ]^{-1} >0. \]
Let $R_n = \min(\tilde{R}_n,R_G)$, for any $\|\theta_k - \hat\theta_n\| \leq R_n$, we have $\|\theta_{k+1} - \hat\theta_n\| \leq (1-\tilde{\gamma})\|\theta_{k} - \hat\theta_n\| \leq R_n$. By recursion, we then have for any $\|\theta_0 - \hat\theta_n\| \leq R_n$:
\[ \|\theta_{k+1} - \hat\theta_n\| \leq (1-\tilde{\gamma})\|\theta_{k} - \hat\theta_n\| \leq \dots \leq (1-\tilde{\gamma})^{k+1}\|\theta_{0} - \hat\theta_n\|,  \]
as stated in (\ref{eq:local_cv}).
\qed

\paragraph{Proof of Proposition \ref{prop:local_cv} (General Case).} Take $\theta_k \in \mathbb{R}^{d_\theta}$,  the update (\ref{eq:update}) can be re-written as:
\begin{equation}
  \begin{aligned}
    \theta_{k+1} - \hat\theta_n = &\Big( I_d - \gamma P_{k,n} G_n(\theta_k)^\prime W_n G_n(\theta_k)\Big)(\theta_k-\hat\theta_n)\\ &- \gamma P_{k,n} G_n(\theta_k)^\prime W_n [\overline{g}_n(\theta_k) - G_n(\theta_k)(\theta_k - \hat\theta_n)].
  \end{aligned}   \tag{\ref{eq:update2}}
\end{equation}
Taking norms on both sides this identity yields:
\begin{equation}
  \begin{aligned}
    \|\theta_{b+1}-\hat\theta_n\| \leq &\sigma_{\max} \Big[ I_d - \gamma P_{k,n} G_n(\theta_k)^\prime W_n G_n(\theta_k) \Big] \| \theta_b-\hat\theta_n \| \\ &+ \gamma \|P_{k,n} G_n(\theta_k)^\prime W_n [\overline{g}_n(\theta_k) - G_n(\theta_k)(\theta_k - \hat\theta_n)]\|,
  \end{aligned}  \tag{\ref{eq:update2}'}
\end{equation}
where $\sigma_{\max}$ returns the largest singular value.
We will now bound each of these two terms. First, note that $\sigma_{\max} [ I_d - \gamma P_{k,n} G_n(\theta_k)^\prime W_n G_n(\theta_k) ] = \sigma_{\max} [ I_d - \gamma P_{k,n}^{1/2} G_n(\theta_k)^\prime W_n G_n(\theta_k) P_{k,n}^{1/2}] = \max_{j=1,\dots,d} | \lambda_{j}  [ I_d - \gamma P_{k,n}^{1/2} G_n(\theta_k)^\prime W_n G_n(\theta_k) P_{k,n}^{1/2}] |$, where $\lambda_j$ are the eigenvalues. Because this is a difference of Hermitian matrices, Weyl's perturbation inequality \citep[Corollary III.2.2]{bhatia2013} implies the following bounds:
\begin{align*}
  1 - \gamma \lambda_{\max}[P_{k,n}^{1/2} G_n(\theta_k)^\prime W_n G_n(\theta_k) P_{k,n}^{1/2}] &\leq \lambda_{\min}[ I_d - \gamma P_{k,n}^{1/2} G_n(\theta_k)^\prime W_n G_n(\theta_k) P_{k,n}^{1/2}] \\ &\leq \lambda_{\max}[ I_d - \gamma P_{k,n}^{1/2} G_n(\theta_k)^\prime W_n G_n(\theta_k) P_{k,n}^{1/2}] \\ &\leq 1 - \gamma \lambda_{\min}[P_{k,n}^{1/2} G_n(\theta_k)^\prime W_n G_n(\theta_k) P_{k,n}^{1/2}].
\end{align*}
Let $\overline{\sigma} = \max_{\theta \in \Theta} \sigma_{\max}[G_n(\theta)]$, suppose $ 0 < \gamma < [\overline{\lambda}_P \overline{\lambda}_W \overline{\sigma}^2]^{-1}$, we then have:
\[ 0 \leq 1 - \gamma \lambda_{\max}[P_{k,n}^{1/2} G_n(\theta_k)^\prime W_n G_n(\theta_k) P_{k,n}^{1/2}] \leq 1 - \gamma \lambda_{\min}[P_{k,n}^{1/2} G_n(\theta_k)^\prime W_n G_n(\theta_k) P_{k,n}^{1/2}],\]
so that we are only concerned with the upper bound. From Assumption \ref{ass:1}, $\|\theta-\hat\theta_n\| \leq R_G \Rightarrow \sigma_{\min}[G_n(\theta)] \geq \underline{\sigma}$. Combine with the bound for $\gamma$ to find: 
\[ 0 \leq \sigma_{\max} [ I_d - \gamma P_{k,n} G_n(\theta_k)^\prime W_n G_n(\theta_k) ] \leq 1- \gamma \underline{\lambda}_P \underline{\lambda}_W \underline{\sigma}^2 < 1, \]
for any choice of $\gamma \in (0,[\overline{\lambda}_P \overline{\lambda}_W \overline{\sigma}^2]^{-1})$. For the second term in (\ref{eq:update2}), using the identity $G_n(\hat\theta_n)^\prime W_n \overline{g}_n(\hat\theta_n) = 0$ and Lemma \ref{lem:OMV}:
\begin{align*}
  P_{k,n} G_n(\theta_k)^\prime W_n [\overline{g}_n(\theta_k) - G_n(\theta_k)(\theta_k - \hat\theta_n)] = &P_{k,n} G_n(\theta_k)^\prime W_n [\overline{G}_n(\theta_k) - G_n(\theta_k)](\theta_k - \hat\theta_n) \\&+ P_{k,n} [G_n(\theta_k) - G_n(\hat\theta_n)]^\prime W_n \overline{g}_n(\hat\theta_n),
\end{align*}
where $\overline{G}_n(\theta_k) = \int_0^1 \{ G_n(\omega \theta_k + (1-\omega)\hat{\theta}_n) \}d\omega$. Since $G_n$ is Lipschitz continuous with constant $L \geq 0$:
\begin{align*}
  \|(\ref{eq:update2}')\| &\leq (1-\gamma \underline{\lambda}_P \underline{\lambda}_W \underline{\sigma}^2)\|\theta_b-\hat\theta_n\| + \gamma \overline{\lambda}_P \overline{\lambda}_W \overline{\sigma} L \|\theta_b - \hat\theta_n\|^2 + \gamma \overline{\lambda}_P \overline{\lambda}_W^{1/2} L \|\overline{g}_n(\hat\theta_n)\|_{W_n} \|\theta_b - \hat\theta_n\|\\
  &= \Big( 1-\gamma \underline{\lambda}_P \underline{\lambda}_W \underline{\sigma}^2 + \gamma \Big[\overline{\lambda}_P \overline{\lambda}_W \overline{\sigma} L \|\theta_b - \hat\theta_n\| + \overline{\lambda}_P \overline{\lambda}_W^{1/2} L \|\overline{g}_n(\hat\theta_n)\|_{W_n} \Big] \Big)\|\theta_b-\hat\theta_n\|.
\end{align*}
Let  $c_1 = \overline{\lambda}_P \overline{\lambda}_W \overline{\sigma} L$, $c_2 = \overline{\lambda}_P \overline{\lambda}_W^{1/2} L$, pick $\tilde{\gamma} \in (0,  \gamma \underline{\lambda}_P \underline{\lambda}_W \underline{\sigma}^2)$, and assume:
\begin{align}
  \|\theta_k - \hat\theta_n\| \leq \frac{\gamma \underline{\lambda}_P \underline{\lambda}_W \underline{\sigma}^2- \tilde{\gamma}}{\gamma c_1} - \frac{c_2}{c_1}\|\overline{g}_n(\hat\theta_n)\|_{W_n}:= \tilde{R}_n. \label{eq:Rn}
\end{align}
Take $R_n = \min(R_G,\tilde{R}_n)$, $\|\theta_k-\hat\theta_n\| \leq R_n$ implies that, by construction:
\[ \|\theta_{k+1} - \hat\theta_n\| \leq (1-\tilde{\gamma})\|\theta_{k} - \hat\theta_n\| \leq \dots \leq (1-\overline{\gamma})^{k+1}\|\theta_0 - \hat\theta_n\|, \]
by recursion, if $\|\theta_0 - \hat\theta_n\| \leq R_n$. \qed

\paragraph{Proof of Proposition \ref{prop:local_cv_ms} (Gauss-Newton):}
The proof is similar to the proof of Proposition \ref{prop:local_cv} with the difference that $\|\overline{g}_n(\hat{\theta}_n)\|_{W_n} \overset{p}{\to} \sqrt{\varphi/2} > 0$. The radius is convergence is asymptotically non-zero for $0 < \tilde{\gamma} < \gamma < 1$ small enough if:
$\sqrt{\varphi} < \frac{\underline{\sigma}^2 \sqrt{\underline{\lambda}_W}}{L \sqrt{\kappa_W}}$.
\qed

\paragraph{Proof of Proposition \ref{prop:local_cv_ms} (General Case):}
The proof is similar to the proof of Proposition \ref{prop:local_cv} with the difference that $\|\overline{g}_n(\hat{\theta}_n)\|_{W_n} \overset{p}{\to} \sqrt{\varphi/2} > 0$. The radius is convergence is asymptotically non-zero for $0 < \tilde{\gamma} < \gamma < 1$ small enough if:
$\sqrt{\varphi} < \frac{\underline{\sigma}^2 \sqrt{\overline{\lambda}_W}}{L \kappa_W \kappa_P}$,
where $\kappa_P =  \overline{\lambda}_P / \underline{\lambda}_P$.
\qed

\section{Commonly used methods and their properties} \label{sec:survey_properties}
%Before introducing the results, the following provides an overview of empirical practice and the properties of the algorithms that are commonly used.
\subsection{A survey of empirical practice} \label{sec:survey}
\paragraph{Survey methodology:} The survey covers empirical papers published in the American Economic Review (AER) between 2016 and 2018. The focus on this specific outlet is driven by the mandatory data and code policy enacted in 2005. Indeed, since a number of papers provide little or no detail in the paper on the methodology used to compute estimates numerically, it is important to read the replication codes to determine what was implemented. The search function in JSTOR was used to find the papers matching the survey criteria. The database did not include more recent publications at the time of the survey.\footnote{The search function in JSTOR allows to search for keywords within the title, abstract, main text, and supplemental material of a paper. Further screening ensures that each paper in the search results actually implements at least one of the estimations considered. The search criteria include keywords: ``Method of Moments," ``Indirect Inference," ``Method of Simulated Moments," ``Minimum Distance," and ``MM."} Table \ref{tab:survey} was constructed by reading through the main text, supplemental material, and all available replication codes of the selected papers.
\begin{table}[th]  \caption{American Economic Review 2016-2018: GMM and related empirical estimations} \label{tab:survey}
  \centering
%  \begingroup
  \setlength\tabcolsep{4.5pt}
    \renewcommand{\arraystretch}{0.9} 
  {\small
  \begin{tabular}{l|ccc} \hline \hline
    Method & \# Papers & \# Parameters (p) & Data available \\ \hline
    Nelder-Mead - one starting value                              & 7 & 2,6 ($\times$2),11,13 ($\times$2),147 & 3\phantom{$^\dagger$} \\
    Simulated Annealing + Nelder-Mead          & 2 & 4,13 & 1\phantom{$^\dagger$} \\
    Nelder-Mead - multiple starting values & 2 & ?,6 & 1$^{\ddagger}$ \\
    Pattern Search                             & 2 & 6,147 & 1$^\dagger$ \\
    Genetic Algorithm                          & 2 & 9,14 & 1\phantom{$^\dagger$}  \\
    Simulated Annealing                        & 2 & 4,13 & 2$^\dagger$  \\
    MCMC                                       & 1 & 15 & 1\phantom{$^\dagger$} \\
    Grid Search                                & 1 & 5 & 1\phantom{$^\dagger$} \\\hline
    No description                     & 3 & - & -\phantom{$^\dagger$} \\
    Stata/Mata default                     & 4 & 3,6 ($\times$2),38 & 3$^\star$ \\\hline \hline
  \end{tabular}}\\
  \notes{ \textbf{Legend:} \# Parameters correspond to the size of the largest specification. Data avail. reports if the dataset is included with the replication files. Estimations surveyed include: Generalized Method of Moments (GMM), Minimum Distance (MD), Simulated Method of Moments (SMM), and Indirect Inference. ?: information not available due to the lack of replication codes. $^\star$: one of the 3 papers reported to include data requires to download the PSID dataset separately. $^\dagger$: two papers in total also rely separately on Nelder-Mead, so they are also reported under Nelder-Mead. $\ddagger$: one paper provides data without codes. }
\end{table}
\paragraph{Survey results:}
Table \ref{tab:survey} provides an overview of the quantitative results of the survey. Additional details on the algorithms in the table are given below. There are 23 papers in total, a little over 7 papers per year. Excluding the estimation with 147 parameters, the average estimation has around 10 coefficients, and the median is 6. 3 papers used more than one starting value, and the remaining 20 papers either used the solver default or typed in a specific value in the replication code. There is generally no information provided on the origin of these specific starting values. Of the papers using multiple starting values, one did not provide replication codes, and the other two used 12 and 50 starting points. Some of the estimations are very time-consuming. For instance, \citet{Lise2017} use MCMC for estimation (but not inference) and report that each evaluation of the moments takes 45s. In total, their estimation takes more than a week to run in a 96-core cluster environment. 

As mentioned in the introduction, although convex optimizers such as (stochastic) gradient-descent and quasi-Newton methods are commonly used to solve large scale convex minimization problems, they are virtually absent from the survey. Overall, 11 papers rely on the Nelder-Mead algorithm, alone or in combination with another method, making it the most popular optimizer in this survey. Pattern search, used in 2 papers, belongs to the same family of algorithms as Nelder-Mead. The following provides a brief overview of the properties of the main Algorithms found used in Table \ref{tab:survey}. %Although it is very popular, Nelder-Mead was shown analytically to have local convergence issues for convex problems \citep{mckinnon1998}. It is not guaranteed to converge to a local optimum. Section \ref{sec:algos} gives a brief overview of methods in Table \ref{tab:survey} and their convergence properties. There are few formal results for Genetic algorithms, so they are not discussed. Appendix \ref{apx:optizs} provides further information about these methods.

\subsection{A brief summary of the Algorithms' properties} \label{sec:algos}

The following briefly discussed the properties of four algorithms from Table \ref{tab:survey}: Nelder-Mead, Grid Search, Multi-Start, and Simulated Annealing. Further discussion, descriptions, and references can be found in Appendix \ref{apx:optizs}.

Nelder-Mead (\textsc{nm}) is the most popular method in the survey, it can be used even if $Q_n$ is discontinuous. Its convergence properties, which measure its ability to find valid estimates, are somewhat limited however. For some smooth convex problems, it can be shown to converge to values that are neither locally nor globally optimal. 
The grid-search converges to the solution under weak conditions, unlike \textsc{nm}. It is very slow, however, and often not practical when estimating three or more coefficients. 
Simulated annealing (\textsc{sa}) is not deterministic. Still it converges, in probability, under weak conditions to the solution. Albeit, the convergence is predicted to be slower than grid search. 
A common approach to improve the convergence of a given algorithm is to combine it with multiple starting values. The required number of starting values depends on $Q_n$ and the choice of algorithm. \citet{andrews1997} provides an asymptotically valid stopping rule for correctly specified GMM models.

When $Q_n$ is strongly convex, several gradient-based methods discussed below are rapidly, globally convergent and do not suffer from a curse of dimensionality. This implies that it is possible to estimate a large number of parameters in a reasonable amount of time. Similar convergence properties are derived in this paper, under rank conditions instead of convexity.

\section{R Code for the MA(1) Example} \label{apx:Rcode}
\lstset{basicstyle=\ttfamily\small,columns=fullflexible,lineskip=-2pt}
{\small
\begin{lstlisting}
library(stats)  # fit an AR(p) model
library(pracma) # compute jacobian

n = 200        # sample size n
theta = -1/2   # MA(1) coefficient

set.seed(123)  # set the seed for random numbers
e = rnorm(n+1)              # draw innovations
y = e[2:(n+1)] - theta*e[1:n]   # generate MA(1) data
p = 12          # number of lags for the AR(p) models

beta <- function(theta) {   
    # computes the p-limit of the OLS estimates
    # V = covariance matrix of (y_{t-1},...,y_{t-p})
    V = diag(p+1)*(1+theta^2)   # variances on the diagonal
    diag(V[,-1]) = -theta       # autocovariance
    V = t(V)                    # transpose
    diag(V[,-1]) = -theta       # autocovariance
    return( 
        solve( V[2:(p+1),2:(p+1)], V[1,2:(p+1)] ) 
        # p-limit = inv(V)*( vector of autocovariances )
    )
}

# Fit the AR(p) auxiliary model:
ols_p = c(ar.ols( y, aic = FALSE, order.max = p, demean = FALSE, intercept = FALSE )$ar)

moments <- function(theta) { 
    # computes the sample moments gn
    return( ols_p - beta(theta) ) # gn = psi_n - psi(theta)
}

objective <- function(theta,disp = FALSE) { 
    # compute the sample objective Qn
    if (disp == TRUE) {
        print(round(theta,3)) # print to tack R's optimization paths
    }
    mm = moments(theta)       # compute sample moments gn
    return( t(mm)%*%mm )      # compute Qn = gn'*gn (W = Id)
}

dQ <- function(theta,disp=FALSE) { 
    # compute the derivative of Qn
    # gradient of Qn = -2*d psi(theta)/ d theta' * gn(theta)
    return(-2*t(jacobian(beta,theta))%*%moments(theta))
}

# L-BFGS-B: with bound constraints
o1 = optim(0.95,objective,gr=dQ,method="L-BFGS-B",lower=c(-1),upper=c(1),disp=TRUE)
# BFGS: without bound constraints
o2 = optim(0.95,objective,gr=dQ,method="BFGS",disp=TRUE)

# *********************************
#           Gauss-Newton
# *********************************
gamma = 0.1 # learning rate
coefsGN = rep(0,150) # 150 iterations in total
coefsGN[1] = 0.95    # starting value: theta = 0.95

for (b in 2:150) { # main loop for Gauss-Newton
    Gn  = -jacobian(beta,coefsGN[b-1])  # 1. compute Jacobian
    mom =  moments(coefsGN[b-1])        # 2. compute moments
    coefsGN[b] = coefsGN[b-1] - gamma*solve(t(Gn)%*%Gn,t(Gn)%*%mom) # 3. update
}   # repeat for each b

# Put the results into a table:
results = matrix(NA,2,3)
colnames(results) = c('L-BFGS-B','BFGS','GN')
results[1,] = c(o1$par,o2$par,coefsGN[150])
results[2,] = sapply(results[1,],objective)
rownames(results) = c('theta','Qn(theta)')

print(results,digits=3)
# Output should look like this:
#           L-BFGS-B   BFGS     GN
# theta         -1.0 -6.979 -0.626
# Qn(theta)      1.7  0.397  0.101
\end{lstlisting}}

\section{Additional Empirical Results} \label{apx:addise}

\subsection{Demand for Cereal} \label{apx:cereal_additional}

\begin{table}[ht]
  \centering \caption{Demand for Cereal: \textsc{gn} with different learning rates}  \label{tab:BLPgamma}
  \setlength\tabcolsep{4.25pt}
    \renewcommand{\arraystretch}{0.9}
{\footnotesize \begin{tabular}{cr|cccc|cccc|c|c}
  \hline \hline
  & & \multicolumn{4}{c|}{\textsc{stdev}} & \multicolumn{4}{c|}{\textsc{income}}  & \multirow{2}{*}{crash} & \multirow{2}{*}{objs} \\
    &  & const. & price & sugar & mushy  & const. & price & sugar & mushy &   &    \\ 
  \hline
%  \multicolumn{4}{l}{\textsc{True}} &  &  &  &  &  &  &   \\
\multirow{2}{*}{\textsc{true}} & est & 0.28 & 2.03 & -0.01 & -0.08 & 3.58 & 0.47 & -0.17 & 0.69 & 33.84 & \multirow{2}{*}{-} \\ 
& se & 0.11 & 0.76 & 0.01 & 0.15 & 0.56 & 3.06 & 0.02 & 0.26 & - &  \\
   \hline
%\multicolumn{4}{l}{\textsc{gn}} &  &  &  &  &  &  &\\ 
\rowcolor{gray}
 & avg & 0.28 & 2.03 & -0.01 & -0.08 & 3.58 & 0.47 & -0.17 & 0.69 & 33.84 &    \\  \rowcolor{gray}
 \multirow{-2}{*}{$\gamma = 0.1$}   &   std  & 0.00 & 0.00 & 0.00 & 0.00 & 0.00 & 0.00 & 0.00 & 0.00 & 0.00 & \multirow{-2}{*}{0}  \\ 
      \hline
%\multicolumn{4}{l}{\textsc{bfgs}} &  &  &  &  &  &  &   \\
%\rowcolor{gray}
 & avg & 0.28 & 2.03 & -0.01 & -0.08 & 3.58 & 0.47 & -0.17 & 0.69 & 33.84 &    \\  %\rowcolor{gray}
 \multirow{-2}{*}{$\gamma = 0.2$}   &   std  & 0.00 & 0.00 & 0.00 & 0.00 & 0.00 & 0.00 & 0.00 & 0.00 & 0.00 & \multirow{-2}{*}{0}  \\ 
      \hline
%\multicolumn{4}{l}{\textsc{nm}} &  &  &  &  &  &  &\\
\rowcolor{gray}
 & avg & 0.28 & 2.03 & -0.01 & -0.08 & 3.58 & 0.47 & -0.17 & 0.69 & 33.84 &    \\  \rowcolor{gray}
 \multirow{-2}{*}{$\gamma = 0.4$}   &   std  & 0.00 & 0.00 & 0.00 & 0.00 & 0.00 & 0.00 & 0.00 & 0.00 & 0.00 & \multirow{-2}{*}{0}  \\ 
      \hline
%\multicolumn{4}{l}{\textsc{sa+nm}} &  &  &  &  &  &  &\\
%\rowcolor{gray}
 & avg & 0.28 & 2.03 & -0.01 & -0.08 & 3.58 & 0.47 & -0.17 & 0.69 & 33.84 &    \\  %\rowcolor{gray}
 \multirow{-2}{*}{$\gamma = 0.6$}   &   std  & 0.00 & 0.00 & 0.00 & 0.00 & 0.00 & 0.00 & 0.00 & 0.00 & 0.00 & \multirow{-2}{*}{0}  \\ 
\hline
\rowcolor{gray}
& avg & 0.28 & 2.03 & -0.01 & -0.08 & 3.58 & 0.47 & -0.17 & 0.69 & 33.84 &    \\  \rowcolor{gray}
\multirow{-2}{*}{$\gamma = 0.8$}   &   std  & 0.00 & 0.00 & 0.00 & 0.00 & 0.00 & 0.00 & 0.00 & 0.00 & 0.00 & \multirow{-2}{*}{0}  \\ 
\hline
%\rowcolor{gray}
& avg & 0.28 & 2.03 & -0.01 & -0.08 & 3.58 & 0.47 & -0.17 & 0.69 & 33.84 &    \\  %\rowcolor{gray}
\multirow{-2}{*}{$\gamma = 1$}   &   std  & 0.00 & 0.00 & 0.00 & 0.00 & 0.00 & 0.00 & 0.00 & 0.00 & 0.00 & \multirow{-2}{*}{0}  \\ 
\hline\hline
\end{tabular}  } 
  \notes{ \textbf{Legend:} Comparison for 50 starting values where $[0,10] \times \dots \times [0,10]$ for standard deviations and $[-10,10] \times \dots \times [-10,10]$ for income coefficients. Avg, Std: sample average and standard deviation of optimizer outputs. \textsc{true}: full sample estimate (est)  and standard errors (se). Objs: avg and std of minimized objective value. crash: optimization terminated because the objective function returned an error. \textsc{gn} run with $\gamma\in\{0.1,0.2,0.4,0.6,0.8,1\}$ for $k=150$ iterations for all starting values.  } 
\end{table}

\subsection{Impulse Response Matching} \label{apx:IRF_additional}

The following tables report results for \textsc{gn} using a range of tuning parameters $\gamma$. Since Assumption \ref{ass:conds} does not hold towards the lower bound for $\eta,\nu$, \textsc{gn} alone can crash and/or fail to converge. Following \citet{Forneron2022b}, we can introduce a global step:
\begin{align}
  &\theta_{k+1} = \theta_k - \gamma P_{k,n} G_n(\theta_k)^\prime W_n \overline{g}_n(\theta_k) \tag{\ref{eq:update}}\\
  &\text{if } \|\overline{g}_n(\theta^{k+1})\|_{W_n} < \|\overline{g}_n(\theta_{k+1})\|_{W_n}, \text{ set } \theta_{k+1} = \theta^{k+1} \notag
\end{align}
where the sequence $(\theta^{k})_{k \geq 0}$ is predetermined and dense in $\Theta$. The results rely on the Sobol sequence, independently randomized for each of the 50 starting values.\footnote{We take $(s_{k})_{k \geq 0}$ in $[0,1]^p$, $p \geq 1$ is the number of parameters, draw one vector $(u_1,\dots,u_p) \sim \mathcal{U}_{[0,1]^p}$, for each starting value, and compute $\tilde{s}_k = ( s_k + u ) \text{ modulo } 1$, then map $\tilde{s}_k$ to the bounds for $\theta = (\theta_1,\dots,\theta_p)$. The randomization is used to create independent variation in the global step between starting values to emphasize that convergence does not rely on a specific value in the sequence $(\theta^k)_{k \geq 0}$; this is called a random shift \citep[see][Ch6.2.1]{lemieux2009}.} Results are reported with and without the global step. Also, the former implements error-handling (try-catch).

\begin{table}[H]
  \centering \caption{\textsc{gn} with different learning rates}  \label{tab:imp_glob1p}
  \setlength\tabcolsep{4.25pt}
    \renewcommand{\arraystretch}{0.9}
{\footnotesize \begin{tabular}{cr|cccc|c|c||cccc|c|c}
  \hline \hline
  & & \multicolumn{6}{c||}{\textsc{without reparameterization}} & \multicolumn{6}{c}{\textsc{with reparameterization}} \\ \hline
  &  & $\eta$ & $\nu$ & $\rho_s$ & $\sigma_s$  & objs  & crash & $\eta$ & $\nu$ & $\rho_s$ & $\sigma_s$  & objs  & crash \\  \hline
\multirow{1}{*}{\textsc{true}} & est & 0.30 & 0.29 & 0.39 & 0.17 & 4.65 &   - & 0.30 & 0.29 & 0.39 & 0.17 & 4.65 &   - \\ 
\hline
$\gamma = $ & & \multicolumn{12}{c}{\textsc{gn without global step}} \\
\hline %\rowcolor{gray}
  %\multicolumn{4}{l}{\textsc{sims}} &    &  &\\
& avg & 0.30 & 0.29 & 0.39 & 0.17 & 4.65 &   &  0.30 & 0.29 & 0.39 & 0.17 & 4.65    &  \\ %\rowcolor{gray}
\multirow{-2}{*}{$0.1$} & std & 0.00 & 0.00 & 0.00 & 0.00 & 0.00  &   \multirow{-2}{*}{1} & 0.00 & 0.00 & 0.00 & 0.00 & 0.00  &   \multirow{-2}{*}{9} \\ 
\hline %\rowcolor{gray}
  %\multicolumn{4}{l}{\textsc{sims}} &    &  &\\
& avg & 0.30 & 0.29 & 0.39 & 0.17 & 4.65    &  & 0.30 & 0.29 & 0.39 & 0.17 & 4.65    &  \\ %\rowcolor{gray}
\multirow{-2}{*}{$0.2$} & std & 0.00 & 0.00 & 0.00 & 0.00 & 0.00  &   \multirow{-2}{*}{1} & 0.00 & 0.00 & 0.00 & 0.00 & 0.00  &   \multirow{-2}{*}{16} \\ 
\hline %\rowcolor{gray}
  %\multicolumn{4}{l}{\textsc{sims}} &    &  &\\
& avg & 0.30 & 0.29 & 0.39 & 0.17 & 4.65    &  & 0.30 & 0.29 & 0.39 & 0.17 & 4.65    &   \\ %\rowcolor{gray}
\multirow{-2}{*}{$0.4$} & std & 0.00 & 0.00 & 0.00 & 0.00 & 0.00  &   \multirow{-2}{*}{1} & 0.00 & 0.00 & 0.00 & 0.00 & 0.00  &   \multirow{-2}{*}{20}\\ 
\hline %\rowcolor{gray}
  %\multicolumn{4}{l}{\textsc{sims}} &    &  &\\
& avg & 0.30 & 0.29 & 0.39 & 0.17 & 4.65    &  & 0.30 & 0.29 & 0.39 & 0.17 & 4.65    &  \\ %\rowcolor{gray}
\multirow{-2}{*}{$0.6$} & std & 0.00 & 0.00 & 0.00 & 0.00 & 0.00  &   \multirow{-2}{*}{1} & 0.00 & 0.00 & 0.00 & 0.00 & 0.00  &   \multirow{-2}{*}{21} \\ 

\hline %\rowcolor{gray}
  %\multicolumn{4}{l}{\textsc{sims}} &    &  &\\
& avg & 0.30 & 0.29 & 0.39 & 0.17 & 4.65    &  & 0.30 & 0.29 & 0.39 & 0.17 & 4.65    & \\ %\rowcolor{gray}
\multirow{-2}{*}{$0.8$} & std & 0.00 & 0.00 & 0.00 & 0.00 & 0.00  &   \multirow{-2}{*}{1} & 0.00 & 0.00 & 0.00 & 0.00 & 0.00  &   \multirow{-2}{*}{27}\\ 
\hline
  %\multicolumn{4}{l}{\textsc{sims}} &    &  &\\
& avg & 0.30 & 0.29 & 0.39 & 0.17 & 4.65    &  & 0.30 & 0.29 & 0.39 & 0.17 & 4.65    &   \\ 
\multirow{-2}{*}{$1.0$} & std & 0.00 & 0.00 & 0.00 & 0.00 & 0.00  &   \multirow{-2}{*}{10} & 0.00 & 0.00 & 0.00 & 0.00 & 0.00  &   \multirow{-2}{*}{29} \\ 
\hline
   % --------
   %
   %
   % --------
 & &  \multicolumn{12}{c}{\textsc{gn with global step}} \\ 
\hline %\rowcolor{gray}
  %\multicolumn{4}{l}{\textsc{sims}} &    &  &\\
& avg & 0.30 & 0.29 & 0.39 & 0.17 & 4.65    & & 0.30 & 0.29 & 0.39 & 0.17 & 4.65    &    \\ %\rowcolor{gray}
\multirow{-2}{*}{$0.1$} & std & 0.00 & 0.00 & 0.00 & 0.00 & 0.00  &   \multirow{-2}{*}{0} & 0.00 & 0.00 & 0.00 & 0.00 & 0.00  &   \multirow{-2}{*}{0} \\ 
\hline %\rowcolor{gray}
  %\multicolumn{4}{l}{\textsc{sims}} &    &  &\\
& avg & 0.30 & 0.29 & 0.39 & 0.17 & 4.65    &  & 0.30 & 0.29 & 0.39 & 0.17 & 4.65    &   \\ %\rowcolor{gray}
\multirow{-2}{*}{$0.2$} & std & 0.00 & 0.00 & 0.00 & 0.00 & 0.00  &   \multirow{-2}{*}{0} & 0.00 & 0.00 & 0.00 & 0.00 & 0.00  &   \multirow{-2}{*}{0} \\ 
\hline %\rowcolor{gray}
  %\multicolumn{4}{l}{\textsc{sims}} &    &  &\\
& avg & 0.30 & 0.29 & 0.39 & 0.17 & 4.65    &  & 0.30 & 0.29 & 0.39 & 0.17 & 4.65    &   \\ %\rowcolor{gray}
\multirow{-2}{*}{$0.4$} & std & 0.00 & 0.00 & 0.00 & 0.00 & 0.00  &   \multirow{-2}{*}{0} & 0.00 & 0.00 & 0.00 & 0.00 & 0.00  &   \multirow{-2}{*}{0} \\ 
\hline %\rowcolor{gray}
  %\multicolumn{4}{l}{\textsc{sims}} &    &  &\\
& avg & 0.30 & 0.29 & 0.39 & 0.17 & 4.65    &   & 0.30 & 0.29 & 0.39 & 0.17 & 4.65    &   \\ %\rowcolor{gray}
\multirow{-2}{*}{$0.6$} & std & 0.00 & 0.00 & 0.00 & 0.00 & 0.00  &   \multirow{-2}{*}{0} & 0.00 & 0.00 & 0.00 & 0.00 & 0.00  &   \multirow{-2}{*}{0} \\ 

\hline %\rowcolor{gray}
  %\multicolumn{4}{l}{\textsc{sims}} &    &  &\\
& avg & 0.30 & 0.29 & 0.39 & 0.17 & 4.65    & & 0.30 & 0.29 & 0.39 & 0.17 & 4.65    &    \\ %\rowcolor{gray}
\multirow{-2}{*}{$0.8$} & std & 0.00 & 0.00 & 0.00 & 0.00 & 0.00  &   \multirow{-2}{*}{0} & 0.00 & 0.00 & 0.00 & 0.00 & 0.00  &   \multirow{-2}{*}{0} \\ 
\hline
  %\multicolumn{4}{l}{\textsc{sims}} &    &  &\\
& avg & 0.30 & 0.29 & 0.39 & 0.17 & 4.65    &  & 0.30 & 0.29 & 0.39 & 0.17 & 4.65    &    \\ 
\multirow{-2}{*}{$1.0$} & std & 0.00 & 0.00 & 0.00 & 0.00 & 0.00  &   \multirow{-2}{*}{0} & 0.00 & 0.00 & 0.00 & 0.00 & 0.00  &   \multirow{-2}{*}{0} \\ 
   \hline
\multicolumn{2}{c|}{lower bound} & 0.05 & 0.01 & -0.95 & 0.01 & - & - & 0.05 & 0.01 & -0.95 & 0.01 & - & -\\
\multicolumn{2}{c|}{upper bound} & 0.99 & 0.90 &  0.95 & 12  & - & - & 0.99 & 0.90 &  0.95 & 12  & - & -\\ \hline \hline
\end{tabular}  } 
  \notes{ \textbf{Legend:} Comparison for 50 starting values.  \textsc{true}: full sample estimate (est). \textsc{gn with global step}: Gauss-Netwon augmented with a global sequence. Both are run for $k = 150$ iterations in total, for all starting values. Objs: avg and std of minimized objective value. \# of crashes: optimization terminated because objective returned error. Lower/upper bound used for the estimation and reparameterization.}
\end{table}

\subsection{Sensitivity of Numerical Derivatives} \label{apx:sensitivity}

In some of the applications, the moments are computed using numerical routines, using e.g. fixed point iterations, which evaluate the moments up to some tolerance level $\eta$. This can affect the optimization as the precision of first and second order numerical derivatives can be sensitive to this approximation. The following gives an brief overview for a scalar moment and parameter. Suppose we can only compute $g_\eta(\theta)$ such that $|g_\eta(\theta)-g(\theta)| \leq \eta$, for all $\theta$. In order to implement a derivative-based optimizer, the derivative $\partial_\theta g(\theta)$ is approximated by finite differences: $G_{\epsilon}(\theta) = \frac{1}{\epsilon} (g(\theta+\epsilon)-g(\theta))$ with some tuning parameter $\epsilon$, the default in R is $\epsilon = 6 \cdot 10^{-6}$. The approximation error for this derivative is at most: $|G(\theta) - G_\epsilon(\theta)| \leq  \epsilon L$ where $L$ is the Lipschitz constant of $G$. Since $g$ itself is not available, a further approximation is needed: $G_{\eta,\epsilon}(\theta) = \frac{1}{2 \epsilon} (g_\eta(\theta+\epsilon)-g_\eta(\theta-\epsilon))$. This has a larger approximation error: $|G_{\eta,\epsilon}(\theta)-G(\theta)| \leq \epsilon L + \frac{\eta}{\epsilon}$. 

In the BLP application, the fixed-point tolerance level is set to $\eta = 10^{-12}$, this yields an approximation error of order $10^{-6}$ for the Jacobian $G_n$, when the inner loop did not terminate because of the limit on the number of iterations (set at $2000$). \textsc{bfgs} further approximates second derivatives using finite differences. The second-order derivative can be computed as: $\partial_\theta G_\epsilon (\theta) = \frac{1}{\epsilon^2}[g(\theta+\epsilon)+g(\theta-\epsilon) - 2 g(\theta)] = \frac{1}{\epsilon}[G_\epsilon(\theta+\epsilon)-G_\epsilon(\theta-\epsilon)]$ which has an approximation error: $|\partial_\theta G_\epsilon (\theta)-\partial_\theta G(\theta)| \leq L_2 \epsilon$, where $L_2$ is the Lipschitz constant of $\partial_\theta G$. Again, since $g$ is not available we need a further approximation error: $\partial_\theta G_{\eta,\epsilon} (\theta) = \frac{1}{\epsilon^2}[g_\eta(\theta+\epsilon)+g_\eta(\theta-\epsilon) - 2 g_\eta(\theta)]$ which has an error of size $|\partial_\theta G_{\eta,\epsilon} (\theta)-\partial_\theta G(\theta)| \leq \epsilon L_2 + \frac{\eta}{\epsilon^2}$. In the BLP application, $\eta = 10^{-12}$ and $\epsilon^{-2} = 1/36 \cdot 10^{12}$ are of the same order of magnitude so that the approximation error, for second-order derivatives is likely to be large.

\section{Additional Material for Section \ref{sec:algos}} \label{apx:optizs}

\subsection{General overview of Algorithms properties}

The following describes three of the algorithms in Table \ref{tab:survey}: Nelder-Mead, Grid Search, Multi-Start, and Simulated Annealing. The goal is to give a brief overview of their known convergence properties; further description for each method is given in Appendix \ref{apx:optizs}.

\paragraph*{Notation:} $Q_n$ is a continuous objective function to be minimized over $\Theta$, a convex and compact subset of $\mathbb{R}^p$, $p \geq 1$, $\hat\theta_n$ denotes the solution to this minimization problem. 

\paragraph{Nelder-Mead.} Also called the simplex algorithm, the \citet[\textsc{nm}]{nelder1965} algorithm comes out as a standard choice for empirical work in our survey. Notably, it was used in \citet[Sec6.5]{Berry1995} to estimate the BLP model for the automobile industry. Its main feature is that it can be used even if $Q_n$ is not continuous. It is often referred to as a \textit{local derivative-free} optimizer. It belongs to the direct search family, which includes pattern search seen in Table \ref{tab:survey} above. 

Despite being widely used, formal convergence results for the simplex algorithm are few. Notably, \citet{lagarias1998} proved convergence for strictly convex continuous functions for $p=1$, and a smaller class of functions for $p=2$ parameters. \citet{mckinnon1998} gave counter-examples for $p=2$ of smooth, strictly convex functions for which the algorithm converges to a point that is neither a local nor a global optimum, i.e. does not satisfy a first-order condition.\footnote{\citet{powell1973} gives additional counter-examples for the class of direct search algorithms which includes \textsc{nm} and Pattern Search.} Using the algorithm once may not produce consistent estimates in well-behaved problems so it is sometimes combined with a multiple starting value strategy, described below. The \textsc{tiktak} Algorithm of \citet{arnoud2019} builds on \textsc{nm} with multiple starting values. Despite these potential limitations, \textsc{nm} remains popular in empirical work. %It is commonly with multiple starting values \citep[e.g.][Sec2.1]{arnoud2019} to increase  chances of convergence. %The choice of starting values is related to grid search discussed below.

\paragraph{Grid-Search.} As the name suggests, a grid-search returns the minimizer of $Q_n$ over a finite grid of points. 
%Unlike NM, a grid search is guaranteed to converge under weak conditions, albeit very slowly. 
In Economics, it is sometimes used to estimate models where the number of parameters $p$ is not too large. One notable example is \citet{donaldson2018}, who estimates $p=3$ non-linear coefficients in a gravity model. 

Contrary to \textsc{nm} above, grid-search has global convergence guarantees. However, convergence is very slow. Suppose we want the minimizer $\tilde \theta_k$ over a grid of $k$ points to satisfy: $Q_n(\tilde \theta_k) - Q_n(\hat\theta_n) \leq \varepsilon$. Then the search requires at least $k \geq C \varepsilon^{-p}$ grid points where $C$ depends on $Q_n$ and the bounds used for the grid. Suppose $C=1$, $p=3$, $\varepsilon = 10^{-2}$, at least $k \geq 10^{6}$ grid points are needed, which is quite large. If each moment evaluation requires 45s, as in \citet{Lise2017}, this translates into 1.5 years of computation time.

\paragraph{Simulated Annealing.} Unlike the methods above, Simulated Annealing (\textsc{sa}) is not a deterministic but a Monte Carlo based optimization method. Along with \textsc{nm}, \textsc{sa} stands out as the standard choice in empirical work. Like the grid-search, \textsc{sa} is guaranteed to converge, with high probability, as the number of iterations increases for an appropriate choice of tuning parameters. The main issue is that tuning parameters for which convergence results have been established result in very slow convergence: $\|\theta_k - \hat\theta_n\| \leq O_p(1/\sqrt{\log[k]}),$ after $k$ iterations. As a result, \textsc{sa} could - in theory - converge more slowly than a grid-search. \citet{chernozhukov2003} consider the frequentist properties of a GMM-based quasi-Bayesian posterior distribution. Draws can be sampled using the random-walk Metropolis-Hastings algorithm, which is closely related to \textsc{sa}.

\paragraph{Multiple Starting Values.} To accommodate some of the limitations of optimizers, especially the lack of global convergence guarantees, it is common to run a given algorithm with multiple starting values. Setting the starting values is similar to choosing a grid for a grid-search. \citet{andrews1997} provides a stopping rule which can be used to determine if sufficiently many starting values were used or not. The required number of starting values depends on the objective function $Q_n$, the choice of the optimizer, and the properties of the sequence used to generate starting values.

\subsection{Implementation of the algorithms}

\paragraph{The Nelder-Mead algorithm.}
The following description of the algorithm is based on \citet[Ch14]{nash1990} which R implements in the optimizer \textit{optim}.  The first step is to build a simplex for the $p$-dimensional parameters, i.e. $p+1$ distinct points $\theta_1,\dots,\theta_{p+1}$ ordered s.t. $Q_n(\theta_1) \leq \dots \leq Q_n(\theta_{p+1})$. The simplex is then transformed at each iteration using four operations called \textit{reflection}, \textit{expansion}, \textit{reduction}, and \textit{contraction}. 
The algorithm also repeatedly computes the centroid $\theta_c$ of the best $p$ points, to do so: take the best $p$ guesses $\theta_1,\dots,\theta_p$ and compute their average: $\theta_c = 1/p \sum_{\ell=1}^p \theta_\ell$. Once this is done, go to step \textbf{R} below.
\begin{tcolorbox}[%enhanced jigsaw,
  colback=white!30,%gray background
  sharp corners,
  %colframe=Maroon,% black frame colour
  %width=25cm,% Use 5cm total width,
  %arc=3mm, auto outer arc,
  %boxrule=5pt,
  %drop shadow={Maroon!50!gray!80}
]  \textbf{Nelder-Mead Algorithm:} \vspace{0.15cm}
 { \small 
\begin{enumerate}
  \item[] \textbf{Inputs:} Initial simplex $\theta_1,\dots,\theta_{p+1}$, parameters $\alpha,\gamma,\beta,\beta^\prime$. NM suggest to use $\alpha=1,\gamma=2,\beta=\beta^\prime = 1/2$.
  \item[] Re-order the points so that $Q_n(\theta_1) \leq \dots \leq Q_n(\theta_{p+1})$, compute the centroid $\theta_c = 1/p \sum_{\ell=1}^p \theta_\ell$ (average of the best $p$ points)
  \item[] Start at \textbf{R} and run until convergence:
  \begin{enumerate}
      \item[\textbf{R}:] The \textit{reflection} step computes $\theta_r = \theta_c + \alpha (\theta_c - \theta_{p+1}) = 2\theta_c - \theta_{p+1}$ for $\alpha=1$. There are now several possibilities: 
    \begin{itemize}
      \item If $Q_n(\theta_r) < Q_n(\theta_1)$ got to step \textbf{E}. 
      \item If $Q_n(\theta_1) \leq Q_n(\theta_r) \leq Q_n(\theta_p)$, replace $\theta_{p+1}$ with $\theta_r$, re-order the points, compute the new $\theta_c$, and do \textbf{R} again. 
      \item By elimination: $Q_n(\theta_r) > Q_n(\theta_p)$.  If $Q_n(\theta_r) < Q_n(\theta_{p+1})$, replace $\theta_{p+1}$ with $\theta_{r}$. Either way, go to step \textbf{R'}.
    \end{itemize}
    \item[\textbf{E}:] The \textit{expansion} step computes $\theta_e = \theta_r + (\gamma-1)(\theta_r-\theta_c) = 2\theta_r - \theta_c$ for $\gamma=2$. If $Q_n(\theta_e) < Q_n(\theta_r)$, then $\theta_e$ replaces $\theta_{p+1}$. Otherwise, $\theta_r$ replaces $\theta_{p+1}$. Once $\theta_{p+1}$ is replaced, re-order the points, compute the new $\theta_c$, and go to \textbf{R}.
    \item[\textbf{R'}:] The \textit{reduction} step computes $\theta_s = \theta_c + \beta(\theta_{p+1}-\theta_c) = (\theta_c + \theta_{p+1})/2$ for $\beta=1/2$. If $Q_n(\theta_s)<Q_n(\theta_{p+1})$, $\theta_s$ replaces $\theta_{p+1}$, then re-order the points, compute the new $\theta_c$, and go to \textbf{R}. Otherwise, go to \textbf{C}.
    \item[\textbf{C}:] The \textit{contraction} step updates $\theta_2,\dots,\theta_{p+1}$ using $\theta_\ell = \theta_1 + \beta^\prime (\theta_\ell - \theta_1) = (\theta_\ell+\theta_1)/2$ for $\beta^\prime = 1/2$. Re-order the points, compute the new $\theta_c$, and go to \textbf{R}.
  \end{enumerate}
\end{enumerate}
}
\end{tcolorbox}
Clearly, the choice of initial simplex can affect the convergence of the algorithm. Typically, one provides a starting value $\theta_1$ and then the software picks the remaining $p$ points of the simplex without user input.  NM proposed their algorithm with statistical estimation in mind, so they considered using the standard deviation $\sqrt{ \sum_{\ell=1}^{n+1} (Q_n(\theta_\ell)-\bar{Q}_n)^2/n } < \text{tol}$ as a convergence criterion, setting $\text{tol} = 10^{-8}$ and $\bar{Q}_n$ the average of $Q_n(\theta_\ell)$ in their application. Here convergence occurs when the simplex collapses around a single point. 

\paragraph{The Grid-Search algorithm.}
The procedure is very simple, pick a grid of $k$ points $\theta_1,\dots,\theta_k$, and compute: \[ \tilde \theta_k = \argmin_{\ell=1,\dots,k} Q_n(\theta_\ell).\] The optimization error $\|\tilde \theta_k-\hat\theta_n\|$ depends on both $k$ and the choice of grid. The following gives an overview of the approximation error and feasible error rates.

For simplicity, suppose that the parameter space is the unit ball in $\mathbb{R}^p$: $\Theta = \mathcal{B}_2^p$, and $Q_n$ is continuous. Under these assumptions, there is an $L \geq 0$ such that $|Q_n(\theta_1) - Q_n(\theta_2)| \leq L \|\theta_1-\theta_2\|$. $L>0$, unless $Q_n$ is constant. This implies: $|Q_n(\tilde \theta_k) - Q_n(\hat\theta_n)| \leq L (\inf_{1 \leq \ell \leq k}\|\theta_\ell-\hat\theta_n\|)$. Suppose we want to ensure $|Q_n(\tilde \theta_k) - Q_n(\hat\theta_n)| \leq \varepsilon$, then we need $\inf_{1 \leq \ell \leq k}\|\theta_\ell-\hat\theta_n\| \leq \varepsilon /L$. Packing arguments give a lower bound for $k$ over all grids, and all possible $\hat\theta_n$: $k \geq \text{vol}( \mathcal{B}_2^p )/\text{vol}( [\varepsilon/L] \mathcal{B}_2^p ) = [\varepsilon/L]^{-p}$, where $\text{vol}$ is the volume. 

For the choice of grid, \citet[Theorem 3]{niederreiter1983} shows that low-discrepancy sequences, e.g. the Sobol or Halton points sets, can  achieve this rate, up to a logarithmic term.\footnote{In comparison, using uniform random draws in a grid search would require $O([\varepsilon/L]^{-2p})$ iterations to achieve the same level of accuracy with high-probability. \citet[Ch3.1]{fang1993} give a review of these results.} This is indeed a common choice for multi-start and grid search optimization.

In practice, $Q_n(\tilde \theta_k) - Q_n(\hat\theta_n)$ is typically not the quantity of interest for empirical estimations, rather we are interested in $\|\tilde \theta_k - \hat\theta_n\|$. Suppose, in addition, that $\hat\theta_n \in \text{int}(\Theta)$, and $Q_n$ is twice continuously differentiable with positive definite Hessian $H_n(\hat\theta_n)$, a local identification condition. Then there exists $0<\underline{\lambda} \leq \overline{\lambda} < \infty$ and $\varepsilon_1 >0$ s.t. $\|\theta-\hat\theta_n\| \leq \varepsilon_1$ implies: 
\begin{align} \underline{\lambda} \|\theta-\hat\theta_n\|^2 \leq Q_n(\theta)-Q_n(\hat\theta_n) \leq \overline{\lambda} \|\theta-\hat\theta_n\|^2, \label{eq:local_str_cvx}\end{align}
i.e. $Q_n$ is locally strictly convex.\footnote{The three $\varepsilon_1,\underline{\lambda},\overline{\lambda}$ only depend on $H_n(\cdot)$.} If $\hat\theta_n$ is the unique minimizer of $Q_n$, there is a $0 < \varepsilon_2\leq \varepsilon_1$ such that $\inf_{\|\theta-\hat\theta_n\| \geq \varepsilon_1 } Q_n(\theta) > Q_n(\hat\theta_n) + \overline{\lambda}\varepsilon_2^2$, using a global identification condition. Now, by local identification: $\|\theta-\hat\theta_n\| \leq \varepsilon_2 \Rightarrow Q_n(\theta) \leq Q_n(\hat\theta_n) + \overline{\lambda}\varepsilon_2^2 < \inf_{\|\theta-\hat\theta_n\| \geq \varepsilon_1 } Q_n(\theta)$. As soon as $k \geq k_0$ where $\inf_{1 \leq \ell \leq k_0}\|\theta_\ell-\hat\theta_n\| \leq \varepsilon_2$, we have $\|\tilde \theta_k - \hat\theta_n\| \leq \varepsilon_1$. Then, for any $k \geq k_0$: $\underline{\lambda}\|\tilde\theta_k - \hat\theta_n\|^2 \leq Q_n(\tilde \theta_k) - Q_n(\hat\theta_n) \leq \overline{\lambda} (\inf_{1\leq\ell\leq k} \|\theta_\ell - \hat\theta_n\|^2)$ and $\|\tilde\theta_k - \hat\theta_n\| \leq [\overline{\lambda}/\underline{\lambda}]^{1/2} (\inf_{1\leq\ell\leq k} \|\theta_\ell - \hat\theta_n\|).$ 

This reveals the interplay between the identification conditions and the optimization error. The best value $\tilde \theta_k$ is only guaranteed to be near $\hat\theta_n$ when $k \geq \varepsilon_2^{-p}$ iterations (using packing arguments for the unit ball), where $\varepsilon_2$ depends on the global identification condition. Local convergence depends on the ratio $\overline{\lambda}/\underline{\lambda} \geq 1$ which is infinite when $H_n(\hat\theta_n)$ is singular. The main drawback of a grid search is its slow convergence. To illustrate, \citet[pp3443-3445]{colacito2018} estimate $p=5$ parameters using a grid search with $k=1551$ points. For simplicity, suppose $\overline{\lambda}/\underline{\lambda} = 1$, $k_0 < k$, and $\Theta = \mathcal{B}_2^p$, the unit ball, then the worst-case optimization error is $\sup_{\hat\theta_n \in \Theta}(\inf_{1\leq\ell\leq k} \|\theta_\ell - \hat\theta_n\|) \geq k^{-1/p} \simeq 0.23$. This is ten times larger than all but one of the standard errors reported in the paper.

%To stress the important of the lower bound in (\ref{eq:local_str_cvx}), suppose that the Hessian $H_n(\hat\theta_n)$ is now singular and that for some $\kappa>1$: $\underline{\lambda}\|\theta-\hat\theta_n\|^{2 \kappa} \leq Q_n(\theta)-Q_n(\hat\theta_n) \leq \overline{\lambda}\|\theta-\hat\theta_n\|^{2}$ when $\|\theta-\hat\theta_n\|\leq \varepsilon_1$.\footnote{If $H_n(\hat\theta_n)$ is singular, the lower bound with $\kappa =1$ only holds for $\underline{\lambda}=0$. Consider for example: $Q_n(\theta) = |\theta_1-\hat\theta_{1n}|^4 + |\theta_2-\hat\theta_{2n}|^2$ with $\theta = (\theta_1,\theta_2)$, and set $\kappa=2$ and $\varepsilon_1 \leq 1$.} Using the same calculations as above: $\|\tilde\theta_k - \hat\theta_n\| \leq [\overline{\lambda}/\underline{\lambda}]^{1/[2 \kappa]} (\inf_{1\leq\ell\leq k} \|\theta_\ell - \hat\theta_n\|^{1/\kappa})$, for $k \geq k_0$. While $Q_n(\tilde \theta_k)$ converges at the rate given above, $\tilde\theta_k$ converges at a slower rate when the objective is not locally strictly convex, i.e. when (\ref{eq:local_str_cvx}) does not hold. 

\paragraph{Simulated Annealing.}
Implementations can vary across software, the following will focus on the implementation used in R's \textit{optim} function.  

\begin{tcolorbox}[%enhanced jigsaw,
  colback=white!30,%gray background
  sharp corners,
  %colframe=Maroon,% black frame colour
  %width=25cm,% Use 5cm total width,
  %arc=3mm, auto outer arc,
  %boxrule=5pt,
  %drop shadow={Maroon!50!gray!80}
]  \textbf{Simulated Annealing Algorithm:} \vspace{0.15cm}
 { \small 
\begin{enumerate}
  \item[] \textbf{Inputs:} Starting value $\theta_1 \in \Theta$, temperature schedule $\infty > T_2 \geq T_3 \geq \dots >0$, a sequence $\infty > \eta_2 \geq \eta_3 \geq \dots >0$, and maximum number of iterations $k$. Common choice: $T_\ell = T_1/\log(\ell)$ for $\ell \geq 2$ and $\eta_\ell$ proportional to $T_\ell$.
  \item[] For $\ell \in \{2,\dots,k\}$, repeat:
  \begin{enumerate}
    \item[1.] Draw $\theta^\star \sim \mathcal{N}(\theta_{\ell-1},\eta_\ell I_d)$, and $u_\ell \sim \mathcal{U}_{[0,1]}$
    \item[2.] Set $\theta_\ell = \theta^\star$ if $u_\ell \leq \exp(-[Q_n(\theta^\star)-Q_n(\theta_{\ell-1})]/T_\ell)$, otherwise set $\theta_\ell = \theta_{\ell-1}$
  \end{enumerate}
  \item[] \textbf{Output:} Return $\tilde \theta_{k} = \argmin_{1 \leq \ell \leq k} Q_n(\theta_\ell)$
\end{enumerate}
}
\end{tcolorbox}
The implementation described above relies on the random-walk Metropolis update. Notice that if $Q_n(\theta^\star) \leq Q_n(\theta_{\ell-1})$, the exponential term in step 2 is greater than $1$ and $\theta^\star$ is always accepted as the next $\theta_\ell$, regardless of $u_\ell$. \citet{belisle1992} gave sufficient condition for $\tilde \theta_k \overset{a.s.}{\to} \hat\theta_n$ when $k \to \infty$ and $Q_n$ is continuous. In practice, the performance of the Algorithm can be measured by its convergence rate. To get some intuition, we give some simplified derivations below which highlight the role of $T_k$ and several quantities which appeared in our discussion of the grid search. 

First, notice that for each $k$, steps 1-2 implement the Metropolis algorithm also used for Bayesian inference using random-walk Metropolis-Hastings. The invariant distribution of these two steps is: \[ f_k(\theta) = \frac{\exp(-[Q_n(\theta)-Q_n(\hat\theta_n)]/T_k)}{\int_{\Theta} \exp(-[Q_n(\theta)-Q_n(\hat\theta_n)]/T_k) d\theta},\]
this is called the Gibbs-Boltzmann distribution. When $T_\infty = +\infty$, $f_\infty$ puts all the probability mass on the unique minimum $\hat\theta_n$. To build intuition, suppose that $k \geq 1$: $\theta_k \sim f_k$. Because SA is a stochastic algorithm, the approximation error $\|\theta_k - \hat\theta_n\|$ is random, but can be quantified using $\mathbb{P}(\|\theta_k - \hat\theta_n\| \geq \varepsilon)$. In the following we will assume the temperature schedule to be $T_k = T_1/\log(k)$, as implemented in R. 

The following relies on the same setting, notation and assumptions as the grid search above. First, we can bound the probability that $\theta_k$ is outside the $\varepsilon_1$-local neighborhood of $\hat\theta_n$ where $Q_n$ is approximately quadratic: $\mathbb{P}(\|\theta_k - \hat\theta_n\| \geq \varepsilon_1)$. Using the global identification condition: \[ \exp(-[Q_n(\theta)-Q_n(\hat\theta_n)]/T_k) \leq \exp( -\overline{\lambda}\varepsilon_2^2/T_k ) = k^{- \overline{\lambda}\varepsilon_2^2/T_1 }, \text{ if } \|\theta-\hat\theta_n\| \geq \varepsilon_1,\]
where $\varepsilon_1,$ $\varepsilon_2$ were defined in the grid search section above.  
This gives an upper bound for the numerator in $f_k(\theta_k)$. A lower bound is also required for the denominator. Using (\ref{eq:local_str_cvx}) and the change of variable $\theta = \hat\theta_n + \sqrt{T_k}h$, we have:
\[ \exp(-\overline{\lambda}\|h\|^2) \leq \exp(-[Q_n(\hat\theta_n + \sqrt{T_k}h)-Q_n(\hat\theta_n)]/T_k) \leq \exp(-\underline{\lambda}\|h\|^2), \text{ if } \|\sqrt{T_k}h\| \leq \varepsilon_1.\]
Suppose $T_k \leq \varepsilon_1^2$, the two inequalities give us the bound:
\[ \mathbb{P}(\|\theta_k - \hat\theta_n\| \geq \varepsilon_1) \leq \frac{k^{- \overline{\lambda}\varepsilon_2^2/T_1} \text{vol}(\Theta) }{ |T_k|^{p/2} \int_{\|h\| \leq 1} \exp(-\overline{\lambda}\|h\|^2) dh } = C [\log(k)]^{d/2} k^{-\overline{\lambda}\varepsilon_2^2/T_1}. \]
This upper bound declines more slowly than for the grid search when $\overline{\lambda}\varepsilon_2^2/T_1 < 1/p$, which can be the case if $T_1$ large and/or $\varepsilon_2$ is small. 
For the lower bound, pick any $\varepsilon \in (0, \varepsilon_1/\sqrt{T_k}) $:
\[ \mathbb{P}(\|\theta_k - \hat\theta_n\| \leq \sqrt{T_k} \varepsilon ) \geq \frac{ \int_{\|h\| \leq \varepsilon } \exp(-\overline{\lambda}\|h\|^2)dh }{ \int_{\|h\| \in \mathbb{R} } \exp(-\underline{\lambda}\|h\|^2)dh + |T_k|^{-p/2} \text{vol}(\Theta) k^{-\overline{\lambda}\varepsilon_2^2/T_1}}, \]
which has a strictly positive limit. This implies that $\sqrt{\log(k)}\|\theta_k - \hat\theta_n\| \geq O_p(1)$, since $T_k = T_1/\log(k)$. This $\sqrt{\log(k)}$ rate is slower than the grid search. To get faster convergence, some authors have suggested using $T_k = T_1/k$ and, by default, Matlab sets $T_k = T_1 \cdot 0.95^k$. However,  theoretical guarantees to have $\theta_k \overset{p}{\to} \hat\theta_n$, as $k \to \infty$ are only available when $T_k= T_1/\log(k)$.\footnote{See \citet[Ch8.4-8.6]{spall2005} for additional details and references.}

\end{appendices}
\end{document}